%% file: main_v2.tex
\documentclass[12pt,a4paper]{scrartcl}
\pdfoutput=1

\usepackage[utf8]{inputenc}
\usepackage[T1]{fontenc}
\usepackage{booktabs}
\usepackage{amsmath,amssymb,amsbsy,amstext, amsthm, amsfonts,mathtools}
\usepackage{slashed}
\usepackage{physics}
\usepackage{graphicx}
\usepackage{booktabs,multicol,multirow}
\usepackage[shortlabels]{enumitem}
\usepackage{hyperref}
\usepackage[capitalize]{cleveref}
\usepackage{bm}
\usepackage{tikz}
\usepackage{xspace}
\usepackage[noblocks]{authblk}
\usepackage{xcolor}
\usepackage{soul}
\usepackage[textsize=tiny]{todonotes}
\usepackage{subcaption}
\usepackage{bbold}
\usepackage{ulem}
\usepackage{comment}
\usepackage{nicefrac}
\numberwithin{equation}{section}

\allowdisplaybreaks

\DeclareMathOperator*{\sumint}{%
\mathchoice%
  {\ooalign{$\displaystyle\sum$\cr\hidewidth$\displaystyle\int$\hidewidth\cr}}
  {\ooalign{\raisebox{.14\height}{\scalebox{.7}{$\textstyle\sum$}}\cr\hidewidth$\textstyle\int$\hidewidth\cr}}
  {\ooalign{\raisebox{.2\height}{\scalebox{.6}{$\scriptstyle\sum$}}\cr$\scriptstyle\int$\cr}}
  {\ooalign{\raisebox{.2\height}{\scalebox{.6}{$\scriptstyle\sum$}}\cr$\scriptstyle\int$\cr}}
}

\newcommand*{\ctikz}[2][]{\hbox{\mathsurround=3pt$\vcenter{\hbox{\tikz[#1]{#2}}}$}}
\usetikzlibrary{arrows.meta}

\usepackage[sorting=none,%
  citestyle=numeric-comp,%
  bibstyle=mynumeric,%
  giveninits=true]{biblatex}
\input{abbr.tex}

\addtokomafont{disposition}{\rmfamily}

\title{\Large Improved Thermal Resummation \\ for Multi-Field Potentials}
\author[1]{Henning Bahl\footnote{\href{mailto:bahl@thphys.uni-heidelberg.de}{bahl@thphys.uni-heidelberg.de}}}
\author[2,3,4]{Marcela Carena\footnote{\href{mailto:carena@fnal.gov}{carena@fnal.gov}}}
\author[2]{\\Aurora~Ireland\footnote{\href{mailto:anireland@uchicago.edu}{anireland@uchicago.edu}}}
\author[2,4,5]{Carlos E.M. Wagner\footnote{\href{mailto:cwagner@uchicago.edu}{cwagner@uchicago.edu}}}
\affil[1]{Institut für Theoretische Physik, Universität Heidelberg, Philosophenweg 16, 61920~Heidelberg, Germany} 
\affil[2]{Department of Physics and Enrico Fermi Institute, University of Chicago, 5720~South~Ellis~Avenue, Chicago, IL~60637~USA}
\affil[3]{Fermi National Accelerator Laboratory, P.O. Box 500, Batavia, Illinois, 60510, USA}
\affil[4]{Kavli Institute for Cosmological Physics, 5640 South Ellis Avenue, University of Chicago, Chicago, IL 60637}
\affil[5]{HEP Division, Argonne National Laboratory, 9700 Cass Ave., Argonne, IL 60439, USA}
\date{}
\titlehead{\hfill \parbox{5cm}{\vspace{-2cm}\begin{flushright} \texttt{EFI 24-2} \\ \texttt{FERMILAB-PUB-24-0040-T} \end{flushright}}}

\begin{document}
\maketitle

\begin{abstract}\noindent
    The resummation of large thermal corrections to the effective potential is mandatory for the accurate prediction of phase transitions. We discuss the accuracy of different prescriptions to perform this resummation at the one- and two-loop level and point out conceptual issues that appear when using a high-temperature expansion at the two-loop level. Moreover, we show how a particular prescription called partial dressing, which does not rely on a high-temperature expansion, consistently avoids these issues. We introduce a novel technique to apply this resummation method to the case of multiple mixing fields. Our approach significantly extends the range of applicability of the partial dressing prescription, making it suitable for phenomenological studies of beyond the Standard Model extensions of the Higgs sector.
\end{abstract}
\setcounter{footnote}{0}

\newpage

\tableofcontents

\newpage


\section{Introduction}\label{sec:intro}

After the discovery of the Higgs boson a little over ten years ago~\cite{ATLAS:2012yve,CMS:2012qbp}, it is one of the main goals of the Large Hadron Collider (LHC) as well as potential future colliders~\cite{Papaefstathiou:2020iag,Ramsey-Musolf:2019lsf,Curtin:2014jma} to work towards understanding the dynamics of electroweak (EW) symmetry breaking. Besides collider experiments, gravitational wave observatories~\cite{Punturo_2010,Yagi:2011wg,Caprini:2019egz,Sesana:2019vho,Hild:2010id,AEDGE:2019nxb} will start to probe the EW phase transition within the next decades. This research could provide substantial insights into the thermal history of the Universe, which is modified in many extensions of the Standard Model (SM) of particle physics. This, in particular, includes extensions of the SM Higgs sector, which may lead to a first order electroweak phase transition, which would enable the scenario of electroweak baryogenesis~\cite{Cohen:1993nk, Riotto:1999yt, Morrissey:2012db, Wagner:2023vqw} or to
phenomena like vacuum trapping, inverse symmetry breaking, or EW symmetry non-restoration (EWSNR)~\cite{Weinberg:1974hy,Meade:2018,Servant:2018,Cline:1999wi,Baum:2020vfl,Biekotter:2021ysx,Biekotter:2022kgf,Ireland:2022quc,Chang:2022psj,Aoki:2023}. In addition to the electroweak sector, phase transitions can also appear outside the electroweak sector in many BSM scenarios, including hidden sectors~\cite{Schwaller:2015tja,Baldes:2018emh,Breitbach:2018ddu,Croon:2018erz,Hall:2019ank,Baldes:2017rcu,Geller:2018mwu,Croon:2019rqu,Hall:2019rld,Chao:2020adk,Dent:2022bcd} and high-scale models (e.g., grand-unified theories)~\cite{Huang:2017laj,Croon:2018kqn,Hashino:2018zsi,Brdar:2019fur}.

In order to fully exploit the available and incoming experimental data, precise theoretical predictions are crucial. In the context of phase transitions, this necessitates an accurate\footnote{See Ref.~\cite{Croon:2020} for a comprehensive overview of theoretical uncertainties in perturbative calculations of fist-order phase transitions.} determination of the finite temperature effective potential, calculated using perturbation theory. It is well known that for high temperatures large corrections occur which exacerbate the behaviour of the perturbative expansion. To resum these large corrections, various resummation schemes have been developed in the literature~\cite{Parwani:1992,ArnoldEspinosa:1993,Pilaftsis:2013xna,Pilaftsis:2015cka,Funakubo:2023cyv,Funakubo:2023eic}. These include schemes employing a diagrammatic expansion --- most notably the Parwani~\cite{Parwani:1992} and Arnold-Espinosa~\cite{ArnoldEspinosa:1993} resummation schemes --- as well as schemes which involve solving the gap equation --- the full dressing (FD) and partial dressing (PD) procedures \cite{Boyd:1993,Curtin:2016,Curtin:2022}. Moreover, the large thermal corrections can also be resummed in an effective field theory (EFT) framework reducing the spacetime dimensions from four to three, an approach called dimensional reduction (DR)~\cite{Farakos:1994kx,Kajantie:1995dw,Braaten:1995cm}.

While the Parwani and Arnold-Espinosa schemes have the benefit of being easy to implement --- explaining their widespread application in the literature ---, they on the other hand suffer from several disadvantages. First, they only allow to resum the leading thermal corrections. In many scenarios, the resummation of subleading thermal corrections is, however, also important for an accurate prediction. Moreover, they intrinsically rely on the high-temperature expansion for calculating the thermal masses and the thermal counterterms (see detailed discussion below). This implies that their accuracy is questionable in the regime where the temperature is close to the order of the relevant masses, which is exactly the interesting region for phase transitions (see \ccite{Curtin:2016} for a more detailed discussion).

Similar issues appear in the DR approach. While DR offers a conceptionally well-defined and systematic way to resum large thermal corrections, it is technically challenging (for recent steps towards automation of the required calculation, see \ccite{Ekstedt:2022bff}). Moreover, as an EFT approach, it intrinsically relies on the separation of the scales. This means that it is difficult to go beyond the high-temperature expansion and that each EFT is only applicable to a pre-defined hierarchy of scales. This makes DR in particular unsuited for parameter scans in BSM models for which many different hierarchies of the masses and the temperature occur --- requiring to work a different EFT for each hierarchy of scales.

Partial dressing promises to resolve the issues of the FD and DR approaches. It includes subleading thermal corrections and can be easily applied beyond the high-temperature expansion making it particularly suited for studying phase transitions. It has been applied to the singlet-extended SM~\cite{Curtin:2016}. Its accuracy and perturbative convergence have been studied recently in~\ccite{Curtin:2022}.

In the first part of the present work, we present a detailed comparison of partial dressing with the Arnold-Espinosa and Parwani resummation schemes. Afterwards, we discuss the application of PD to the phenomenon of EWSNR. EWSNR occurs if the thermal corrections to a particle mass dominate over the tree-level mass turning the overall thermal mass squared negative. Thus, thermal corrections are by definition large and large differences between the FD schemes have been found in the literature~\cite{Biekotter:2021ysx,Biekotter:2022kgf}. In this paper, we extend this comparison to the two-loop level finding unphysical predictions originating from large imaginary contributions to the effective potential. We demonstrate that these issues do not occur if PD is used leading to a more reliable prediction.

In the last part of this paper, we discuss the case of mixing scalar fields. So far, the PD approach is restricted to scenarios in which only one scalar takes a non-zero value effectively forbidding the description of models with mixing scalars. This strongly limits the applicability of PD for phenomenological studies. We demonstrate a new method to extend PD for scenarios with more than one non-zero scalar field.

Our paper is structured as follows. In \cref{sec:resummation}, we review the need for thermal resummation. Then, we discuss resummation in non-mixing one- and two-field models in \cref{sec:single_field}. In \cref{sec:servant_model}, we compare the FD and PD models for a toy model for EWSNR. In \cref{sec:multi_field_mixing}, we demonstrate the application of PD to models with mixing scalar fields. We provide conclusions in \cref{sec:conclusions}.

\section{Perturbative Breakdown \& Thermal Resummation}
\label{sec:resummation}

Bosonic field theories at finite temperature suffer from various issues in the infrared, principal among them being that the usual perturbation expansion breaks down. To demonstrate how this breakdown comes about, consider a simple $\phi^4$ theory with quartic self-coupling $\lambda$. Working in the imaginary-time formalism \cite{Kapusta:2006}, the 1-loop correction to the bosonic propagator corresponds to the expression
\begin{equation}
    \ctikz{\draw(0,0)--(1,0); \draw(0.5,0.3)circle(0.3);} = \frac{\lambda}{2} \mathcal{I}[m] \equiv \frac{\lambda}{2} \, T \sum_{\omega_n} \int \frac{d^3k}{(2\pi)^3} \frac{1}{K^2+m^2} \,,
\end{equation}
where $K=(\omega_n,\vec{k})$ is the Euclidean four-momentum and $\omega_n = 2 \pi n T$ is the bosonic Matsubara frequency. There are two useful ways to decompose this expression. One option would be to split $\mathcal{I}$ into a zero-temperature piece $\mathcal{I}_{0}[m]$ and a finite-temperature piece $\mathcal{I}_T[m]$ \cite{Laine:2016},
\begin{equation}
    \mathcal{I}[m] = \underbrace{\int \frac{d^4 k}{(2\pi)^4} \frac{1}{k^2 + m^2}}_{\mathcal{I}_{0}[m]} + \underbrace{\int \frac{d^3k}{(2\pi)^3} \frac{1}{E_k} \frac{1}{e^{E_k/T} - 1}}_{\mathcal{I}_{T}[m]} \,,
\end{equation}
where $E_k^2 = \vec{k}^{\,2} + m^2$. The zero-temperature $\mathcal{I}_{0}[m]$ is UV-divergent, and we choose to regularize it using dimensional regularization\footnote{Note that for convenience we display all equations with $\epsilon \rightarrow 0$, such that $D = 4 - 2 \epsilon \rightarrow 4$.} and the $\overline{\text{MS}}$-scheme with renormalization scale $\mu_R$. Meanwhile the finite-temperature piece $\mathcal{I}_T[m]$ is UV-finite, but sensitive to the IR. A more convenient decomposition to reveal this would be to split the Matsubara sum appearing in $\mathcal{I}[m]$ into a ``soft'' zero-mode piece with $\omega_n = 0$ and a ``hard'' non-zero mode piece with $\omega_n \neq 0$, 
\begin{equation}
    \mathcal{I}[m] = \underbrace{T \int \frac{d^3 k}{(2\pi)^3} \frac{1}{\vec{k}^{\,2} + m^2}}_{\mathcal{I}_{\text{soft}}[m]} + \underbrace{T \sum_{n \neq 0} \int \frac{d^3 k}{(2\pi)^3} \frac{1}{\omega_n^2 + \vec{k}^{\,2} + m^2}}_{\mathcal{I}_{\text{hard}}[m]} \,.
\end{equation}
Working in the high-temperature limit $m/T \sim \lambda \ll 1$, the zero-mode contribution can be evaluated explicitly as \cite{Laine:2016}
\begin{equation}
    \mathcal{I}_{\text{soft}}[m] = - \frac{1}{4\pi} m T \,,
\end{equation}
while the hard modes give a contribution 
\begin{equation}
    \mathcal{I}_{\text{hard}}[m] = \frac{T^2}{12} - \frac{m^2}{16\pi^2} \left( \frac{1}{\epsilon} + \ln \left( \frac{\mu_R^2 e^{2 \gamma_E}}{16\pi^2 T^2} \right) \right) + \frac{\zeta(3)}{128 \pi^4} \frac{m^4}{T^2} + \mathcal{O}\left( \frac{m^6}{T^4} \right) \,.
\end{equation}
Thus we see that in this limit, the correction to the zero-mode mass scales like $\delta m_{\text{soft}}^2 \sim m T \sim \lambda T^2$ while the mass correction for the hard modes scale like $\delta m_{\text{hard}}^2 \sim T^2$ in the high-$T$ limit. Since the latter is parametrically larger, excitations of non-zero modes in the thermal plasma will act to screen the zero mode.

\begin{figure}
\centering
\includegraphics[width=0.25\textwidth]{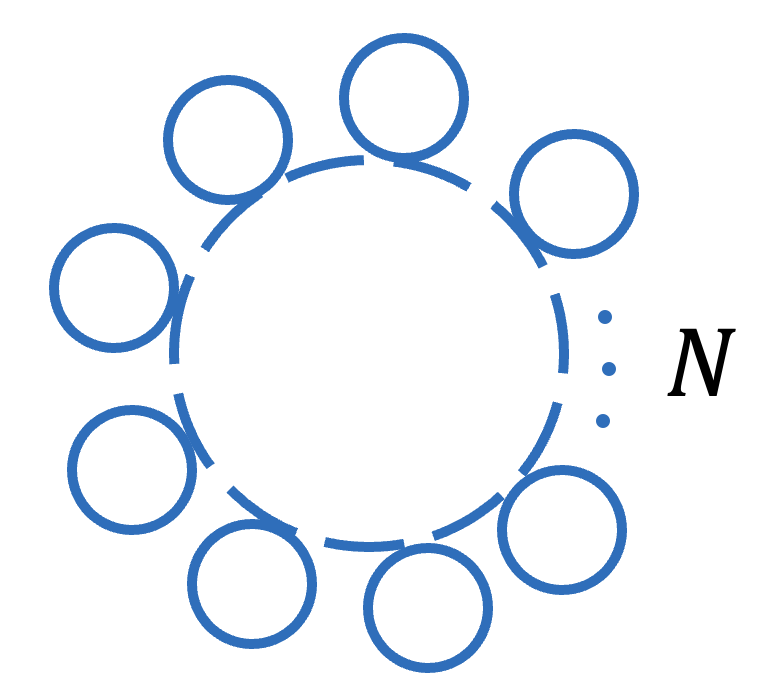}
\hspace{2mm}
\caption{A daisy diagram featuring a zero-mode inner loop (dashed) surrounded by $N$ hard external loops (solid).}
\label{fig:daisy}
\end{figure}

The IR problem manifests when considering higher loop ``daisy diagrams'', like that shown in \cref{fig:daisy}.
This diagram features a zero-mode inner loop (dashed line) surrounded by $N$ hard outer loops (solid lines). Ignoring the overall symmetry factor, the contribution to the effective potential coming from such a diagram is
\begin{equation}
    V_{N}^{\text{daisy}} \sim \left( T \int \frac{d^3 k}{(2\pi)^3} \frac{1}{(\vec{k}^2 + m^2)^N} \right) \left( \lambda T \sum_{n \neq 0} \int \frac{d^3 k}{(2\pi)^3} \frac{1}{\omega_n^2 + \vec{k}^2 + m^2} \right)^N \,,
\end{equation}
where the quantity in the first parenthesis comes from the $N$ soft propagators in the inner loop and scales as $m^{3-2N}T$ while that in the second parenthesis comes from the $N$ hard external loops and scales as $\lambda^N T^{2N}$. The result is
\begin{equation}\label{Vdaisyscaling}
\begin{split}
    V_{N}^{\text{daisy}} \sim \big( m^{3-2N}T \big) \big( \lambda^N T^{2N}\big) = m^3 T \left( \frac{\lambda T^2}{m^2} \right)^N \,.
\end{split}
\end{equation}
There are two potential issues here. First comparing with the contribution from an $(N+1)$-loop daisy diagram, we see that each new hard thermal loop comes at a cost 
\begin{equation}
    \alpha \equiv \frac{V_{N+1}^{\text{daisy}}}{V_{N}^{\text{daisy}}} = \frac{\lambda T^2}{m^2} \,.
\end{equation}

The issue is that this expansion parameter $\alpha$ is not parametrically small at all times. In particular when the system exhibits a phase transition, the expansion parameter becomes $\mathcal{O}(1)$ at the critical temperature, since here $m^2 \sim \lambda T^2$. This signals a breakdown of the perturbative expansion; diagrams which formally appear to be higher order may actually contribute with a magnitude equal to formally lower-order diagrams due to the contributions from these hard thermal loops. Intuitively, this perturbative breakdown occurs because at high temperatures, IR bosonic modes become highly occupied, leading to an enhanced expansion parameter \cite{Linde:1980}. 

The second issue occurs for fields with vanishing mass, for which daisy diagrams with $N \geq 2$ are IR divergent. Clearly this divergence is not physical, as the thermodynamic properties of a plasma of weakly interacting massless bosons (such as photons) are observed to be finite, and so these divergent contributions must cancel amongst one another when all terms in the expansion are summed. This is just another way to see that at finite temperature, the fixed-order perturbative expansion fails. 

A natural solution\footnote{There are also alternative strategies to thermal resummation, such as dimensional reduction \cite{Kajantie:1995,Schicho:2021}, which we will not review here. Another option would be to just treat the problem non-perturbatively, using appropriate lattice techniques.}, then, would be to reorganize the expansion using a new parameter in terms of which the series is convergent --- \textit{thermal resummation}. Consider, for example, how one would resum the daisy diagrams of Fig.~\ref{fig:daisy}. Computing the contribution from an $N$-loop daisy diagram while more carefully keeping track of the combinatorial factors, we would find
\begin{equation}
    V_N^{\text{daisy}} = - \frac{T}{12 \pi} \frac{1}{N!} \left( \frac{\lambda T^2}{4} \right)^N \left( \frac{d}{dm^2} \right)^N m^3 \,.
\end{equation}
One can check that this correctly reproduces the scaling of Eq.~(\ref{Vdaisyscaling}) by using the fact that $m^{3-2N} = \frac{4\sqrt{\pi}}{3} \frac{(-1)^N}{(N-1)!} \frac{\Gamma(N)}{\Gamma(N-3/2)} \left( \frac{d}{dm^2} \right)^N m^3$. If we now sum over all such diagrams, we find
\begin{equation}
\begin{split}
    \sum_{N=0}^\infty V_N^{\text{daisy}} & = - \frac{T}{12\pi} \sum_{N=0}^\infty \frac{1}{N!} \left( \frac{\lambda T^2}{4} \frac{d}{dm^2} \right)^N m^3 \\
    & = - \frac{T}{12\pi} \exp \left( \frac{\lambda T^2}{4} \frac{d}{dm^2} \right) m^3 \,.
\end{split}
\end{equation}
Finally letting $x=m^2$ and noting that $\exp\left(c \frac{d}{dx}\right)f(x) = f(x+c)$, we find
\begin{equation}
    \sum_{N=0}^\infty V_N^{\text{daisy}} = - \frac{T}{12 \pi} \left( m^2 + \frac{\lambda}{4} T^2 \right)^{3/2} \,.
\end{equation}
This is a rather significant result; after summing all contributions, we find an expression where the limit $m^2 \rightarrow 0$ can meaningfully be taken without running into IR divergences. 

The quantity appearing in parenthesis is the thermally corrected mass
\begin{equation}
    \overline{M}^2(\phi,T) \equiv m^2(\phi) + \frac{\lambda}{4} T^2 \,,
\end{equation}
and to leading order, daisy resummation amounts to replacing instances of $m^2$ in the effective potential with the thermally corrected version $\overline{M}^2$. Of course, daisy diagrams are not the only problematic diagrams that appear in finite temperature field theory; there are also so-called ``super-daisy'' diagrams as well as other sub-leading diagrams which demonstrate IR-sensitivity and so should be resummed. Given the questions of which class of diagrams to resum and how to re-order the expansion, there exist several different prescriptions for implementing thermal resummation. 

Historically, the most popular methods employing a diagrammatic approach to thermal resummation are the Parwani \cite{Parwani:1992} and Arnold Espinosa \cite{ArnoldEspinosa:1993} schemes. The major conceptual difference between these methods is that in the Parwani prescription, all modes are resummed, while in the Arnold Espinosa prescription, only the problematic Matsubara zero-modes are resummed. On a technical level, this is equivalent to whether the thermally corrected mass $\overline{M}^2$ is substituted into all terms of the effective potential or only those non-analytic in $m^2$, which can be shown to correspond to zero-modes. Because the thermally corrected mass is included in different terms, when working to fixed order in perturbation theory these methods feature different higher-order terms, resulting in different degrees of convergence.

In addition to these diagrammatic approaches to thermal resummation, which in complicated theories quickly become impractical at higher-loop order, there is also a non-diagrammatic method which we will refer to as \textit{gap resummation} \cite{Espinosa:1992,Espinosa:1993,Quiros:1992}. Rather than computing the contributions from many higher loop diagrams, which quickly becomes impractical in complicated theories, in gap resummation one need simply compute the one-loop effective potential $V_\text{eff}^{(1)}$ and then solve the gap equation for the thermally corrected mass. This gap equation includes the dominant contributions from many\footnote{Notably, contributions from superdaisy diagrams and other sub-leading diagrams like the bosonic sunset are not automatically included.} higher-order diagrams, in particular daisy diagrams to all orders in the effective potential. In a theory with $i=1...n$ bosonic species $\{ \phi_i \}$, the gap equation for the thermal mass of species $i$ reads
\begin{equation}
    M_i^2 = \frac{\partial^2}{\partial \phi_i^2} V_\text{eff}^{(1)}(M_j^2) \,,
\end{equation}
where masses appearing in the effective potential on the right-hand side are the thermally corrected masses for all species in the plasma $M_j^2$. Because the thermal mass appears on both sides of this equation, it must generally be solved numerically. For convenience, it is common to 
solve the truncated gap equation
\begin{equation}\label{eq:M2at1L}
     \left. M_i^2 \right|_\text{trunc.} = \frac{\partial^2}{\partial \phi_i^2} V_\text{eff}^{(1)}(m_j^2) \,,
\end{equation}
where now the right hand side is evaluated on the field dependent effective masses $m_j^2$. When combined with the high temperature expansion, this truncated treatment only resums the leading order hard thermal loops, and results in a solution of the form $M_i^2|_\text{trunc.} \overset{\text{high-}T}{=} m_i^2 + c T^2 \equiv \overline{M}_i^2$, with $c$ some constant function of the couplings.

After solving the gap equation, one usually proceeds by replacing the background field dependent masses $m_i^2(\phi_j)$ with the thermally corrected versions $M_i^2(\phi_j,T)$ in the effective potential $V_{\text{eff}}^{(1)}$. This prescription is called \textit{full dressing} (FD), or \textit{truncated full dressing} (TFD) if one uses the thermal mass obtained by solving the truncated gap equation, and diagrammatically it corresponds to dressing both the propagator and vertex in 1-loop tadpole diagrams. 

Interestingly at the 1-loop level, the Arnold-Espinosa and Parwani prescriptions coincide with special cases of truncated gap resummation. As we will soon see, the 1-loop potential can be factorized into a zero-temperature Coleman-Weinberg (CW) piece $V_{\text{CW}}$ and a finite temperature piece $V_T$. In the Parwani prescription, we replace $m_i^2 \rightarrow \overline{M}_i^2$ in both $V_{\text{CW}}$ and $V_T$ while in the Arnold-Espinosa prescription we replace $m_i^2 \rightarrow \overline{M}_i^2$ only in the non-analytic term appearing in the thermal piece $V_T$, corresponding to resumming only the Matsubara zero-modes. The former then coincides with TFD at 1-loop while the latter corresponds to a special case of TFD.

While the FD prescription has the obvious benefit of not needing to evaluate leading order diagrams analytically, it also faces several difficulties. Beginning at 2-loop order, certain higher-order diagrams such as the sunset diagram are not automatically included and need to be added by hand. More concerningly, the FD prescription has been shown to miscount daisy and superdaisy diagrams starting at 2-loop order \cite{Dine:1992,Boyd:1992}. An alternative procedure which reliably resums the dominant contributions to higher order is the \textit{partial dressing} (PD) prescription, first introduced in \ccite{Boyd:1993} under the name of \textit{tadpole resummation}. Rather than substituting $m_i^2 \rightarrow M_i^2$ directly in the effective potential, the PD prescription instructs us to perform the replacement in the first derivative of the effective potential $V_{\text{eff}}'$, which can then be integrated to obtain $V_{\text{eff}}$ as
\begin{equation}
    V_{\mathrm{eff}} = \int d\phi \, \frac{\partial V_{\mathrm{eff}}}{\partial \phi} \bigg|_{m^2 \rightarrow M^2} \,.
\end{equation}
This scheme corresponds to dressing just the propagator and has been demonstrated by explicit calculation to 4-loop order to give the right counting of daisy and superdaisy diagrams \cite{Boyd:1993}. A variant of PD resummation proposed in \cite{Curtin:2016}, \textit{optimized partial dressing}, has been shown to yield an even better degree of convergence. 

Despite its promise as a resummation candidate, the PD prescription is not without its challenges. PD omits a class of subleading diagrams starting at 2-loops. These are the lollipop diagrams (obtained from the vacuum sunset diagram by attaching one external leg to one of the vertices). Moreover, sunset-type tadpole diagrams are miscounted (obtained from the vacuum sunset diagram by attaching one external leg to one of the propagators). These issues can be resolved by adding the lollipop diagrams by hand and adjusting the gap equation to fix the miscounting of the sunset diagrams (see \ccite{Boyd:1992,Curtin:2016} as well as \cref{app:sunset}).

Another more pressing issue pointed out in \cite{Curtin:2016} is that it is unclear how to implement PD in multi-field scenarios where field excursions can occur along more than one direction. Given that this is the case in a variety of beyond the Standard Model (BSM) extensions capable of yielding a strongly first order electroweak phase transition, it is crucial that the formalism be extended to accommodate this situation. We propose a multi-field generalization of PD resummation which can be applied in scenarios where the Higgs mixes with BSM scalars and then go on to compare the convergence of this scheme with that of other resummation techniques.


\section{Resummation in \texorpdfstring{$\phi^4$}{phi\^4} theory}
\label{sec:single_field}

Before getting into these technicalities, we will review the various resummation prescriptions in the context of a simple $\phi^4$ theory. We begin in \cref{subsec:unresummed1} by computing the effective potential at 1- and 2-loop order in the context of a single field $\phi^4$ theory, to later generalize to the multi-field case (without mixing) in Sec. \ref{subsec:unresummed2}. In Secs. \ref{subsec:Parwani}, \ref{subsec:AE}, \ref{subsec:fulldressing}, and \ref{subsec:tadpole}, we resum the effective potential using the Parwani, Arnold-Espinosa, full dressing, and partial dressing schemes, respectively. Finally in \cref{subsec:numcompare} we compare for a few benchmark points and comment on the differences.

\subsection{Unresummed \texorpdfstring{$V_{\text{eff}}(\phi)$}{Veff(phi)} at 1- and 2-loops}\label{subsec:unresummed1}

 We consider a single-field $\phi^4$ theory with tree-level potential
\begin{equation}\label{Vtree}
    V_0 = \frac{\mu^2}{2} \phi^2 + \frac{\lambda}{4} \phi^4 \,.
\end{equation}
At 1-loop, the effective potential receives radiative and finite temperature corrections captured by the sum-integral \cite{Dolan:1974}
\begin{equation}\label{Jdef}
    V_{1\text{-loop}} = \mathcal{J}[m] \equiv \frac{1}{2} \sumint_K \ln (K^2 + m^2) \,,
\end{equation}
where $K=(\omega_n, \vec{k})$ are Euclidean 4-momenta, $\omega_n = 2 \pi n T$ are bosonic Matsubara modes, and the symbol $\sumint$ is shorthand for Euclidean integration
\begin{equation}
    \sumint_K \equiv T \sum_{\omega_n} \int \frac{d^3k}{(2\pi)^3} \,. 
\end{equation}
The mass entering into this 1-loop correction is the field-dependent effective mass
\begin{equation}
    m^2 = \frac{\partial^2 V_0}{\partial \phi^2} = \mu^2 + 3 \lambda \phi^2 \,,
\end{equation}
where $\phi$ here is understood to take its background field value. In many contexts, it is convenient to decompose the bosonic $\mathcal{J}$-function into a zero temperature Coleman-Weinberg piece\footnote{We work in the $\overline{\text{MS}}$ scheme with renormalization scale $\mu_R$. Note that this prescription at finite temperature does not eliminate all factors of $4\pi$ and $\gamma_E$, as it does at zero temperature.} 
\begin{equation}
    V_{\mathrm{CW}} = \frac{m^4}{64 \pi^2} \left( \ln \left( \frac{m^2}{\mu_R^2} \right) - \frac{3}{2} \right) \,,
\end{equation}
and a finite temperature piece
\begin{equation}
    V_{T} = \frac{T^4}{2\pi^2} J_B \left( \frac{m^2}{T^2} \right) \,, \,\,\, \text{with} \,\,\, J_B(y^2) = \int_0^\infty dx\, x^2 \ln \left( 1 - e^{- \sqrt{x^2 + y^2}} \right) \,,
\end{equation}
such that $V_{1\text{-loop}} = V_{\mathrm{CW}} + V_T$. In the high- and low-temperature limits, $J_B(y^2)$ admits expansions given in Eqs.~(\ref{JBhighT}) and (\ref{JBlowT}) of Appendix A. We will often have cause to work in the high-temperature regime, in which the full 1-loop correction reads:
\begin{equation}\label{V1loop}
    V_{1\text{-loop}} = \mathcal{J}[m] \simeq \frac{1}{24} m^2 T^2 - \frac{1}{12 \pi} m^3 T - \frac{L_R}{64 \pi^2} m^4 + \mathcal{O}\left(\frac{m^6}{T^2}\right)\,,
\end{equation}
where $L_R = \ln \left(\mu_R^2/T^2 \right) + 2 (\gamma_E - \ln 4 \pi)$ and we have ignored the field-independent constant. Note that the logarithmic terms have cancelled between the Coleman-Weinberg and finite temperature contributions, such that $- \frac{1}{12\pi} m^3 T$ is the only term originating from the Matsubara zero-mode, as evidenced by the fact that it is non-analytic in $m^2$. Combining with the tree level piece, the effective potential at 1-loop $V_{\text{eff}}^{(1)} = V_0 + V_{1\text{-loop}}$ in the high-temperature expansion is
\begin{equation}\label{Veff1singlephihighT}
    V_{\text{eff}}^{(1)} = \frac{1}{2} (\mu^2 + c_\phi T^2) \phi^2 + \frac{\lambda}{4} \phi^4 - \frac{1}{12 \pi} m^3 T - \frac{L_R}{64 \pi^2} m^4 + \mathcal{O}\left(\frac{m^6}{T^2}\right)\,,
\end{equation}
where we have defined $c_\phi = \lambda/4$. 

\begin{figure}
\centering
\includegraphics[width=0.75\textwidth]{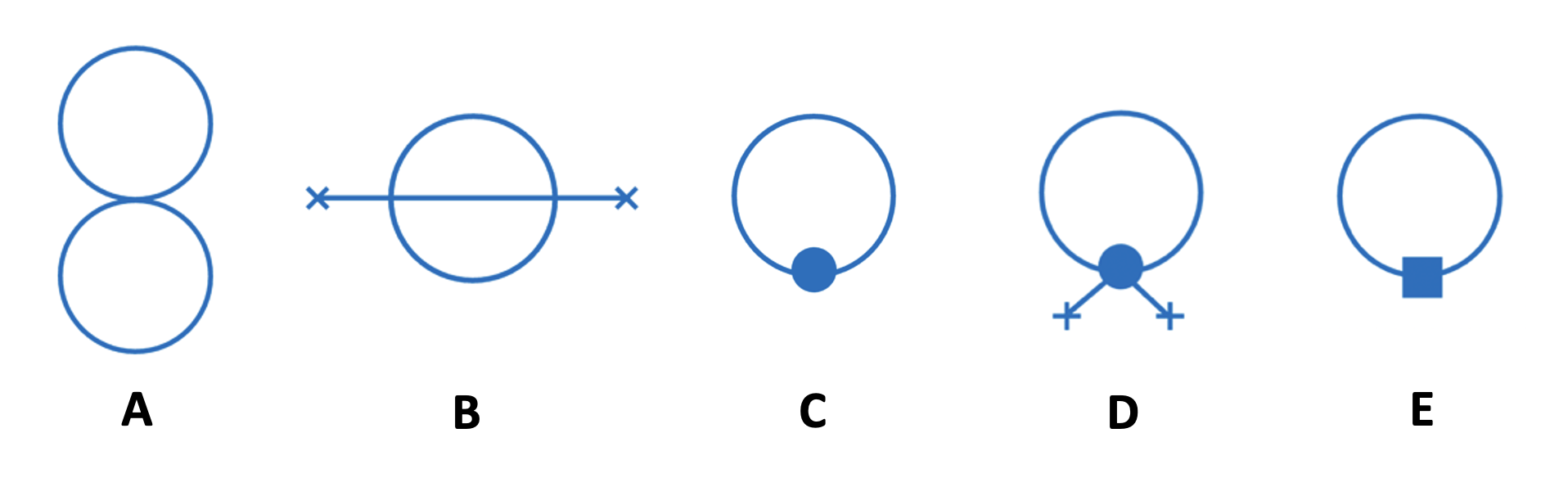}
\hspace{2mm}
\caption{2-loop corrections to the effective potential in $\phi^4$ theory, including (A) the figure-8 diagram, (B) the sunset diagram, (C) the 1-loop mass counterterm diagram, and (D) the 1-loop vertex counterterm diagram. Solid circles denote 1-loop counterterms and x's indicate explicit field insertions. Diagram (E) should be added to the effective potential upon performing thermal resummation and features a 1-loop ``thermal counterterm'' (solid square).}
\label{phi42loop}
\end{figure}

At 2-loops, the corrections to the effective potential are summarized diagramatically in \cref{phi42loop}. 
The figure-8 diagram (A) corresponds to the contribution
\begin{equation}
    V_{2\text{-loop}}^A = \frac{3 \lambda}{4} \mathcal{I}[m]^2 \,,
\end{equation}
where
\begin{equation}\label{Idef}
    \mathcal{I}[m] \equiv \sumint_K \frac{1}{K^2 + m^2} \,.
\end{equation}
This sum-integral is related to the $\mathcal{J}$ of Eq.~(\ref{Jdef}) as $\mathcal{I}=m^{-1} \frac{d \mathcal{J}}{d m}$, and so admits the high-temperature expansion given in Eq.~(\ref{eq:IhighT}) of Appendix A.
Using this expansion, $V_{2\text{-loop}}^A$ explicitly evaluates to
\begin{equation}\label{V2loopA}
    V_{2\text{-loop}}^A = - \frac{\lambda}{32 \pi} m T^3 + \frac{(3-L_R) \lambda}{64 \pi^2} m^2 T^2 + \frac{3 L_R \lambda}{128 \pi^3} m^3 T + \frac{(\zeta(3) + 3 L_R^2)\lambda}{1024 \pi^4} m^4 \,.
\end{equation}
The sunset diagram (B) corresponds to the contribution
\begin{equation}
    V_{2\text{-loop}}^B = - 3 \lambda^2 \phi^2 \mathcal{H}[m,m,m] \,, 
\end{equation}
where, for three arbitrary masses $m_1$, $m_2$, $m_3$
\begin{equation}
    \mathcal{H}[m_1, m_2, m_3] = \sumint_P \sumint_Q \frac{1}{(P^2 + m_1^2)(Q^2 + m_2^2)((P+Q)^2 + m_3^2)} \,.
\end{equation}
Using the high-temperature expansion of $\mathcal{H}$ in Eq.~(\ref{HhighT}), the contribution $V_{2\text{-loop}}^B$ is explicitly
\begin{equation}\label{V2loopB}
\begin{split}
    V_{2\text{-loop}}^B & = - \frac{3 \lambda^2}{32 \pi^2} \left( \ln \left( \frac{\mu_R^2}{m^2} \right) - 2 \ln 3 + 1 \right) T^2 \phi^2 \\
    & + \frac{9 \lambda^2}{64 \pi^3} \left( \ln \left( \frac{\mu_R^2}{m^2} \right) + L_R - 2 \ln 2+ 2 \right) m T \phi^2 \\
    & + \frac{9 \lambda^2}{256 \pi^4} \left( L_R^2 + L_R - 2 \gamma_E^2 - 4 \gamma_1 + \frac{\pi^2}{4} + \frac{3}{2} \right) m^2 \phi^2 \,.
\end{split}
\end{equation}
Next we have diagrams (C) and (D), which feature 1-loop mass and vertex counterterms, respectively. The former corresponds to an expression of the form
\begin{equation}\label{eq:V2loopC}
    V_{2\text{-loop}}^C = \frac{3\lambda}{32 \pi^2} \mu^2 \mathcal{I}[m] \frac{1}{\epsilon} \,,
\end{equation}
while the latter is
\begin{equation}\label{eq:V2loopD}
    V_{2\text{-loop}}^D = \frac{27 \lambda^2}{32 \pi^2} \phi^2 \mathcal{I}[m] \frac{1}{\epsilon} \,.
\end{equation}
These diagrams feature divergent $\mathcal{O}(1/\epsilon)$ pieces that cancel against the divergences in the 2-loop diagrams of Eqs.~(\ref{V2loopA}) and (\ref{V2loopB}), which we have already suppressed for simplicity of presentation since they are taken care of by renormalization. We include the explicit divergences in Eqs.~(\ref{eq:V2loopC}) and (\ref{eq:V2loopD}) since, due to the $\mathcal{O}(\epsilon)$ pieces in\footnote{See Eq.~(3.12) of Ref.~\cite{ArnoldEspinosa:1993} for the full expression.} $\mathcal{I}[m] \supset \frac{\epsilon}{12} L_R T^2$, they can also give a finite contribution to the effective potential. For $V_{2\text{-loop}}^C$, the leading order field-dependent contribution first comes in at $\mathcal{O}(\lambda^{5} T^4)$, which is higher than the order to which we work, and so can be neglected. For $V_{2\text{-loop}}^D$, the leading order finite contribution is
\begin{equation}\label{V2loopD}
    V_{2\text{-loop}}^D = \frac{9 L_R \lambda^2}{128 \pi^2} T^2 \phi^2 \,.
\end{equation}
The 2-loop correction to the effective potential is the sum of Eqs.~(\ref{V2loopA}), (\ref{V2loopB}), and (\ref{V2loopD}),
\begin{equation}
\begin{split}
    V_{2\text{-loop}} & = - \frac{\lambda}{32 \pi} m T^3 + \frac{(12-3L_R) \lambda}{256 \pi^2} m^2 T^2 + \frac{3 L_R \lambda}{128 \pi^3} m^3 T + \frac{(\zeta(3) + 3 L_R^2)\lambda}{1024 \pi^4} m^4\\ 
    & - \frac{3 \lambda^2}{128 \pi^2} \left( 4 \ln \left( \frac{\mu_R^2}{m^2} \right) - 3 L_R - 8 \ln 3 + 4 \right) T^2 \phi^2 \\
    & + \frac{9 \lambda^2}{64 \pi^3} \left( \ln \left( \frac{\mu_R^2}{m^2} \right) + L_R - 2 \ln 2+ 2 \right) m T \phi^2 \\
    & + \frac{9 \lambda^2}{256 \pi^4} \left( L_R^2 + L_R - 2 \gamma_E^2 - 4 \gamma_1 + \frac{\pi^2}{4} + \frac{3}{2} \right) m^2 \phi^2 \,.
\end{split}
\end{equation}
Combining with the 1-loop correction of Eq.~(\ref{V1loop}) and the tree-level potential of Eq.~(\ref{Vtree}) yields the 2-loop effective potential $V_\text{eff}$ in the high-temperature approximation.

\subsection{Unresummed \texorpdfstring{$V_\text{eff}(\phi_1,\phi_2)$}{Veff(phi1,phi2)} at 1- and 2-loops}\label{subsec:unresummed2}

Interesting finite-temperature effects can arise upon the addition of a second scalar field. Consider the tree-level potential
\begin{equation}
    V_0 = \frac{\mu_1^2}{2} \phi_1^2 + \frac{\mu_2^2}{2} \phi_2^2 + \frac{\lambda_1}{4} \phi_1^4 + \frac{\lambda_2}{4} \phi_2^4 + \frac{\lambda_{12}}{4} \phi_1^2 \phi_2^2 \,.
\end{equation}
For now, we will restrict to the case where only one of the fields develops a vacuum expectation value, such that we need not worry about mixing. In this case there will be no off-diagonal terms in the field-dependent mass matrix, and the effective masses for $\phi_1$ and $\phi_2$ are
\begin{subequations}
\begin{equation}
    m_1^2 = \mu_1^2 + 3 \lambda_1 \phi_1^2 + \frac{\lambda_{12}}{2} \phi_2^2 \,,
\end{equation}
\begin{equation}
    m_2^2 = \mu_2^2 + 3 \lambda_2 \phi_2^2 + \frac{\lambda_{12}}{2} \phi_1^2 \,.
\end{equation}
\end{subequations}
The 1-loop contribution to the effective potential is
\begin{equation}
    V_{1\text{-loop}} = \mathcal{J}[m_1] + \mathcal{J}[m_2] \,,
\end{equation}
with the bosonic $\mathcal{J}$-function defined in Eq.~(\ref{Jdef}). In the high-temperature approximation, the effective potential at 1-loop is then
\begin{equation}\label{Veff1loop2phi4}
\begin{split}
    V_\text{eff}^{(1)} = & \frac{1}{2} (\mu_1^2 + c_1 T^2) \phi_1^2 + \frac{1}{2} (\mu_2^2 + c_2 T^2) \phi_2^2 + \frac{\lambda_1}{4} \phi_1^4 + \frac{\lambda_2}{4} \phi_2^4 + \frac{\lambda_{12}}{4} \phi_1^2 \phi_2^2 \\
    & \hspace{45mm} - \frac{1}{12 \pi} (m_1^3 + m_2^3) T - \frac{L_R}{64 \pi^2} (m_1^4 + m_2^4) \,,
    \end{split}
\end{equation}
where we have defined the thermal mass parameters
\begin{equation}\label{c1c2def}
    c_1 = \frac{1}{24} (6 \lambda_1 + \lambda_{12}) \,, \,\,\, c_2 = \frac{1}{24} (6 \lambda_2 + \lambda_{12}) \,.
\end{equation}
When the mixed quartic coupling is negative, $\lambda_{12}<0$, one\footnote{The boundedness-from-below condition on $\lambda_1$ and $\lambda_2$ prevents both $c_1$ and $c_2$ from being negative.} of these thermal mass parameters $c_i$ can be negative. Then at high temperatures when the thermal contribution dominates the bare mass, one of the thermal masses
\begin{align}\label{eq:M2highT}
   \overline{M}_{i}^2 = m_{i}^2 + c_{i} T^2
\end{align} may become negative --- indicative of spontaneous symmetry breaking. Occasionally in the literature (see e.g.\ \ccite{Servant:2018}), the thermal masses are modified by inserting an additional Boltzmann factor,
\begin{align}\label{eq:thermal_masses_Boltzmann}
    \overline{M}_{\text{Boltzmann},i}^2 = m_{i}^2 + c_{i} T^2 e^{-m_i/T} \,,
\end{align}
which allows for a better approximation of the full thermal loop function for low temperatures. 

\begin{figure}[ht!]
\centering
\includegraphics[width=0.65\textwidth]{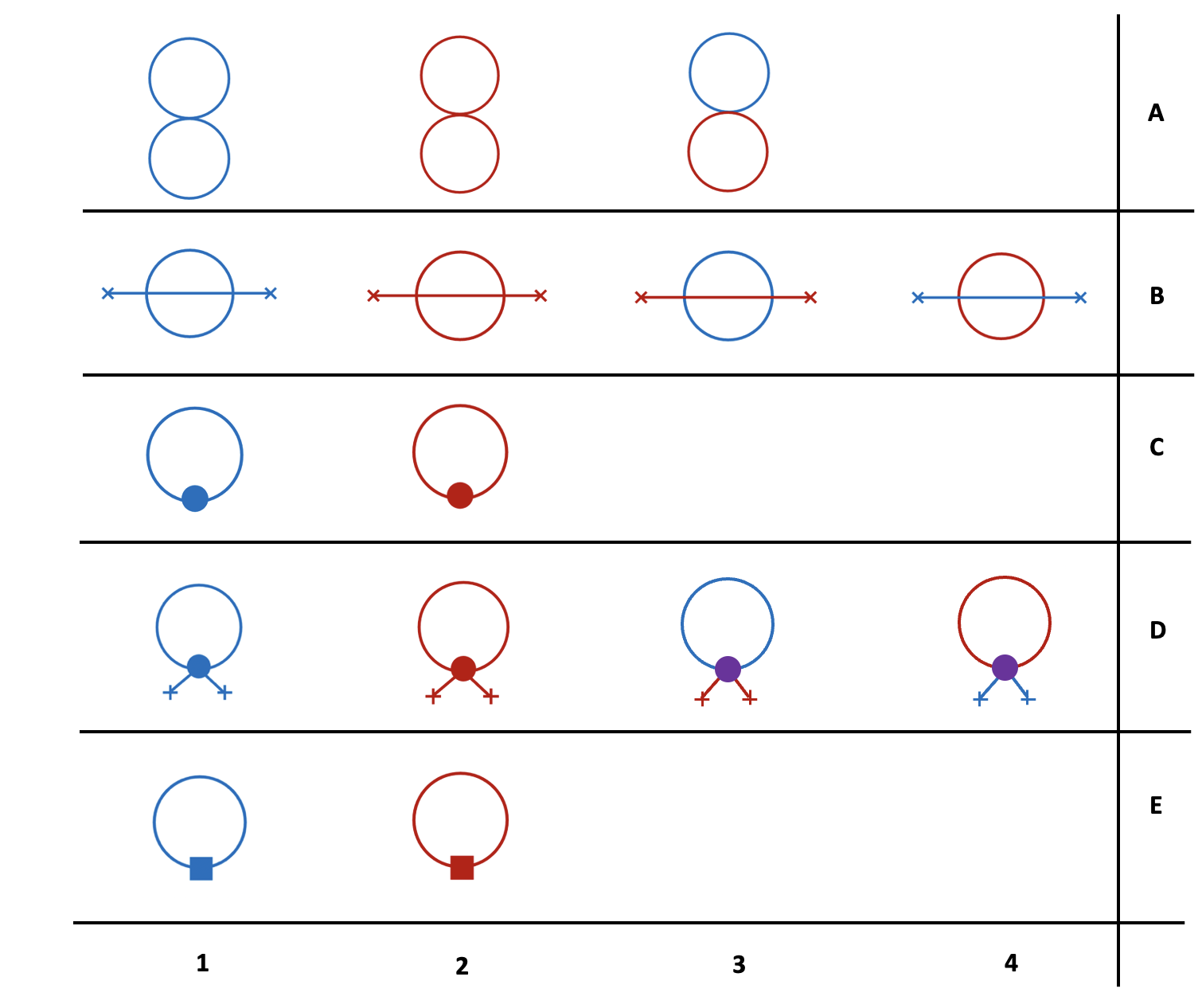}
\hspace{2mm}
\caption{2-loop corrections to the effective potential, including (A) figure-8 diagrams, (B) the sunset diagrams, (C) 1-loop mass counterterm diagrams, and (D) 1-loop vertex counterterm diagrams. Blue propagators correspond to $\phi_1$ while red correspond to $\phi_2$. Solid circles indicate 1-loop counterterms and x's indicate explicit field insertions. In particular, the blue, red, and purple circles of row (D) are $\lambda_1$, $\lambda_2$, and $\lambda_{12}$ vertex counterterms, respectively. Finally, the 1-loop ``thermal counterterm'' (solid square) diagrams of row (E) should be included upon performing thermal resummation.}
\label{2phi42loop}
\end{figure}

The 2-loop contributions to the effective potential are summarized diagrammatically in \cref{2phi42loop}.
Now in addition to figure-8, sunset, and counterterm diagrams for each species, we must also consider mixed diagrams. The first row of figure-8 diagrams corresponds to the expressions
\begin{subequations}\label{Adiagrams}
\begin{equation}
    V_{2\text{-loop}}^{A1} = \frac{3 \lambda_1}{4} \mathcal{I}[m_1]^2 \,,
\end{equation}
\begin{equation}
    V_{2\text{-loop}}^{A2} = \frac{3 \lambda_2}{4} \mathcal{I}[m_2]^2 \,,
\end{equation}
\begin{equation}
    V_{2\text{-loop}}^{A3} = \frac{\lambda_{12}}{4} \mathcal{I}[m_1] \, \mathcal{I}[m_2] \,,
\end{equation}
\end{subequations}
which, using the high-temperature expansion of $\mathcal{I}$ in Eq.~(\ref{eq:IhighT}), may be written explicitly in this limit as
\begin{subequations}
\begin{equation}
    V_{2\text{-loop}}^{A1} = \frac{\lambda_1}{32 \pi} \bigg[ - m_1 T^3 + \frac{(6-L_R)}{4 \pi} m_1^2 T^2 + \frac{3 L_R}{4 \pi^2} m_1^3 T + \frac{(\zeta(3) + 3 L_R^2)}{32 \pi^3} m_1^4 \bigg] \,,
\end{equation}
\begin{equation}
    V_{2\text{-loop}}^{A2} = \frac{\lambda_2}{32 \pi} \bigg[ - m_2 T^3 + \frac{(6-L_R)}{4 \pi} m_2^2 T^2 + \frac{3 L_R}{4 \pi^2} m_2^3 T + \frac{(\zeta(3) + 3 L_R^2)}{32 \pi^3} m_2^4 \bigg] \,,
\end{equation}
\begin{equation}
\begin{split}
    V_{2\text{-loop}}^{A3} & = \frac{\lambda_{12}}{64 \pi^2} \bigg[ - \frac{\pi}{3} (m_1+ m_2) T^3 + m_1 m_2 T^2 - \frac{L_R}{12} (m_1^2 + m_2^2) T^2 \\
    & \hspace{17mm} + \frac{L_R}{4\pi} (m_1 + m_2) m_1 m_2 T + \frac{\zeta(3)}{96\pi^2} (m_1^4 + m_2^4) + \frac{L_R^2}{16\pi^2} m_1^2 m_2^2 \bigg] \,.
\end{split}
\end{equation}
\end{subequations}
The second row of sunset diagrams corresponds to expressions
\begin{subequations}
\begin{equation}
    V_{2\text{-loop}}^{B1} = - 3 \lambda_1^2 \phi_1^2 \mathcal{H}[m_1, m_1, m_1] \,,
\end{equation}
\begin{equation}
    V_{2\text{-loop}}^{B2} = - 3 \lambda_2^2 \phi_2^2 \mathcal{H}[m_2, m_2, m_2] \,,
\end{equation}
\begin{equation}
    V_{2\text{-loop}}^{B3} = - \frac{\lambda_{12}^2}{4} \phi_2^2 \mathcal{H}[m_1, m_1, m_2] \,,
\end{equation}
\begin{equation}
    V_{2\text{-loop}}^{B4} = - \frac{\lambda_{12}^2}{4} \phi_1^2 \mathcal{H}[m_1, m_2, m_2] \,.
\end{equation}
\end{subequations}
Note that since these expressions feature external $\phi_i$ insertions, they vanish unless $\phi_i$ has non-zero vacuum expectation value. Suppose we take $\phi_1$ to be the field which develops a vacuum expectation value; then $V_{2\text{-loop}}^{B2} = V_{2\text{-loop}}^{B3} = 0$. Using the high-temperature expansion of $\mathcal{H}$ in Eq.~(\ref{HhighT}), the remaining $V_{2\text{-loop}}^{B1}$ and $V_{2\text{-loop}}^{B4}$ are explicitly
\begin{subequations}\label{Bdiagrams}
\begin{equation}
\begin{split}
    V_{2\text{-loop}}^{B1} = & - \frac{3 \lambda_1^2}{32 \pi^2} \left( \ln \left( \frac{\mu_R^2}{9 m_1^2} \right) + 1 \right) \phi_1^2 T^2 \\
    & + \frac{9 \lambda_1^2}{64 \pi^3} \left( \ln \left( \frac{\mu_R^2}{m_1^2} \right) + L_R - 2 \ln 2+ 2 \right) m_1 \phi_1^2 T \\
    & + \frac{9 \lambda_1^2}{256 \pi^4} \left( L_R^2 + L_R - 2 \gamma_E^2 - 4 \gamma_1 + \frac{\pi^2}{4} + \frac{3}{2} \right) m_1^2 \phi_1^2 \,,
\end{split}
\end{equation}
\begin{equation}
\begin{split}
    V_{2\text{-loop}}^{B4} = & - \frac{\lambda_{12}^2}{64 \pi^2} \left( \ln \left( \frac{\mu_R}{m_1 + 2 m_2} \right) + \frac{1}{2} \right) \phi_1^2 T^2 \\
    & + \frac{\lambda_{12}^2}{256 \pi^3} \left( \ln \left( \frac{\mu_R^2}{4 m_1^2} \right) + L_R + 2 \right) m_1 \phi_1^2 T \\
    & + \frac{\lambda_{12}^2}{128 \pi^3} \left( \ln \left( \frac{\mu_R^2}{4 m_2^2} \right) + L_R + 2 \right) m_2 \phi_1^2 T \\
    & + \frac{\lambda_{12}^2}{1024 \pi^4} \left( L_R^2 + L_R - 2 \gamma_E^2 - 4 \gamma_1 + \frac{\pi^2}{4} + \frac{3}{2} \right) (m_1^2 + 2 m_2^2) \phi_1^2 \,.
\end{split}
\end{equation}
\end{subequations}
As in the single $\phi^4$ case, the leading order, field-dependent contribution from the diagrams of row (C) are higher order than that to which we work, and so can be neglected. Of the diagrams of row (D), corresponding to expressions
\begin{subequations}
\begin{equation}
    V_{2\text{-loop}}^{D1} = \frac{3}{128 \pi^2} (36 \lambda_1^2 + \lambda_{12}^2) \phi_1^2 \, \mathcal{I}[m_1] \frac{1}{\epsilon} \,,
\end{equation}
\begin{equation}
    V_{2\text{-loop}}^{D2} = \frac{3}{128 \pi^2} (36 \lambda_2^2 + \lambda_{12}^2) \phi_2^2 \, \mathcal{I}[m_2] \frac{1}{\epsilon} \,,
\end{equation}
\begin{equation}
    V_{2\text{-loop}}^{D3} = \frac{1}{64 \pi^2} \left( 3(\lambda_1 + \lambda_2) \lambda_{12} + 2 \lambda_{12}^2 \right) \phi_2^2 \, \mathcal{I}[m_1] \frac{1}{\epsilon} \,,
\end{equation}
\begin{equation}
    V_{2\text{-loop}}^{D4} = \frac{1}{64 \pi^2} \left( 3(\lambda_1 + \lambda_2) \lambda_{12} + 2 \lambda_{12}^2 \right) \phi_1^2 \, \mathcal{I}[m_2] \frac{1}{\epsilon} \,,
\end{equation}
\end{subequations}
we need only consider $V_{2\text{-loop}}^{D1}$ and $V_{2\text{-loop}}^{D4}$, since $\phi_2$'s vanishing background value means $V_{2\text{-loop}}^{D2} = V_{2\text{-loop}}^{D3} = 0$. The explicit contributions from the non-vanishing diagrams in the high-temperature approximation are
\begin{subequations}\label{Ddiagrams}
\begin{equation}
    V_{2\text{-loop}}^{D1} = \frac{L_R}{512 \pi^2} (36 \lambda_1^2 + \lambda_{12}^2) \phi_1^2 T^2 \,,
\end{equation}
\begin{equation}
    V_{2\text{-loop}}^{D4} = \frac{L_R}{768 \pi^2} (3 (\lambda_1 + \lambda_2) \lambda_{12} + 2 \lambda_{12}^2) \phi_1^2 T^2 \,.
\end{equation}
\end{subequations}
Gathering all these terms, the 2-loop contribution to the effective potential is
\begin{equation}\label{V2looplong}
    V_{2\text{-loop}} = V_{2\text{-loop}}^{A1} + V_{2\text{-loop}}^{A2} + V_{2\text{-loop}}^{A3} + V_{2\text{-loop}}^{B1} + V_{2\text{-loop}}^{B4} + V_{2\text{-loop}}^{D1} + V_{2\text{-loop}}^{D4} \,,
\end{equation}
and the effective potential at 2-loop order is
\begin{equation}
    V_\text{eff}^{(2)} = V_\text{eff}^{(1)} + V_{2\text{-loop}} \,,
\end{equation}
with $V_\text{eff}^{(1)}$ in Eq.~(\ref{Veff1loop2phi4}).

\subsection{Parwani}\label{subsec:Parwani}

The Parwani prescription \cite{Parwani:1992} is a diagramatic approach to resummation in which all modes are resummed. On a technical level, it simply amounts to replacing $m_i^2$ with the high-temperature expanded thermal mass $\overline{M}_i^2 = m_i^2 + c_i T^2$ everywhere and adding to the 2-loop effective potential the ``thermal counterterm'' diagrams shown in row E of \cref{2phi42loop}. This has the effect of resumming the dominant parts of ring diagrams.

To see why, consider for the moment a simpler single field $\phi^4$ theory with tree-level potential $V_0 = \frac{\mu^2}{2} \phi^2 + \frac{\lambda}{4} \phi^4$. As discussed in Sec.~\ref{sec:resummation}, resumming daisy diagrams amounts to replacing the $m^2$ appearing in propagators with $\overline{M}^2$. In order to do this consistently, we can add and subtract the thermal contribution to the mass $c_\phi T^2 = \frac{\lambda}{4} T^2$ in a clever way, such that the tree-level potential becomes
\begin{equation}
    V_0 = \frac{1}{2} (\mu^2 + c_\phi T^2) \phi^2 + \frac{\lambda}{4} \phi^4 - \frac{c_\phi}{2} T^2 \phi^2 \,.
\end{equation}
This is equivalent to the original potential, but now the idea is to treat the first two terms as defining the unperturbed theory, and the last term as a perturbation --- a ``thermal counterterm''. Now order-by-order, the $c_\phi T^2$ pieces of quadratically divergent sub-loops will cancel against new diagrams involving thermal counterterms, resulting in a new convergent loop expansion parameter $\lambda T/\overline{M}$ \cite{ArnoldEspinosa:1993}. Returning to the full theory, we see that to implement Parwani resummation we should: 
\begin{enumerate}
    \item Replace all field dependent effective masses $m_i^2$ with leading order thermal masses $\overline{M}_i^2 = m_i^2 + c_i T^2$, with $c_i$ given in Eq.~(\ref{c1c2def}).
    \item Include thermal counterterm diagrams in calculating $V_{\text{eff}}$.
\end{enumerate}
These thermal counterm diagrams do not enter until 2-loop order, so at 1-loop the Parwani-resummed effective potential is simply $V_\text{eff,P}^{(1)} = V_\text{eff}^{(1)} |_{m_i^2 \rightarrow M_i^2}$, or explicitly 
\begin{equation}\label{P1loop}
\begin{split}
    V_\text{eff,P}^{(1)} = & \frac{1}{2} (\mu_1^2 + c_1 T^2) \phi_1^2 + \frac{1}{2} (\mu_2^2 + c_2 T^2) \phi_2^2 + \frac{\lambda_1}{4} \phi_1^4 + \frac{\lambda_2}{4} \phi_2^4 + \frac{\lambda_{12}}{4} \phi_1^2 \phi_2^2 \\
    & \hspace{45mm} - \frac{1}{12 \pi} (\overline{M}_1^3 + \overline{M}_2^3) T - \frac{L_R}{64 \pi^2} (\overline{M}_1^4 + \overline{M}_2^4) \,.
    \end{split}
\end{equation}
The 2-loop diagrams featuring ``thermal counterterms'' are shown in row E of Fig.~\ref{2phi42loop} and correspond to expressions
\begin{subequations}
\begin{equation}
    V_{2\text{-loop},\text{P}}^{E1} = - \frac{1}{2} c_1 T^2 \mathcal{I}[\overline{M}_1] \,,
\end{equation}
\begin{equation}
    V_{2\text{-loop},\text{P}}^{E2} = - \frac{1}{2} c_2 T^2 \mathcal{I}[\overline{M}_2] \,.
\end{equation}
\end{subequations}
Using the high-temperature expansion of $\mathcal{I}[m]$ in \cref{eq:IhighT}, these evaluate to
\begin{subequations}
\begin{equation}
    V_{2\text{-loop},\text{P}}^{E1} = \left( \frac{\lambda_1}{4} + \frac{\lambda_{12}}{24} \right) \left[ \frac{1}{8\pi} \overline{M}_1 T^3 + \frac{L_R}{32 \pi^2} \overline{M}_1^2 T^2 - \frac{\zeta(3)}{256 \pi^4} \overline{M}_1^4 \right] \,,
\end{equation}
\begin{equation}
    V_{2\text{-loop},\text{P}}^{E2} = \left( \frac{\lambda_2}{4} + \frac{\lambda_{12}}{24} \right) \left[ \frac{1}{8\pi} \overline{M}_2 T^3 + \frac{L_R}{32 \pi^2} \overline{M}_2^2 T^2 - \frac{\zeta(3)}{256 \pi^4} \overline{M}_2^4 \right] \,.
\end{equation}
\end{subequations}
We see that these expressions have pieces which exactly cancel the IR sensitive pieces coming from $V_{2\text{-loop}}^{A} |_{m_i^2 \rightarrow M_i^2}$,
\begin{equation}
    V_{2\text{-loop}}^{A} |_{m_i^2 \rightarrow M_i^2} \supset - \frac{\lambda_1}{32 \pi} \overline{M}_1 T^2 - \frac{\lambda_2}{32 \pi} \overline{M}_2 T^2 - \frac{\lambda_{12}}{192 \pi} (\overline{M}_1 + \overline{M}_2)T^3 \,.
\end{equation}
Upon adding these contributions to $V_{2\text{-loop}}$, as given in Eq.~(\ref{V2looplong}), and replacing $m_i^2 \rightarrow \overline{M}_i^2$ everywhere in $V_{2\text{-loop}}$, we arrive at the Parwani-resummed 2-loop effective potential
\begin{equation}
    V_{\text{eff,P}}^{(2)} = V_{\text{eff,P}}^{(1)} + V_{2\text{-loop}}(\overline{M}_i^2) + V_{2\text{-loop},\text{P}}^{E1}(\overline{M}_i^2) + V_{2\text{-loop},\text{P}}^{E2} (\overline{M}_i^2) \,.
\end{equation}
As demonstrated in the discussion above, Parwani resummation intrinsically depends on the high-temperature expansion and also only resum the leading contributions in the high-temperature limit.

\subsection{Arnold-Espinosa}\label{subsec:AE}

Alternatively because only the Matsubara zero mode $\omega_0$ demonstrates problematic behavior in the IR, we could just resum these ``soft'' modes whilst leaving the hard non-zero modes untouched. This is the basic premise behind the Arnold-Espinosa prescription \cite{ArnoldEspinosa:1993}, another popular diagrammatic approach to resummation. Because it requires splitting calculations into soft and hard modes, it is in principle a bit more cumbersome to implement. Consider for example the $\mathcal{J}$-function defined in Eq.~(\ref{Jdef}). Previously we had taken the customary approach of separating this into a zero-temperature Coleman-Weinberg piece and a finite temperature piece \cite{Quiros:1992},
\begin{equation}
    \mathcal{J}[m] = \underbrace{\frac{1}{2} \int \frac{d^4k}{(2\pi)^4} \ln (k^2 + m^2)}_{\mathcal{J}_\text{CW}[m]} - \underbrace{T \int \frac{d^3 k}{(2\pi)^3} \ln \left( 1 \mp n_{\text{B/F}}(E_k, T) \right)}_{\mathcal{J}_T[m]} \,,
\end{equation}
where $n_{\text{B/F}}(E_k,T) = 1/\left( e^{E_k/T} \mp 1 \right)$. A more useful decomposition for the purposes of Arnold-Espinosa resummation would be to isolate the zero mode
\begin{equation}
    \mathcal{J}[m] = \underbrace{\frac{T}{2} \int \frac{d^3 k}{(2\pi)^3} \ln \left( k^2 + m^2 \right)}_{\mathcal{J}_\text{soft}[m]} + \underbrace{\frac{T}{2} \sum_{n \neq 0} \int \frac{d^3 k}{(2\pi)^3} \ln \left( \omega_n^2 + k^2 + m^2 \right)}_{\mathcal{J}_\text{hard}[m]} \,,
\end{equation}
where $\omega_n = 2 \pi n T$. Working in the high-temperature expansion, one can show that this first zero-mode piece evaluates to
\begin{equation}
    \mathcal{J}_\text{soft}[m] \simeq - \frac{1}{12 \pi} m^3 T \,.
\end{equation}
Comparing against the full expression in Eq.~(\ref{V1loop}), we note that the zero-mode contribution is just the term non-analytic in $m^2$. This is a more generic phenomenon; it will turn out to be the case that all terms non-analytic in $m^2$ contain zero-mode contributions. This makes implementing Arnold-Espinosa resummation surprisingly simple in practice when working in the high-temperature expansion, since the terms requiring resummation are readily identifiable. At 1-loop there is only one such term (per scalar field $\phi_i$), and we resum it by replacing $m_i^3 \rightarrow \overline{M}_i^3$. The 1-loop effective potential in the Arnold-Espinosa scheme then reads
\begin{equation}\label{AE1loop}
\begin{split}
    V_\text{eff,AE}^{(1)} = & \frac{1}{2} (\mu_1^2 + c_1 T^2) \phi_1^2 + \frac{1}{2} (\mu_2^2 + c_2 T^2) \phi_2^2 + \frac{\lambda_1}{4} \phi_1^4 + \frac{\lambda_2}{4} \phi_2^4 + \frac{\lambda_{12}}{4} \phi_1^2 \phi_2^2 \\
    & \hspace{45mm} - \frac{1}{12 \pi} (\overline{M}_1^3 + \overline{M}_2^3) T - \frac{L_R}{64 \pi^2} (m_1^4 + m_2^4) \,.
    \end{split}
\end{equation}
This expression is similar to the Parwani-resummed $V_\text{eff,P}^{(1)}$ but differs in the higher order terms $\propto m^4$. Explicitly, the difference between the two is
\begin{equation}
    V_\text{eff,AE}^{(1)} - V_\text{eff,P}^{(1)} = \frac{L_R}{64 \pi^2} \left[(c_1^2 + c_2^2) T^4 + 2(c_1 m_1^2 + c_2 m_2^2) T^2 \right] \,.
\end{equation}
Before moving on to the 2-loop effective potential, we note that often in the literature one speaks of resumming $V_{\text{eff}}^{(1)}$  by adding the daisy ``ring improvement'' term 
\begin{equation}
    V_{\text{daisy}} = - \frac{1}{12 \pi} (\overline{M}^3 - m^3) T \,.
\end{equation}
We see that the result is completely identical to that of the procedure described above. 

At 2-loop order, there are many more terms with non-analytic $m^2$ dependence. It will be useful to define the Arnold-Espinosa resummed $\mathcal{I}$-function, whose high-temperature expansion is given in Eq.~(\ref{eq:AEI}). Then from Eqs.~(\ref{Adiagrams}), the contributions to the resummed $V_{2\text{-loop}}$ coming from the figure-8 diagrams are
\begin{subequations}
\begin{equation}
    V_{2\text{-loop,AE}}^{A1} = \frac{\lambda_1}{32\pi} \left[ - \overline{M}_1 T^3 - \frac{L_R}{4\pi} m_1^2 T^2 + \frac{3}{2\pi} \overline{M}_1^2 T^2 + \frac{3L_R}{4\pi^2} m_1^2 \overline{M}_1 T + \frac{( \zeta(3) + 3 L_R^2)}{32 \pi^3} m_1^4 \right] \,,
\end{equation}
\begin{equation}
    V_{2\text{-loop,AE}}^{A2} = \frac{\lambda_2}{32\pi} \left[ - \overline{M}_2 T^3 - \frac{L_R}{4\pi} m_2^2 T^2 + \frac{3}{2\pi} \overline{M}_2^2 T^2 + \frac{3L_R}{4\pi^2} m_2^2 \overline{M}_2 T + \frac{( \zeta(3) + 3 L_R^2)}{32 \pi^3} m_2^4 \right] \,, 
\end{equation}
\begin{equation}
\begin{split}
    V_{2\text{-loop,AE}}^{A3} & = \frac{\lambda_{12}}{64 \pi^2} \bigg[ - \frac{\pi}{3} (\overline{M}_1 + \overline{M}_2)T^3 + \overline{M}_1 \overline{M}_2 T^2 - \frac{L_R}{12} (m_1^2 + m_2^2) T^2 \\
    & \hspace{32mm} + \frac{L_R}{4\pi} (\overline{M}_1 m_2^2 + \overline{M}_2 m_1^2) T + \frac{L_R^2}{16 \pi^2} m_1^2 m_2^2 + \frac{\zeta(3)}{96\pi^2} (m_1^4 + m_2^4) \bigg] \,.
\end{split}
\end{equation}
\end{subequations}
From Eq.~(\ref{Bdiagrams}), the contribution from the sunset diagrams are
\begin{subequations}
\begin{equation}
\begin{split}
    V_{2\text{-loop,AE}}^{B1} = & - \frac{3 \lambda_1^2}{32 \pi^2} \left( \ln \left( \frac{\mu_R^2}{9 \overline{M}_1^2} \right) + 1 \right) \phi_1^2 T^2 \\
    & + \frac{9 \lambda_1^2}{64 \pi^3} \left( \ln \left( \frac{\mu_R^2}{\overline{M}_1^2} \right) + L_R - 2 \ln 2+ 2 \right) \overline{M}_1 \phi_1^2 T \\
    & + \frac{9 \lambda_1^2}{256 \pi^4} \left( L_R^2 + L_R - 2 \gamma_E^2 - 4 \gamma_1 + \frac{\pi^2}{4} + \frac{3}{2} \right) m_1^2 \phi_1^2 \,,
\end{split}
\end{equation}
\begin{equation}
\begin{split}
    V_{2\text{-loop,AE}}^{B4} = & - \frac{\lambda_{12}^2}{64 \pi^2} \left( \ln \left( \frac{\mu_R}{\overline{M}_1 + 2 \overline{M}_2} \right) + \frac{1}{2} \right) \phi_1^2 T^2 \\
    & + \frac{\lambda_{12}^2}{256 \pi^3} \left( \ln \left( \frac{\mu_R^2}{4 \overline{M}_1^2} \right) + L_R + 2 \right) \overline{M}_1 \phi_1^2 T \\
    & + \frac{\lambda_{12}^2}{128 \pi^3} \left( \ln \left( \frac{\mu_R^2}{4 \overline{M}_2^2} \right) + L_R + 2 \right) \overline{M}_2 \phi_1^2 T \\
    & + \frac{\lambda_{12}^2}{1024 \pi^4} \left( L_R^2 + L_R - 2 \gamma_E^2 - 4 \gamma_1 + \frac{\pi^2}{4} + \frac{3}{2} \right) (m_1^2 + 2 m_2^2) \phi_1^2 \,,
\end{split}
\end{equation}
\end{subequations}
where again we are taking $\phi_1$ as the field which develops a vacuum expectation value. From Eq.~(\ref{Ddiagrams}), we see that the contributions from vertex counterterm diagrams are unaffected by the resummation. Finally, the thermal counterterm diagrams now include only the zero-mode contribution, and read
\begin{subequations}
\begin{equation}
    V_{2\text{-loop},\text{AE}}^{E1} = \frac{1}{192 \pi} \left( 6 \lambda_1 + \lambda_{12} \right) \overline{M}_1 T^3 \,,
\end{equation}
\begin{equation}
    V_{2\text{-loop},\text{AE}}^{E2} =\frac{1}{192 \pi} \left( 6 \lambda_2 + \lambda_{12} \right) \overline{M}_2 T^3 \,.
\end{equation}
\end{subequations}
The 2-loop effective potential resummed in the Arnold-Espinosa scheme is then
\begin{equation}
\begin{split}
    V_\text{eff,AE}^{(2)} = & V_\text{eff,AE}^{(1)} + V_{2\text{-loop,AE}}^{A1} + V_{2\text{-loop,AE}}^{A2} + V_{2\text{-loop,AE}}^{A3}\\
    & + V_{2\text{-loop,AE}}^{B1} + V_{2\text{-loop,AE}}^{B4} + V_{2\text{-loop}}^{D1} + V_{2\text{-loop}}^{D4} + V_{2\text{-loop},\text{AE}}^{E1} + V_{2\text{-loop},\text{AE}}^{E2} \,.
\end{split}
\end{equation}
As for Parwani resummation, the Arnold-Espinosa resummation scheme intrinsically depends on the high-temperature expansion and resums only the leading contributions in the high-temperature limit.

\subsection{Gap resummation}\label{sec:gapresum}

Gap resummation offers an alternative to the diagrammatic approaches to resummation described above, which quickly become cumbersome at higher loop order. Rather than evaluating such diagrams analytically, in gap resummation one need merely compute $V_{\text{eff}}^{(1)}$ and then solve the so-called ``gap equation'' for the thermal mass. This gap equation includes the dominant contributions from many higher-order diagrams, though admittedly it does not include contributions from certain sub-leading diagrams at each order (for example, parts of the 2-loop sunset diagram). After solving the gap equation for the thermal mass, this is substituted into either $V_{\text{eff}}^{(1)}$ in the full dressing (FD) prescription or into $\partial_\phi V_{\text{eff}}^{(1)}$ in the partial dressing (PD) prescription. The latter is sometimes also called tadpole resummation, and has been demonstrated to count daisy and superdaisy diagrams more faithfully to higher order. 

The first step in either procedure is solving the gap equation for the thermal mass,
\begin{equation}
    M_i^2 = \frac{\partial^2}{\partial \phi_i^2} V_{\text{eff}}^{(1)}(M_j^2) \,.
\end{equation}
Note that the thermal mass $M^2$ appears on both the left- and right-hand sides of this equation, which must be solved numerically. Considering for the moment single-field $\phi^4$ theory in the high temperature expansion --- the effective potential for which given in Eq.~(\ref{Veff1singlephihighT}) --- this gap equation is explicitly
\begin{equation}\label{eq:massgap}
    M^2 \overset{\text{high-}T}{=} m^2 + \frac{\lambda T^2}{4} - \frac{3 \lambda M T}{4\pi} - \frac{3\lambda L_R M^2}{16 \pi^2} - \frac{9\lambda^2 \phi^2 T}{4\pi M} - \frac{9 \lambda^2 L_R \phi^2}{8 \pi^2} \,.
\end{equation}
The set of gap equations in the 2-field case is even more complicated given their coupled nature
\begin{subequations}
\begin{equation}
\begin{split}
    M_1^2 & \overset{\text{high-}T}{=} \mu_1^2 + 3 \lambda_1 \phi_1^2 + \frac{\lambda_{12}}{2} \phi_2^2 + c_1 T^2 - \frac{3\lambda_1}{4\pi} M_1 T - \frac{\lambda_{12}}{8 \pi} M_2 T \\
    & \hspace{20mm} - \frac{3\lambda_1}{16 \pi^2} L_R M_1^2 - \frac{\lambda_{12}}{32 \pi^2} L_R M_2^2 - \frac{L_R}{32 \pi^2} (36 \lambda_1^2 + \lambda_{12}^2) \phi_1^2 \,,
\end{split}
\end{equation}
\begin{equation}
\begin{split}
    M_2^2 & \overset{\text{high-}T}{=} \mu_2^2 + 3 \lambda_2 \phi_2^2 + \frac{\lambda_{12}}{2} \phi_1^2 + c_2 T^2 - \frac{3\lambda_2}{4\pi} M_2 T - \frac{\lambda_{12}}{8 \pi} M_1 T \\
    & \hspace{20mm} - \frac{3\lambda_2}{16 \pi^2} L_R M_2^2 - \frac{\lambda_{12}}{32 \pi^2} L_R M_1^2 - \frac{L_R}{32 \pi^2} (36 \lambda_2^2 + \lambda_{12}^2) \phi_2^2 \,.
\end{split}
\end{equation}
\end{subequations}

\subsubsection{Full dressing}\label{subsec:fulldressing}

In the full dressing (FD) prescription, the thermal masses $M_i^2$ obtained by solving the gap equations are substituted directly into the effective potential, $V_\text{eff}^{\text{FD}} = V_\text{eff}^{(1)} |_{m_i^2 \rightarrow M_i^2}$, with the result
\begin{equation}
\begin{split}
    V_\text{eff}^{\text{FD}} = & \frac{1}{2} (\mu_1^2 + c_1 T^2) \phi_1^2 + \frac{1}{2} (\mu_2^2 + c_2 T^2) \phi_2^2 + \frac{\lambda_1}{4} \phi_1^4 + \frac{\lambda_2}{4} \phi_2^4 + \frac{\lambda_{12}}{4} \phi_1^2 \phi_2^2 \\
    & \hspace{45mm} - \frac{1}{12 \pi} (M_1^3 + M_2^3) T - \frac{L_R}{64 \pi^2} (M_1^4 + M_2^4) \,.
    \end{split}
\end{equation}
When one uses the high-temperature expanded truncated thermal masses, this is identical to the 1-loop effective potential in the Parwani scheme $V_\text{eff,P}^{(1)}$, given in Eq.~(\ref{P1loop}). More generally when using the full solutions of the gap equation, however, they differ.

\subsubsection{Partial dressing}\label{subsec:tadpole}

One issue with the FD prescription is that it miscounts certain diagrams starting at 2-loop order \cite{Boyd:1992,Dine:1992}. This shortcoming led to the introduction of the partial dressing (PD) prescription, in which one replaces $m^2 \rightarrow M^2$ on the level of the first derivative of the effective potential $\partial_\phi V_\text{eff}$ and then integrates to obtain the resummed effective potential
\begin{equation}
    V_\text{eff}^\text{PD} = \int d\phi \left( \frac{\partial V_\text{eff}^{(1)}(m_i^2)}{\partial \phi} \right)_{m_i^2 \rightarrow M_i^2} \,,
\end{equation}
where the $M_i$'s are the full solution of the gap equations.
The procedure can be understood by thinking about Dyson resummation in zero-temperature field theory. There, self-energy corrections are resummed into the propagator
\begin{align}
    \frac{i}{p^2 - m^2} \rightarrow  \frac{i}{p^2 - m^2 + \Sigma(p^2)},
\end{align}
where $\Sigma(p^2)$ is the self energy and $p^2$ is the momentum. If $\Sigma$ is not dependent on $p^2$, all corrections to the propagator can be absorbed into the mass via
\begin{align}
    m^2 \rightarrow M^2 = m^2 - \Sigma\,.
\end{align}
This resummation can also be used to absorb thermal corrections since their leading part does not depend on the momentum (see \cref{app:sunset} for a discussion of the momentum-dependent parts). Moreover, we can identify $M^2$ as the solution of the gap equation. To ensure proper resummation, it just needs to be ensured that this replacement is done only for all propagators and not for vertices, which in the effective potential are, however, expressed in terms of the mass. To avoid also dressing the vertex, the replacement is done on the level of the tadpole for which the coupling is explicit (see below).

The PD procedure has been demonstrated to correctly count the most relevant diagrams up to 4-loop order \cite{Boyd:1993}. To compare the two and understand why FD leads to a miscounting while PD does not, we will consider the 1-loop tadpole diagrams which compute $\partial_\phi V_{1\text{-loop}}$ (in contrast to the vacuum diagrams which compute $V_{1\text{-loop}}$). These tadpoles can be formed from the corresponding vacuum diagrams by attaching a zero-momentum truncated external leg to each part of the vacuum diagram. By shifting $m^2 \rightarrow M^2$ at the level of vacuum diagrams, FD is equivalent to dressing both the propagator and 3-point vertex $c_3 \equiv \partial_\phi m^2$ of the corresponding 1-loop tadpole diagrams. In contrast, by shifting $m^2 \rightarrow M^2$ at the tadpole level, PD dresses only the propagator.

To see explicitly that the former dressing of both vertex and propagator leads to a miscounting, let us return for the moment to the single-field $\phi^4$ theory. Following Ref.~\cite{Boyd:1993}, we will presume $\phi^2/T^2 \ll 1$ so that we need only consider diagrams with one 3-point vertex.\footnote{We impose this condition only for the remainder of Section 3.5.2.} Starting with Eq.~(\ref{eq:massgap}) and taking only the leading order terms in $\lambda$, the mass gap equation becomes
\begin{equation}\label{eq:gap_equation_example}
    M^2 \overset{\text{high-}T}{=} m^2 + \frac{\lambda T^2}{4} - \frac{3 \lambda M T}{4\pi} - \frac{9 \lambda^2 \phi^2 T}{4\pi M} \,.
\end{equation}
Note that the last term proportional to $\lambda^2 \phi^2$ is formally subleading. We include it nevertheless since it becomes important when taking the derivative with respect to $\phi$ for computing the 3-point vertex. The solution gives the dressed propagator, and to $\mathcal{O}(\lambda^3)$ reads
\begin{equation}\label{eq:gap_equation_example_sol}
    M \overset{\text{high-}T}{\simeq} m + \frac{\lambda T^2}{8 m} - \frac{3 \lambda T}{8\pi} - \frac{\lambda^2 T^4}{128 m^3} + \frac{9 \lambda^2 T^2}{128 \pi^2 m} + \frac{\lambda^3 T^6}{1024 m^5} - \frac{9 \lambda^3 T^4}{1024 \pi^2 m^3} \,.
\end{equation}
Meanwhile, the dressed 3-point vertex $C_3$ is obtained by differentiating the gap equation, $C_3 = \partial_\phi M^2$. Working to the same order, the solution is
\begin{equation}
    C_3 \overset{\text{high-}T}{=} 6 \lambda \phi \left( 1 - \frac{9 \lambda T}{8 \pi m} + \frac{9 \lambda^2 T^3}{64 \pi m^3} - \frac{27 \lambda^3 T^5}{1024 \pi m^5} + \frac{81 \lambda^3 T^3}{1024 \pi^3 m^3} \right) \,,
\end{equation}
In FD, both the propagator and vertex of the tadpole are improved, and so starting from Eq.~(\ref{V1loop}), the contribution to the derivative of the 1-loop effective potential reads
\begin{equation}
    \partial_\phi V_{1\text{-loop}}^{\text{FD}} \overset{\text{high-}T}{=} \frac{T^2}{24} (6\lambda \phi) - \frac{M T}{8\pi} C_3 + ...
\end{equation}
where the first leading order term comes from the hard thermal loop, the second term comes from the zero mode, and we suppress higher order terms. This is in contrast to PD, for which only the propagator is dressed, leading to
\begin{equation}
    \partial_\phi V_{1\text{-loop}}^{\text{PD}} \overset{\text{high-}T}{=} \frac{T^2}{24} (6\lambda \phi) - \frac{M T}{8\pi} c_3 + ...
\end{equation}
where $c_3 = \partial_\phi m^2 = 6 \lambda \phi$ is the undressed 3-point vertex. Explicitly using the forms of $M$ and $C_3$ above, the zero-mode piece in either case reads
\begin{align}
    - \frac{M T}{8\pi} C_3 &\overset{\text{high-}T}{=} (6 \lambda \phi) \bigg( - \frac{m T}{8\pi} - \frac{\lambda T^3}{64 \pi m} + \frac{3 \lambda T^2}{16 \pi^2} + \frac{\lambda^2 T^5}{1024 \pi m^3} - \frac{63 \lambda^2 T^3}{1024 \pi^3 m} \nonumber\\
    &\hspace{2.4cm}- \frac{\lambda^3 T^7}{8192 \pi m^5} + \frac{63 \lambda^3 T^5}{8192 \pi^3 m^3} \bigg) \,, \\
    - \frac{M T}{8\pi} c_3 &\overset{\text{high-}T}{=} (6 \lambda \phi) \bigg( - \frac{m T}{8\pi} - \frac{\lambda T^3}{64 \pi m} + \frac{3 \lambda T^2}{64 \pi^2} + \frac{\lambda^2 T^5}{1024 \pi m^3} - \frac{9 \lambda^2 T^3}{1024 \pi^3 m} \nonumber\\
    &\hspace{2.4cm}- \frac{\lambda^3 T^7}{8192 \pi m^5} + \frac{9 \lambda^3 T^5}{8192 \pi^3 m^3} \bigg) \,.
\end{align}
The difference between the fully dressed and partially dressed 1-loop tadpole, defined as $\Delta \equiv \partial_\phi V_{1\text{-loop}}^{\text{FD}} - \partial_\phi V_{1\text{-loop}}^{\text{PD}}$, is then
\begin{equation}
    \Delta \overset{\text{high-}T}{=} (6 \lambda \phi) \left( \frac{9 \lambda T^2}{64 \pi^2} - \frac{27 \lambda^2 T^3}{512 \pi^3 m} + \frac{27 \lambda^3 T^5}{4096 \pi^3 m^3} \right) \,.
\end{equation}
By computing contributions from the relevant Feynman diagrams up to 4-loop order, Ref.~\cite{Boyd:1993} finds an expression for $\partial_\phi V_{1\text{-loop}}$ which precisely matches that of $\partial_\phi V_{1\text{-loop}}^{\text{PD}}$ above, and so $\Delta$ quantifies the extraneous contribution due to miscounting in the FD procedure. These extra terms are not present for PD, which automatically includes subleading thermal corrections of super-daisy order.

There is yet another way to see that PD correctly counts daisy and superdaisy diagrams to higher order than FD. We now turn to the regime relevant for the phase transition, $\alpha \equiv \lambda T^2 / m^2 \simeq 1$, where the perturbative expansion breaks down and resummation becomes important. Following Ref.~\cite{Curtin:2016}, we ask what other parameters need to be small in order that the improved perturbative expansion converges. Of course, the usual requirement of perturbative unitarity at zero-$T$ demands $\lambda$ be sufficiently small. Additionally though, convergence of the high-$T$ contributions requires
\begin{equation}
    \beta \equiv \lambda \frac{T}{m} \ll 1 \,.
\end{equation}
To see why, consider the contributions to the mass correction at 1-loop $\delta m_1^2$, from daisy diagrams $\delta m_D^2$, from superdaisy diagrams $\delta m_{SD}^2$, from lollipop diagrams $\delta m_L^2$, and finally from sunset diagrams $\delta m_S^2$. After resummation in $\alpha$, these scale as
\begin{subequations}
\begin{equation}
    \delta m_1^2 \sim \lambda T^2 \,,
\end{equation}
\begin{equation}
    \delta m_D^2 \sim \lambda^2 \frac{T^3}{m} \,,
\end{equation}
\begin{equation}
    \delta m_{SD}^2 \sim \lambda^3 \frac{T^4}{m^2} \,,
\end{equation}
\begin{equation}
    \delta m_L^2 \sim \lambda^2 T^2 \,,
\end{equation}
\begin{equation}
    \delta m_S^2 \sim \lambda^3 \phi^2 \frac{T^2}{m^2} \,.
\end{equation}
\end{subequations}
Thus around the phase transition when $\alpha \sim 1$, the ratios between the diagram classes scale as
\begin{subequations}
\begin{equation}
    \frac{\delta m_D^2}{\delta m_1^2} \sim \beta \,,
\end{equation}
\begin{equation}
    \frac{\delta m_{SD}^2}{\delta m_D^2} \sim \beta \,,
\end{equation}
\begin{equation}
    \frac{\delta m_L^2}{\delta m_D^2} \sim \frac{m}{T} = \frac{\beta}{\alpha} \sim \beta \,,
\end{equation}
\begin{equation}
    \frac{\delta m_S^2}{\delta m_D^2} \sim \frac{\lambda \phi^2}{m T} = \frac{\phi^2}{T^2} \beta \,.
\end{equation}
\end{subequations}
Clearly then convergence requires $\beta \ll 1$. Now we can compare the $\beta$-scaling in the PD and FD resummation approaches. As worked out explicitly in Ref.~\cite{Boyd:1993} by means of a systematic expansion in $\alpha \sim 1$ and $\beta$ (and confirmed in Ref.~\cite{Curtin:2016}), PD is accurate to $\mathcal{O}(\beta^2)$, with the leading neglected contribution $\delta m_N^2$ coming in as
\begin{equation}
    \frac{\delta m_N^2}{\delta m_1^2} \sim \beta^3 \,.
\end{equation}
Meanwhile for FD, which miscounts daisy and superdaisy diagrams starting at 2-loop, neglects sunset diagrams, and includes the wrong prefactor for lollipop diagrams, error comes in as
\begin{equation}
    \frac{\delta m_N^2}{\delta m_1^2} \sim \beta \,.
\end{equation}
Despite its promise, one shortcoming of PD is that an ambiguity arises when field excursions can proceed along multiple directions. Namely it is unclear which field to take the derivative of $V_{\text{eff}}$ with respect to, since in general $V_{\text{eff}}^{\text{PD},1} \neq V_{\text{eff}}^{\text{PD},2}$, where $V_{\text{eff}}^{\text{PD},i} = \int d\phi_i \, \partial_i V_{\text{eff}}\big|_{M^2}$. In Sec.~\ref{sec:mixing_PD} we propose a multi-field generalization which holds even in the case of mixing scalar fields, making PD suitable for a wider range of applications, in particular BSM extensions of the Higgs sector. 


\subsection{Numerical comparison}\label{subsec:numcompare}


\subsubsection*{Single-field $\phi^4$ theory}

For our numerical discussion of the single-field $\phi^4$ theory, we focus on the benchmark point
\begin{align}
    \lambda = 1/3, \hspace{.3cm} \mu = 1\tev\,.
\end{align}
We set the renormalisation scale equal to the temperature. The chosen value for $\lambda$ is of no particular significance. As we will see below, it is well within the regime for which partial dressing is reliable.

\begin{figure}
    \centering
    \includegraphics[width=.65\textwidth,trim={.5cm .5cm .5cm .5cm},clip]{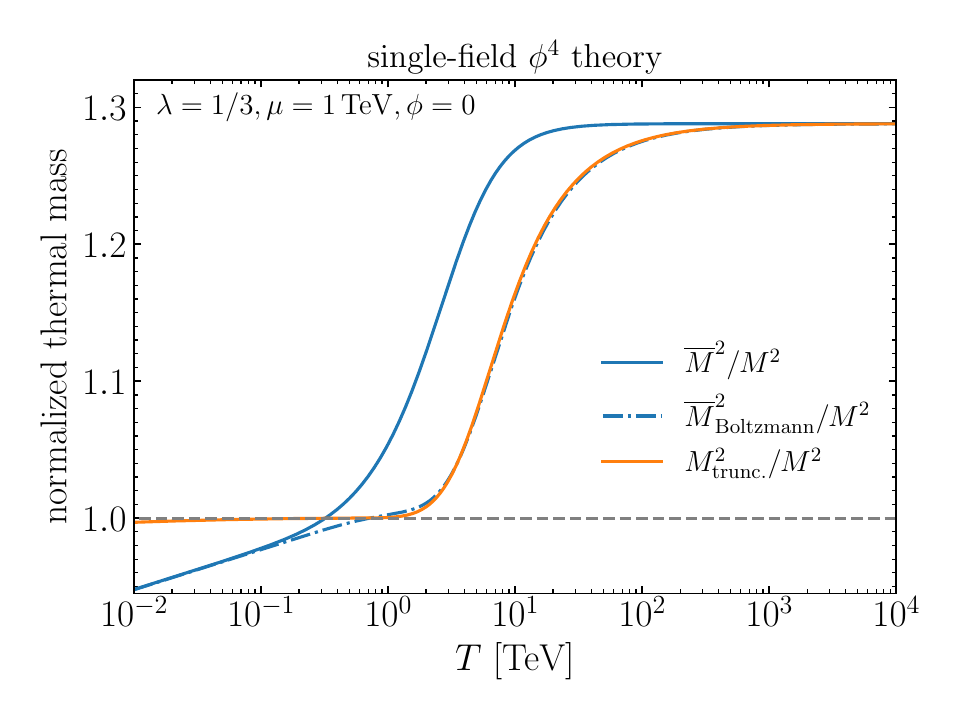}
    \caption{Different approximations of the thermal masses (blue: high-temperature thermal mass, $\overline{M}^2$, see \cref{eq:M2highT}; blue dot-dashed: high-temperature thermal mass with additional Boltzmann factor, $\overline{M}_\text{Boltzmann}^2$, see \cref{eq:thermal_masses_Boltzmann}; orange: tree-level mass plus full one-loop correction, $M^2|_\text{trunc.}$, see \cref{eq:M2at1L}) normalized to the thermal mass obtained by solving the gap equation $M^2$ as a function of the temperature. All masses are evaluated for $\phi=0$.}
    \label{fig:mass_iteration_1singlet}
\end{figure}

We start our numerical discussion by comparing different approximations for the thermal masses as a function of the temperature for $\phi=0$ in \cref{fig:mass_iteration_1singlet}. The thermal masses are normalized to the full thermal mass computed by solving the gap equation.

As expected, the high-temperature expansion of the thermal mass $\overline{M}^2$ is close to the full solution of the gap equation $M^2$ for low temperatures since the overall thermal corrections are negligible (see blue curve and \cref{eq:M2highT}). The small $\sim 5\%$ deviation for temperatures close to zero is explained by loop corrections induced by the Coleman-Weinberg potential, which are not taken into account in the high-temperature expansion. For temperatures above $1\tev$, the ratio of the high-temperature thermal mass to the full thermal mass increases quickly until the curve converges at $\sim 1.3$ for $T\gtrsim 10\tev$ and then stays constant for higher temperatures. The constant off-shift is caused by temperature-dependent higher-loop contributions which are generated by solving the gap equation but not taken into account in the high-temperature thermal mass. This can easily be seen by solving \cref{eq:gap_equation_example} for $\phi=0$ without expanding in $\lambda$ (as done for \cref{eq:gap_equation_example_sol}). Rearranging \cref{eq:gap_equation_example}, we can write
\begin{align}
    \frac{\overline{M}^2}{M^2} = 1 + \frac{3}{4\pi}\lambda \frac{T}{M} \simeq 1 + \frac{3}{4\pi}\lambda \frac{T}{\overline{M}} \overset{T\to\infty}{\simeq} 1 + \frac{3}{2\pi}\sqrt{\lambda} \overset{\lambda=1/3}{\simeq} 1.28\;,
\end{align}
where we solved the equation iteratively in the second step. This approximative first-iteration result is in very good numerical agreement with the exact result in \cref{eq:M2highT}. So, we conclude that a one-loop order high-temperature expansion of the thermal mass is only a good approximation in the high-temperature limit if $\lambda \ll 1$ while PD relying on the full solution of the gap equation holds for much higher $\lambda$ values. As \cref{fig:mass_iteration_1singlet} shows, $\beta \simeq \frac{\overline{M}^2}{M^2} - 1$ is well below corroborating the reliability of the PD prescription (see discussion at the end of \cref{subsec:tadpole}).

If an additional Boltzmann factor is included in the equation for the high-temperature mass (see \cref{eq:thermal_masses_Boltzmann}, blue dot-dashed curve), the ratio stays close to one for a slightly larger temperature range than without the Boltzmann factor. The overall agreement with the full thermal mass is, however, not substantially improved.

If instead the full one-loop correction (without any high-temperature expansion) is used to calculate the thermal mass, $M^2|_\text{ trunc.}$, (see orange curve and \cref{eq:M2at1L}), the low-temperature behaviour of the full thermal mass is very well captured (since now also the one-loop corrections from the Coleman-Weinberg potential are taken into account). At $T\gtrsim 4\tev$, where temperature-dependent higher-loop order corrections start to become relevant, however, also this approximation fails to capture the temperature dependence of the full thermal mass.

\begin{figure}
    \centering
    \includegraphics[width=.49\textwidth,trim={.5cm 0 .5cm 0},clip]{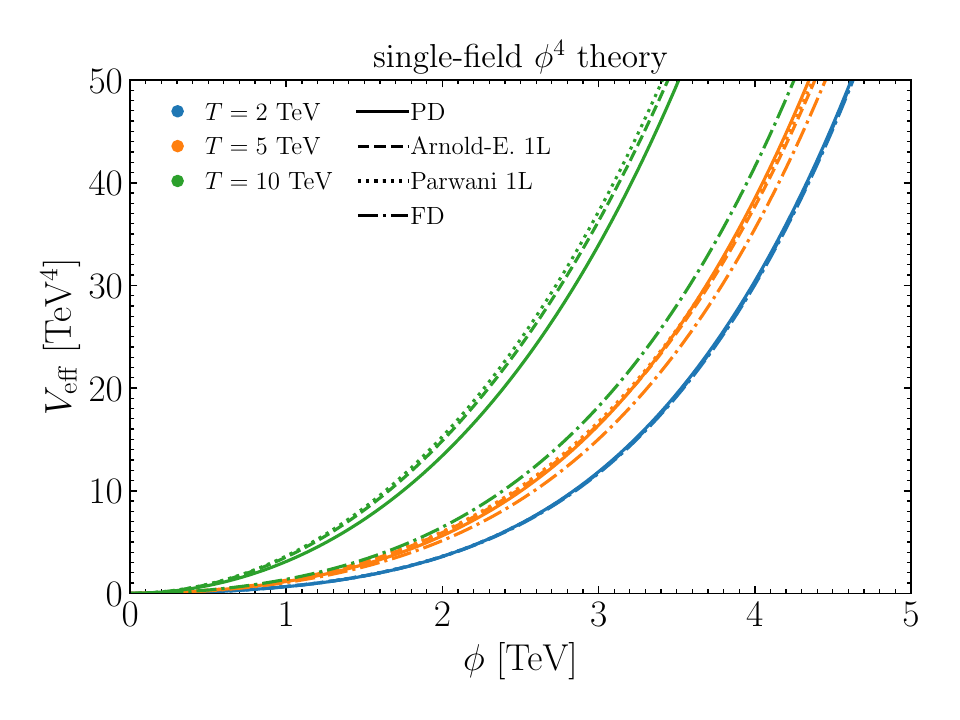}
    \includegraphics[width=.49\textwidth,trim={.5cm 0 .5cm 0},clip]{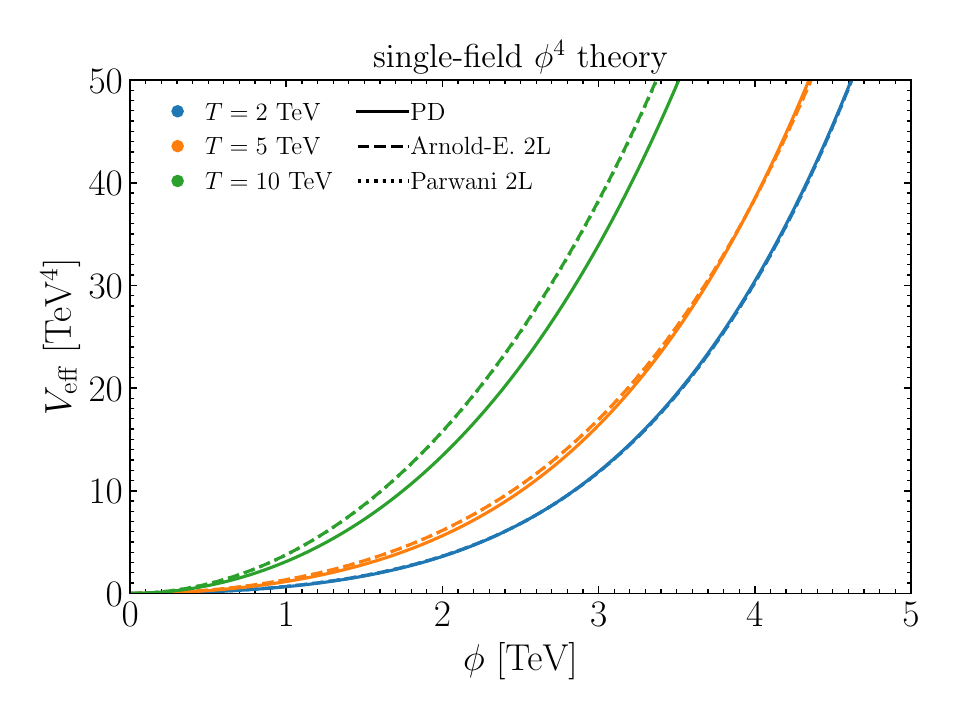}
    \caption{Left: The one-loop effective potential in the one-field $\phi^4$ model evaluated using different resummation methods. Right: Same as left, but two-loop corrections are included for the full-dressing approaches.}
    \label{fig:Veff_1singlet}
\end{figure}

The real parts of the effective potential are compared in \cref{fig:Veff_1singlet}. For low temperatures, the thermal corrections are small and, consequently, the different resummation methods yield almost identical results. Only for higher temperatures of $T \gtrsim 10\tev$, differences between the three methods become visible.

In the left panel of \cref{fig:Veff_1singlet}, showing the effective potential evaluated using various resummation methods at the one-loop level, very small differences between Parwani and Arnold-Espinosa resummation are visible originating from subleading thermal corrections which are partially included for Parwani resummation but not for Arnold-Espinosa resummation. This discrepancy is completely gone in the right panel of \cref{fig:Veff_1singlet}, for which the effective potential is evaluated at the two-loop level for the Parwani and Arnold-Espinosa curves. As a consequence of explicitly including the subleading thermal two-loop corrections, whose leading contribution is $\propto \lambda^2 m^2 T^2$, the curves using Parwani and Arnold-Espinosa resummation lie on top of each other. This seemingly signals a well-behaved perturbative convergence of the subleading thermal corrections.

The result using partial dressing lies below the Parwani and Arnold-Espinosa results for high temperatures. By varying the coupling $\lambda$, we confirmed that the difference between the Parwani/Arnold-Espinosa and partial dressing results originates --- as expected --- mainly from $\mathcal{O}(\beta^2)$ contributions, which are correctly included if using partial dressing but not controlled if using Parwani/Arnold-Espinosa resummation.

Interestingly, the difference is increased when comparing the partial-dressing result to the two-loop full-dressing results in comparison to the one-loop full-dressing results. While in principle a difference is expected since partial dressing correctly includes subleading thermal corrections, one would naively expect this difference to shrink down once these subleading thermal corrections are explicitly included at the two-loop order for the full-dressing methods. To understand why this increases the difference between full and partial dressing, it is instructive to understand the proportionalities of the formally leading terms missed by the full dressing method. While the formally leading missed term is $\propto \lambda^2 m^2 T^2$ if computing the effective potential at the one-loop level, it is $\propto \lambda^3 m T^3$ if computing the effective potential at the two-loop level. For high temperatures, this three-loop term is larger than the respective two-loop term demonstrating that including the full two-loop corrections in the full dressing approach can worsen the result. Also explicitly including the term $\propto \lambda^3 m T^3$ might not improve the result since for high temperatures the four-loop term $\propto \lambda^4 T^5/m$ could be even larger. This demonstrates the necessity of correctly resuming also subleading thermal corrections and is in direct correspondence to the behaviour of the thermal masses at high temperatures (see \cref{fig:mass_iteration_1singlet}).

Moreover, we show in the left panel of \cref{fig:Veff_1singlet} the result using full dressing (see \cref{subsec:fulldressing}). As discussed in \cref{subsec:tadpole}, this resummation scheme miscounts diagrams starting at the two-loop level. This is clearly visible by the large difference to the other resummation methods for $T=10\tev$.


\subsubsection*{Two-field $\phi^4$ theory}

Next, we compare the different resummation methods in the two-field $\phi^4$ theory without mixing. We focus on the benchmark point
\begin{align}
    \lambda_1 = \lambda_2 = 1/3, \hspace{.3cm} \lambda_{12} = 2, \hspace{.3cm} \mu_{11}^2 = -4\tev^2, \hspace{.3cm} \mu_{22}^2 = -1\tev^2\,.
\end{align}
The renormalisation scale is again set equal to the temperature. The chosen values for the couplings are of no particular significance. As discussed below, they are well within the regime for which partial dressing is reliable.

\begin{figure}
    \centering
    \includegraphics[width=.65\textwidth,trim={.5cm .5cm .5cm .5cm},clip]{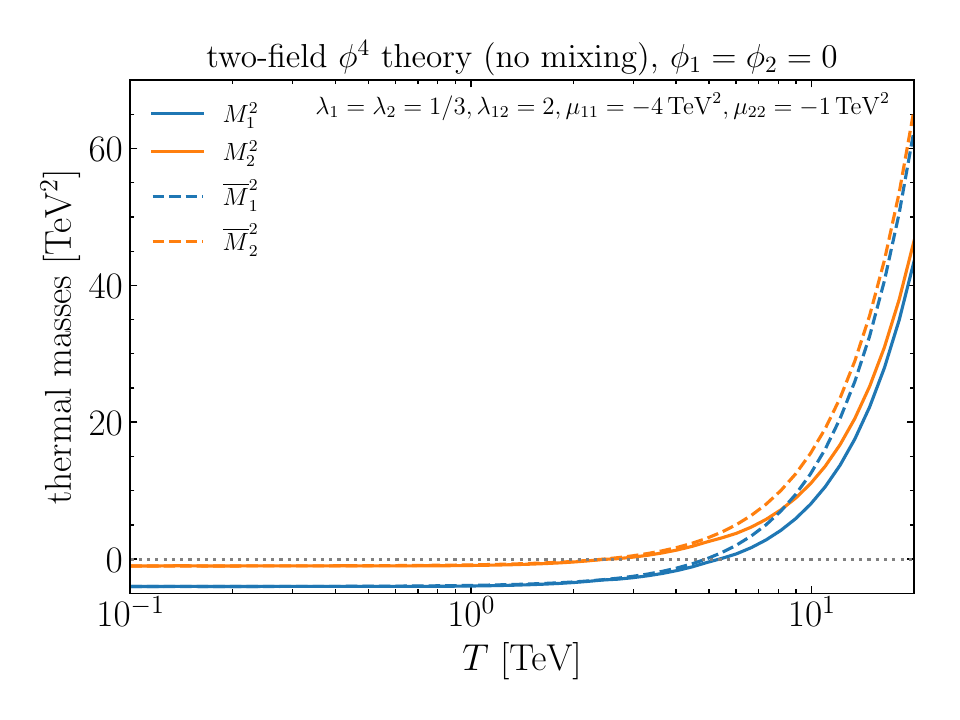}
    \caption{Thermal masses of $\phi_1$ (blue) and $\phi_2$ (orange) in the two-field $\phi^4$ model. The thermal masses are either evaluated by numerically solving the gap equations (solid) or in the high-temperature expansion (dashed).}
    \label{fig:mass_iteration_2singlet_no_mix}
\end{figure}

\cref{fig:mass_iteration_2singlet_no_mix} shows the thermal masses computed either by numerically solving the gap equation (solid lines) or by keeping only the leading term in the high-temperature expansion (dashed lines). As expected, the full and high-temperature versions of the thermal masses agree well for low temperatures since thermal effects are small in general. For very high temperatures, small differences are visible originating from high-order corrections induced by numerically solving the gap equations (see discussion of \cref{fig:mass_iteration_1singlet}). We observe the largest absolute differences for intermediary temperatures, for which the temperature is similar to the tree-level masses. Here, thermal corrections are important but the high-temperature expansion is not yet a good approximation. Moreover, \cref{fig:mass_iteration_2singlet_no_mix} shows that $\beta_{1,2}\sim \frac{\overline{M}_{1,2}^2}{M_{1,2}^2} - 1$ are well below one corroborating the reliability of the partial dressing prescription.

\begin{figure}
    \centering
    \includegraphics[width=.49\textwidth,trim={.5cm 0 .5cm 0},clip]{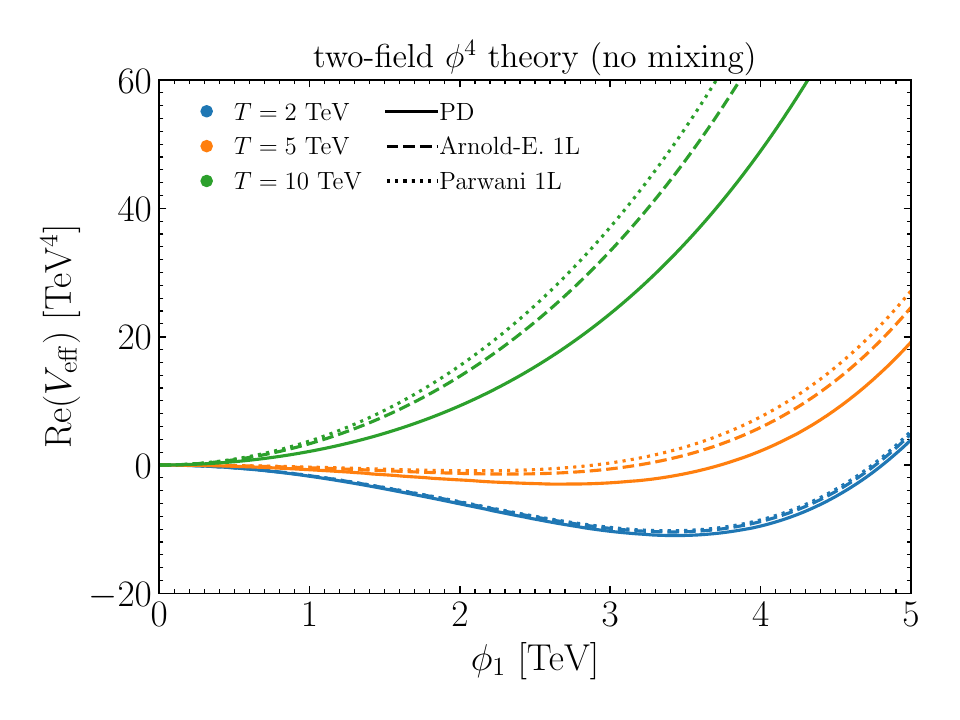}
    \includegraphics[width=.49\textwidth,trim={.5cm 0 .5cm 0},clip]{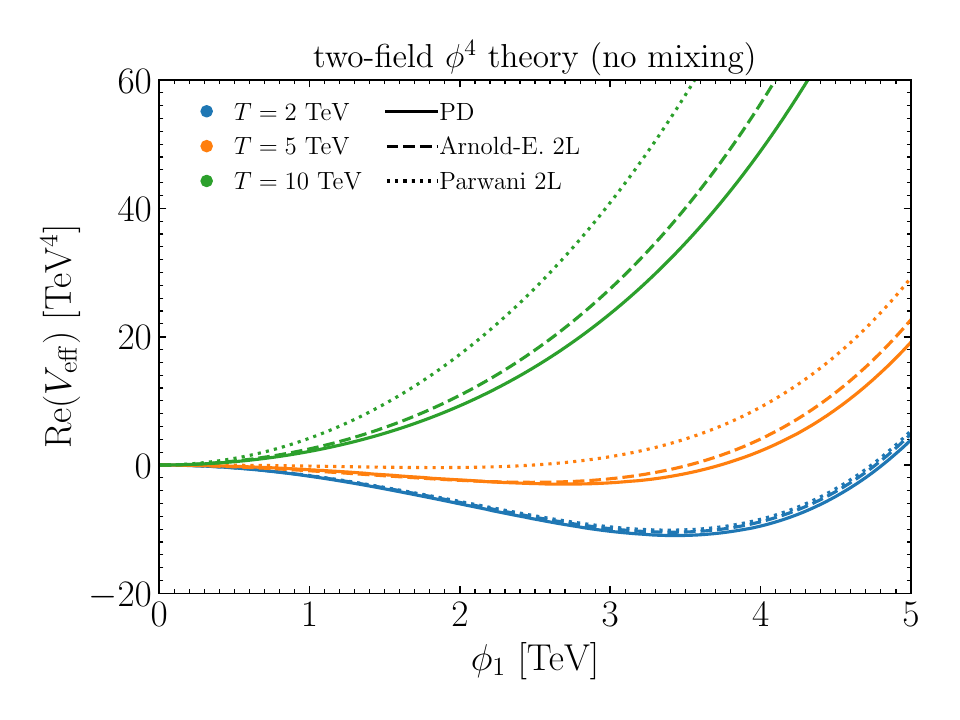}
    \caption{Left: Real part of the one-loop effective potential in the two-field $\phi^4$ model without mixing evaluated using different resummation methods. The imaginary part of the effective potential is negligible compared to the real part. Right: Same as left, but two-loop corrections are included for the full-dressing approaches.}
    \label{fig:Veff_2singlet_no_mix}
\end{figure}

The real part of the effective potential is shown in \cref{fig:Veff_2singlet_no_mix} as a function of $\phi_1$ setting $\phi_2 = 0$, implying that there is no mixing between the fields. As for the one-field $\phi^4$ theory, the Parwani, Arnold-Espinosa, and partial dressing approaches agree well for low temperatures. For higher temperatures, larger differences are visible. In contrast to the single-field case (see \cref{fig:Veff_1singlet}), the difference between the Arnold-Espinosa and Parwani resummation methods is increasing from the one- (left panel of \cref{fig:Veff_2singlet_no_mix}) to the two-loop level (right panel of \cref{fig:Veff_2singlet_no_mix}). This signals that the dominant difference between the two approaches in the given scenario is not of two-loop order but induced by higher-order effects. This is due to the comparably large numerical values for the $\lambda$'s as well as the larger number of fields which enhance the significance of higher-order corrections. This again shows the importance of resuming also subleading effects as achieved in the partial dressing approach.


\section{Toy model for symmetry non-restoration}
\label{sec:servant_model}

Next, we discuss the phenomenon of EWSNR. EWSNR refers to situations in which the EW symmetry is not only broken at low temperatures but also not restored at high temperatures (or the restoration is delayed up to very high energies). This is particularly interesting from the point of resuming thermal corrections. In order for EWSNR to occur, the thermal corrections need to dominate over the tree-level mass turning the squared thermal mass negative and thereby ensuring that the EW symmetry is broken.

We base our discussion on a toy model for symmetry non-restoration which was presented in \ccite{Servant:2018}. Its potential is given by
\begin{align}
    V^{(0)}(\phi,\chi,S) ={}& \frac{1}{2}\mu_S^2 S^2 + \frac{1}{2}\mu_\chi^2 \sum_i \chi_i^2 + \frac{1}{2}\mu_\phi^2 \phi^2 \nonumber\\
    & + \frac{1}{4}\lambda_\phi \phi^4 + \frac{1}{4}\lambda_\chi \sum_i \chi_i^4  + \frac{1}{4}\lambda_S S^4  + \frac{1}{4}\lambda_{\phi\chi} \phi^2\sum_i\chi_i^2 +  \frac{1}{4}\lambda_{\phi S} \phi^2 S^2,
\end{align}
where $S$ and $\chi_i$ are vectors of dimension $N_S$ and $N_{\chi_i}$. The index $i$ is a generation index which runs from 1 to $N_\text{gen}$. We will only evaluate the potential at zero field values for $S$ and the $\chi_i$. Therefore, all $\chi_i$ have the same mass $m_\chi$ and we will drop the sum over $i$ and instead use $N_\chi = N_{\chi_i} N_\text{gen}$. The parameters are chosen such that only $\phi$ develops a non-zero vacuum expectation value. Therefore, we are only interested in the $\phi$ direction and set $\chi_i$ and $S$ to zero for the evaluation of the effective potential.

In the high-temperature limit, the thermal masses are given
\begin{subequations}\label{eq:Servant_thermal_masses_highT}
\begin{align}
    \overline{M}_{\chi}^2 &= m_{\chi}^2 + T^2 c_\chi = m_{\chi_i}^2 + T^2 \bigg[\frac{1}{12}(N_{\chi} + 2)\lambda_\chi + \frac{1}{24}\lambda_{\phi\chi}\bigg]\,, \\
    \overline{M}_{\phi,\text{Boltzmann}}^2 &= m_{\phi}^2 + T^2 c_\phi = \nonumber\\ 
    &= m_{\phi}^2 + T^2 \bigg[\frac{1}{12}(N_{\phi} + 2)\lambda_\phi + \frac{1}{24}N_\chi\lambda_{\phi\chi} + \frac{1}{24}N_S\lambda_{\phi S} e^{-m_S/T}\bigg]\,, \\
    \overline{M}_{S,\text{Boltzmann}}^2 &= m_{S}^2 + T^2 c_S = m_{S}^2 + T^2 \bigg[\frac{1}{12}(N_{S} + 2)\lambda_S e^{-m_S/T} + \frac{1}{24}N_\phi\lambda_{\phi S}\bigg]\,.
\end{align}
\end{subequations}
This also defines the coefficients $\{c_{\chi},c_{S},c_{\phi}\}$, which are used later. Since the parameters are chosen such that $m_\phi\sim m_\chi \ll m_S$, the thermal contributions of $S$ are multiplied by the Boltzmann factor $e^{-m_S/T}$ to better approximate the full thermal loop function for $T\lesssim m_S$ (see also \cref{eq:thermal_masses_Boltzmann}). 

To achieve symmetry non-restoration, $\lambda_{\phi\chi}$ is chosen to be negative such that the thermal mass of $\phi$ becomes negative. For $T\sim m_S$, the thermal contribution of $S$ compensates for the negative contribution of the $\chi_i$ resulting in the eventual symmetry restoration at $T\gtrsim m_S$. 

Since the stability of the potential at the tree level requires 
\begin{align}
    \lambda_{\phi\chi} > - 2 \sqrt{\frac{\lambda_\phi \lambda_\chi}{N_\text{gen}}}\,,
\end{align}
a large number of generations is required to ensure symmetry non-restoration by pushing the thermal mass $\overline{M}_\phi^2$ to become negative,\footnote{In more realistic models, the negative BSM contributions to for instance the thermal mass of the Higgs boson must also overcome the positive contributions of other SM particles.} while still satisfying perturbative unitarity bounds (see \ccite{Servant:2018} for more details).


\subsection{One-loop effective potential}

The one-loop effective potential is given by
\begin{align}
    V^{(1)}(\phi,\chi,S) ={}& \mathcal{J}(m_\phi) +  N_\chi \mathcal{J}(m_{\chi}) + N_S \mathcal{J}(m_S)\,.
\end{align}
The counterterm contributions are
\begin{align}
    V^{(1,\text{CT})}(\phi,\chi_i = 0, S = 0) ={}& \frac{1}{2}\delta^{(1)}\mu_\phi^2 \phi^2 + \frac{1}{4}\delta^{(1)}\lambda_\phi \phi^4 \,.
\end{align}
We choose to renormalize $\lambda_\phi$ in the \MS scheme. For the renormalization of $\mu_\phi$, we include a finite piece to the counterm
\begin{align}
    \delta^{(1)}\mu_\phi^2\Big|_\text{fin} = - \frac{1}{2}\phi^2 \left[\frac{\partial^2}{\partial\phi^2} V^{(1)}\right]_{\phi = \chi_i = S = 0, T = 0}^{\mathcal{O}(\mu_S^2)} = \frac{1}{64\pi^2}N_S \lambda_{\phi S} \phi^2 \mu_S^2 \left(1 - \ln\frac{\mu_S^2}{\mu_R^2}\right)\,,
\end{align}
where the superscript $\mathcal{O}(\mu_S^2)$ denotes that only the leading contribution proportional to $\mu_S^2$ is considered. This counterterm is chosen to absorb the very large zero-temperature loop corrections to $m_\phi$ from $S$ (due to $\mu_S \gg |\mu_\phi|$) into the definition of $\mu_\phi$.


\subsection{Two-loop effective potential}

The genuine two-loop corrections to the potential (with the sunset contributions in the last two lines) are given by
\begin{align}
    V^{(2,\text{gen})}(\phi,\chi_i=0,S=0) ={}& \frac{3}{4}\left[\lambda_\phi\mathcal{I}(m_\phi)^2 + N_S\lambda_S \mathcal{I}(m_S)^2 + N_\chi \mathcal{I}(m_{\chi_i})^2\right] \nonumber\\
    & + \frac{1}{4}\bigg[(N_S^2 - N_S) \lambda_S \mathcal{I}(m_S)^2 + N_\chi(N_{\chi_i} - 1)\lambda_\chi \mathcal{I}(m_\chi)^2 \nonumber\\
    &\hspace{.9cm} + N_S \lambda_{\phi S}\mathcal{I}(m_{\phi})\mathcal{I}(m_{S}) + N_\chi \lambda_{\phi\chi}\mathcal{I}(m_{\phi})\mathcal{I}(m_\chi)\bigg] \nonumber\\
    & - 3\lambda_\phi^2\phi^2\mathcal{H}(m_\phi,m_\phi,m_\phi) - \frac{1}{4}\lambda_{\phi\chi}^2N_\chi\phi^2\mathcal{H}(m_\phi,m_\chi,m_\chi) \nonumber\\
    & - \frac{1}{4}\lambda_{\phi S}^2N_S\phi^2\mathcal{H}(m_\phi,m_S,m_S)\,,
\end{align}
where here we already set $\chi_i = S = 0$. The counterterm contributions at the two-loop level are
\begin{align}
    V^{(2,\text{CT})}(\phi,\chi_i=0,S=0) ={}& \frac{1}{2}\delta^{(2)}\mu_\phi^2 \phi^2 + \frac{1}{4}\delta^{(2)}\lambda_\phi \phi^4 \nonumber\\
    & + \frac{1}{2}\delta^{(1)}\mu_\phi^2 \mathcal{I}(m_\phi) + \frac{1}{2}N_\chi\delta^{(1)}\mu_\chi^2 \mathcal{I}(m_\chi) + \frac{1}{2}N_S\delta^{(1)}\mu_S^2 \mathcal{I}(m_S) \nonumber\\
    & + \frac{3}{2}\delta^{(1)}\lambda_\phi \phi^2 \mathcal{I}(m_\phi) + \frac{1}{4}\delta^{(1)}\lambda_{\phi S} \phi^2 \mathcal{I}(m_S) \nonumber\\
    & + \frac{1}{4}\delta^{(1)}\lambda_{\phi\chi} \phi^2 \mathcal{I}(m_\chi)\;.
\end{align}
The first line contains the needed two-loop counterterm; the two last lines represent the subloop renormalization. We checked analytically that all UV divergencies and $\epsilon^1$ pieces of the loop integrals (with $\epsilon$ being the UV regulator) cancel.

The thermal mass counterterms, relevant for Parwani and Arnold-Espinosa approaches, give additional two-loop contributions:
\begin{align}
    V^{(2,\text{thermal-CT})}(\phi,\chi,S) = -\frac{1}{2}T^2 \left(c_\phi \mathcal{I}(m_\phi^2) + N_\chi c_\chi \mathcal{I}(m_\chi^2) + N_S c_S \mathcal{I}(m_S^2)\right)\,.
\end{align}
We checked explicitly that, in the high-temperature expansion, all \order{T^3} terms cancel. This cross-check was performed separately for Parwani and Arnold/Espinosa resummation in the limits $m_S \ll T$ and $m_S \gg T$. The two-loop counterterm for $\mu_\phi$ is again chosen such that the very large zero-temperature loop corrections to $m_\phi$ from $S$ are absorbed into the definition of $\mu_\phi$. 


\subsection{Numerical comparison}

For our numerical comparison, we choose the benchmark point already used in \ccite{Servant:2018}:
\begin{align}
    & N_\text{gen} = 12,\hspace{.3cm}N_{\chi_i}=4,\hspace{.3cm}N_S=12, \nonumber\\
    & \mu_\phi^2 = - 0.01 \tev^2,\hspace{.3cm} \mu_\chi^2 = 0.01 \tev^2,\hspace{.3cm} \mu_S^2 = 400\tev^2,\nonumber\\
    & \lambda_\phi = 0.1 \tev^2,\hspace{.3cm} \lambda_\chi = 0.5,\hspace{.3cm} \lambda_S = 1,\hspace{.3cm} \lambda_{\phi\chi} = -0.1,\hspace{.3cm} \lambda_{\phi S} = 1\,.
\end{align}
This benchmark is chosen to realize symmetry non-restoration for $T\lesssim 10\tev$. At higher temperatures, the $S$ field eventually ensures symmetry restoration.

We note that this benchmark point is non-perturbative in the sense that the expansion parameter $\beta$ for the scalars $\chi_i$ is larger than one for most parts of the considered parameter space (as visible from \cref{fig:servant_thermal_masses} discussed below). In this way, we ensure comparability with the results of \ccite{Servant:2018}. Moreover, we are mainly interested in the behaviour of the various resummation schemes for large negative squared thermal masses, which drive EWSNR, and not so much in the actual prediction.


\subsubsection*{Thermal masses}

\begin{figure}
    \centering
    \includegraphics[width=.65\textwidth,trim={.5cm .5cm .5cm .5cm},clip]{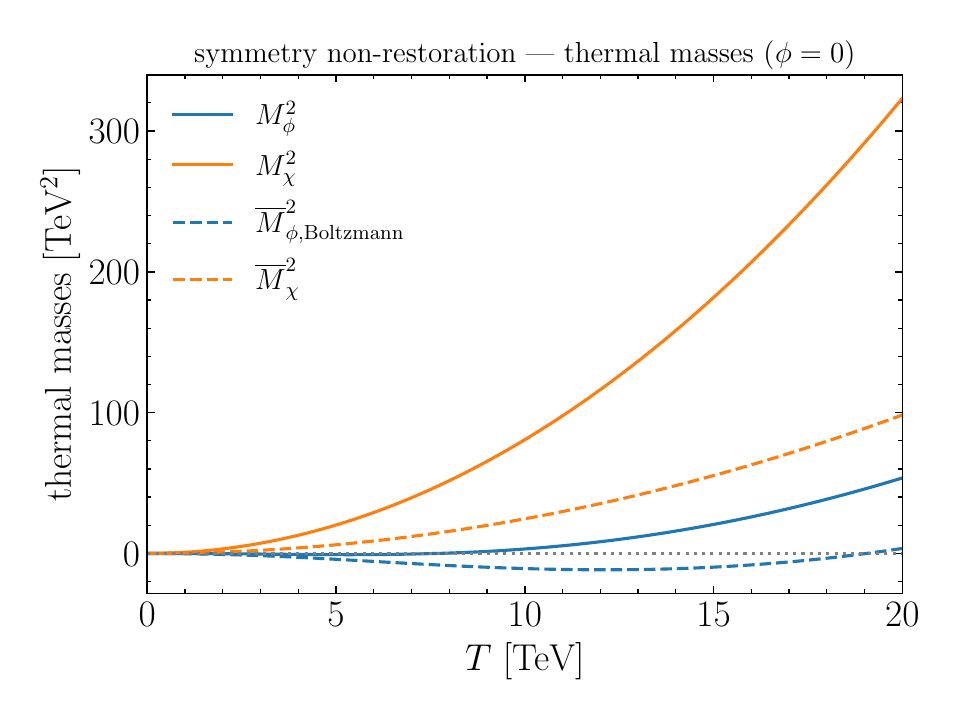}
    \caption{Thermal masses of $\phi$ (blue) and $\chi_i$ (orange) in the symmetry non-restoration toy model of \ccite{Servant:2018}. The thermal masses are either evaluated by numerically solving the gap equations (solid) or in the high-temperature expansion (dashed).}
    \label{fig:servant_thermal_masses}
\end{figure}

We start with a comparison of the thermal masses $M_\phi$ and $M_\chi$ in \cref{fig:servant_thermal_masses}. For the scalar $\phi$, the thermal mass calculated by solving the gap equation $M_\phi^2$ is negative only for small temperatures ($T\lesssim 7\tev$). If instead the high-temperature expansion is used (see \cref{eq:Servant_thermal_masses_highT}), the squared thermal mass stays negative until much higher masses ($T\sim 19\tev$). This difference has two origins: 1) the thermal loop functions appearing in the gap equation is not expanded in the high-temperature limit; 2) solving the gap equation numerically effectively includes higher-order corrections (see discussions in \cref{subsec:numcompare}). For the scalars $\chi$, the differences between the thermal mass obtained by solving the gap equation and the high-temperature expansion of \cref{eq:Servant_thermal_masses_highT} is even more pronounced due to the relatively large number of fields $N_\chi = 48$ coupled to each other.


\subsubsection*{Effective potential}

\begin{figure}
    \centering
    \includegraphics[width=.49\textwidth,trim={0.7cm 0.7cm 0.7cm 0.7cm}, clip]{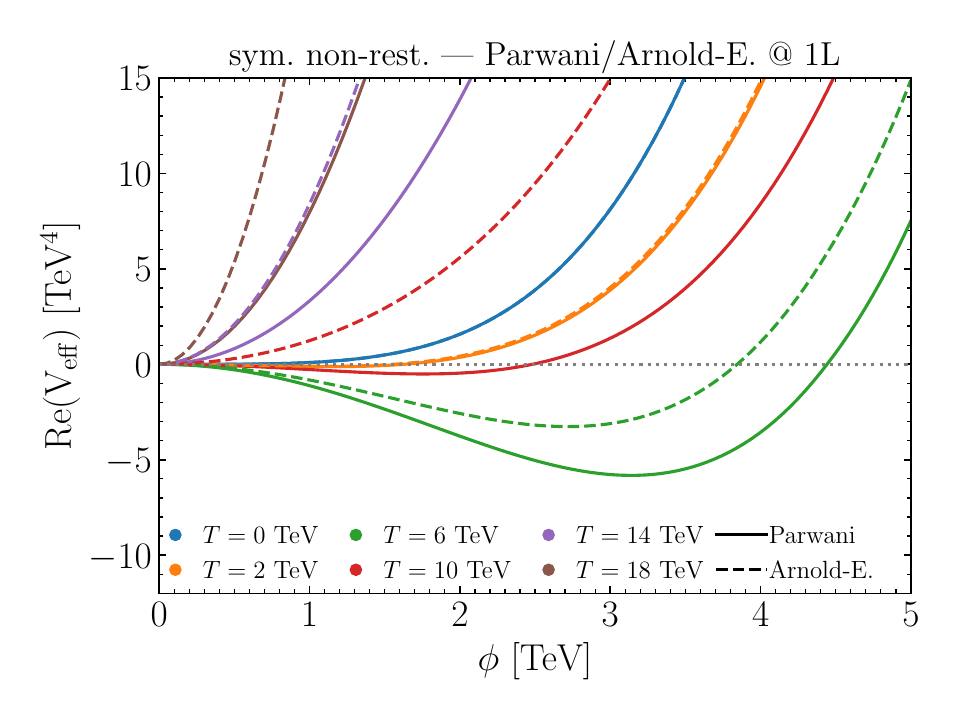}
    \hspace{.01cm}
    \includegraphics[width=.49\textwidth,trim={0.7cm 0.7cm 0.7cm 0.7cm}, clip]{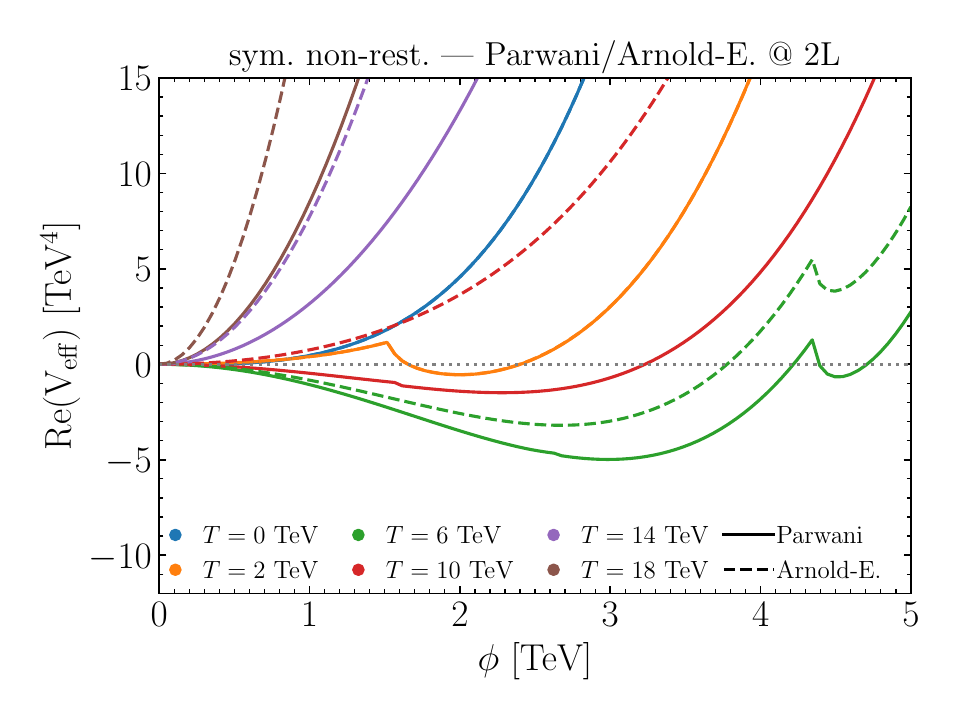}\\
    \vspace{.3cm}
    \includegraphics[width=.49\textwidth,trim={0.7cm 0.7cm 0.7cm 0.7cm}, clip]{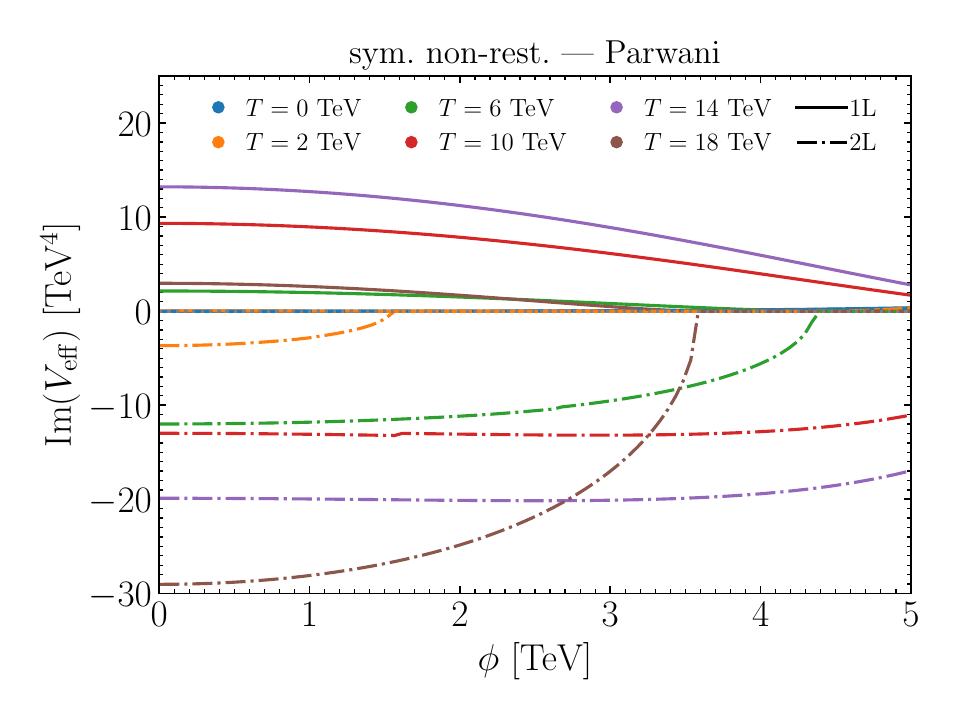}
    \hspace{.01cm}
    \includegraphics[width=.49\textwidth,trim={0.7cm 0.7cm 0.7cm 0.7cm}, clip]{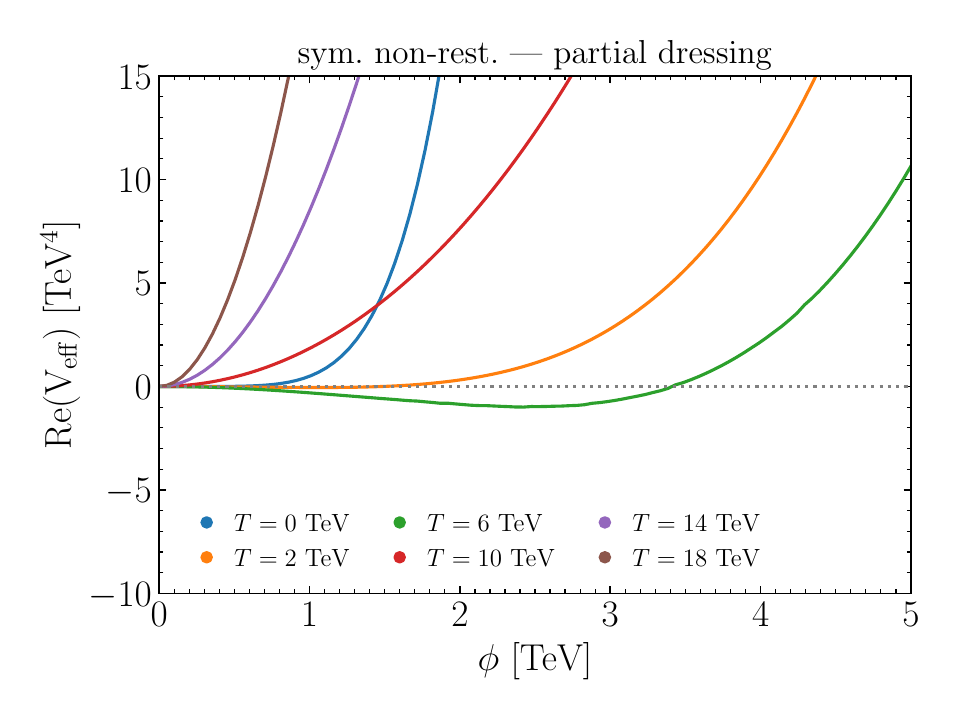}
    \caption{Upper left: Real part of the one-loop effective potential for the symmetry non-restoration model of \ccite{Servant:2018} evaluated using Arnold-Espinosa and Parawni resummation. For the thermal resummation, we employ either the Parwani (solid) or Arnold-Espinsa (dashed) methods. Upper right: Same as upper left, but the effective potential is evaluated at the two-loop level. Lower left: Imaginary part of the effective potential evaluated at the one- and two-loop level using Parwani resummation. Lower right: Real part of the one-loop effective potential evaluated using partial dressing.}
    \label{fig:servant_comparison}
\end{figure}

Next, we study the effective potential itself. \cref{fig:servant_comparison} shows the dependence of the effective potential on the value of $\phi$ for various temperature values.

In the upper left panel, we show the real part of $V_\text{eff}$ calculated at the one-loop level using Arnold-Espinosa and Parwani. For both Parwani resummation (solid lines) and Arnold-Espinosa reummation (dashed lines), electroweak symmetry non-restoration is clearly visible for $T\lesssim 6\tev$. For higher temperatures, the thermal contribution of the $S$ triggers the eventual symmetry restoration. While for low temperatures both methods yield very similar results, there is an increasing difference for higher temperatures. As discussed in \cref{sec:single_field}, this difference arises from subleading super daisy-like contributions which are partially included in the Parwani approach.

One natural way to reduce the difference between both methods and thereby the theoretical uncertainty is to explicitly include the full two-loop corrections as outlined above. The results are shown in the upper right panel of \cref{fig:servant_comparison}. Interestingly, the difference between Parwani and Arnold/Espinosa resummation is not reduced if including the full two-loop corrections but of similar size as at one-loop level. After including the full two-loop corrections, the difference between the two resummation schemes is of three-loop order. The fact that the difference is not decreased when going from the one- to the two-loop level signals that also the three-loop difference is sizeable. This is a consequence of the large multiplicity of fields and demonstrates that a resummation of subleading super-daisy corrections is needed.

It is even more interesting that some of the two-loop results feature unphysical kinks. These kinks appear if one of the squared thermal masses crosses zero (see e.g.\ the scenario of \cref{fig:servant_thermal_masses}). For a negative mass squared, the loop functions develop an imaginary part. While the imaginary parts of a negative tree-level mass squared cancel once large thermal effects are resumed as shown in \ccite{Delaunay:2007wb}, an imaginary part remains if one of the squared thermal masses becomes negative. We show this remaining imaginary part of the effective potential calculated using Parwani resummation (without normalizing the potential to zero at the origin) in the lower left panel of \cref{fig:servant_comparison}.\footnote{If using Arnold-Espinosa resummation, the imaginary parts of similar magnitude.} We see that imaginary parts already occur at the one-loop level. The size of the imaginary parts is, however, enhanced at the two-loop level. This is due to a mismatch between the figure-eight diagrams and the thermal counterterm contributions. While the full thermal loop functions (without any high-temperature expansion) are considered for the former contribution, the thermal counterterms appearing in the latter contribution are by definition derived in the high-temperature expansion. This (unavoidably) different treatment of the two contributions artificially enhances the imaginary part of the effective potential and also induces kinks in the real part via products of two imaginary parts. 

It is actually expected that the perturbatively calculated effective potential should develop an imaginary part in the regions where the classical potential becomes non-convex. As discussed in \ccite{Weinberg:1987vp}, this imaginary part can be understood as corresponding to the decay rate of modes expanded around unstable regions of field space. Large imaginary parts then call into question the validity of the perturbatively calculated effective potential. The real part of the perturbatively calculated effective potential is only trustworthy so long as the imaginary part remains small enough relative to the real part that the field can be considered ``stable''. 


The problems with the perturbative convergence and large imaginary parts are completely avoided if we use partial dressing. We show the corresponding result in the lower right panel of \cref{fig:servant_comparison}. As discussed in \cref{subsec:numcompare}, partial dressing does not only resum daisy but also super-daisy contributions. Since no high-temperature expansion is applied and all contributions are treated on the same footing, the imaginary part of the effective potential (not shown) is negligible compared to the real part of the effective potential. Consequently, no kinks appear in the result and calculations based on the pertubatively calculated effective potential are trustworthy.


\section{Resummation in multi-field \texorpdfstring{$\phi^4$}{phi\^4} theory with mixing}
\label{sec:multi_field_mixing}

After discussing resummation in multi-field theories without mixing, we now turn to the case with mixing between the scalar fields. For simplicity, we focus on a simple toy model consisting of two real scalar fields $\phi_1$ and $\phi_2$ with tree-level potential
\begin{equation}\label{V0}
    V_0 = - \frac{\mu_1^2}{2} \phi_1^2 - \frac{\mu_2^2}{2} \phi_2^2 + \frac{\lambda_1}{4} \phi_1^4 + \frac{\lambda_2}{4} \phi_2^4 + \frac{\lambda_{12}}{4} \phi_1^2 \phi_2^2 \,.
\end{equation}
We allow both fields to potentially develop a zero-temperature vacuum expectation value. While this significantly complicates several formal aspects of the resummation procedure, it is nonetheless important that we allow them to mix in anticipation of concrete BSM applications. 

At the tree level, the scalar mass matrix is given by
\begin{equation}
\begin{split}
    \mathcal{M}^2(\phi_1, \phi_2) & \equiv \begin{pmatrix} m_{11}^2 & m_{12}^2 \\ m_{12}^2 & m_{22}^2 \end{pmatrix} = \begin{pmatrix} -\mu_1^2 + 3 \lambda_1 \phi_1^2 +\frac{\lambda_{12}}{2} \phi_2^2 & \lambda_{12} \phi_1 \phi_2 \\ \lambda_{12} \phi_1 \phi_2 & -\mu_2^2 + 3 \lambda_2 \phi_2^2 +\frac{\lambda_{12}}{2} \phi_1^2 \end{pmatrix} \,,
\end{split}
\end{equation}
where the fields $\phi_{1,2}$ take on their background values. This matrix can be diagonalized as $\mathcal{R}_\theta^{-1} \mathcal{M}^2 \mathcal{R}_\theta = \mathcal{M}^2_{\mathrm{diag}} \equiv \text{diag}(m_+^2, m_-^2)$, with mass eigenvalues
\begin{equation}
    m_\pm^2 = \frac{1}{2} \left(m_{11}^2 + m_{22}^2 \pm D \right) \,, \,\,\, \text{with} \,\,\, D = \sqrt{(m_{11}^2 - m_{22}^2)^2 + 4 m_{12}^4} \,.
\end{equation}
It is convenient to parameterize $\mathcal{R}_\theta$ in terms of the tree level mixing angle $\theta$:
\begin{equation}
    \mathcal{R}_\theta = \begin{pmatrix} \cos \theta & \,\, - \sin \theta \\ \sin \theta & \,\, \cos \theta \end{pmatrix} \,, \,\,\, \sin 2 \theta = \frac{2 m_{12}^2}{\sqrt{(m_{11}^2 - m_{22}^2)^2 + 4 m_{12}^4}} \,,
\end{equation}
which can then be used to relate $\phi_{1,2}$ to the mass eigenstates $\phi_\pm$:
\begin{equation}
    \begin{pmatrix} \phi_1 \\ \phi_2 \end{pmatrix} = \mathcal{R}_\theta \begin{pmatrix} \phi_+ \\ \phi_- \end{pmatrix} \,.
\end{equation}

The background field-dependent mass eigenstates enter into the 1-loop contribution to the effective potential as
\begin{equation}
    V_{1\text{-loop}} = \mathcal{J}[m_+] + \mathcal{J}[m_-] \,,
\end{equation}
where the $\mathcal{J}$ function is defined in \cref{sec:loop_funcs}.

\subsection{High-temperature expansion and truncated full dressing}
\label{sec:TFD}

To gain some intuition, we will first consider the high-temperature limit, in which the field-dependent part of the one-loop effective potential reads
\begin{equation}\label{V1}
    V_{1\text{-loop}} \simeq \frac{T^2}{24} \left(m_+^2 + m_-^2 \right) - \frac{T}{12\pi} \left( m_+^3 + m_-^3 \right) - \frac{L}{64 \pi^2} \left( m_+^4 + m_-^4 \right) \,,
\end{equation}
where $L = \log \left( \frac{\mu_R^2}{T^2} \right) + 2 (\gamma_E - \ln \pi)$ is field independent. Considering just the leading contribution $\sim T^2$, the one-loop corrected effective potential is
\begin{equation}
    V_\text{eff}^{(1)} \xrightarrow{\text{high}\,T} \frac{1}{2} \left( - \mu_1^2 + c_1 T^2 \right) \phi_1^2 + \frac{\lambda_1}{4} \phi_1^4 + \frac{1}{2} \left( - \mu_2^2 + c_2 T^2 \right) \phi_2^2 + \frac{\lambda_2}{4} \phi_2^4 + \frac{\lambda_{12}}{4} \phi_1^2 \phi_2^2 \,,
\end{equation}
where we have defined the coefficients
\begin{equation}
    c_1 = \frac{1}{24}(6 \lambda_1 + \lambda_{12}) \,, \,\,\, c_2 = \frac{1}{24}(6 \lambda_2 + \lambda_{12}) \,.
\end{equation}
Letting $M_i^2(\Phi,T) = m_i^2(\Phi) + \delta m_i^2(\Phi,T)$ be the thermally corrected mass, the truncated gap equation is simply
\begin{equation}
    M_i^2 = \frac{\partial^2}{\partial \phi_i^2} V_{\mathrm{eff}}^{(1)} \,,
\end{equation}
which leads to the finite-temperature mass matrix
\begin{equation}
\begin{split}
    \mathcal{M}_T^2(\phi_1, \phi_2, T) & \equiv \begin{pmatrix} M_{11}^2 & M_{12}^2 \\ M_{12}^2 & M_{22}^2 \end{pmatrix}\\
    & = \begin{pmatrix} -\mu_1^2 + c_1 T^2 + 3 \lambda_1 \phi_1^2 +\frac{\lambda_{12}}{2} \phi_2^2 & \lambda_{12} \phi_1 \phi_2 \\ \lambda_{12} \phi_1 \phi_2 & -\mu_2^2 + c_2 T^2 + 3 \lambda_2 \phi_2^2 +\frac{\lambda_{12}}{2} \phi_1^2 \end{pmatrix} \,.
\end{split}
\end{equation}
Diagonalizing $\mathcal{M}_T^2$ yields the finite-temperature mass eigenstates $M_\pm^2(\Phi,T)$,
\begin{equation}
    M_\pm^2 = \frac{1}{2} (M_{11}^2 + M_{22}^2 \pm \mathcal{D}) \,, \,\,\, \text{with} \,\,\, \mathcal{D} = \sqrt{(M_{11}^2 - M_{22}^2)^2 + 4 M_{12}^2} \,,
\end{equation}
as well as the finite-temperature mixing angle $\Theta(\phi_1,\phi_2,T)$,
\begin{equation}
    \sin 2 \Theta = 2 M_{12}^2/\mathcal{D} \,.
\end{equation}
With these preliminaries out of the way, we now turn to the resummation of the 1-loop effective potential.

In the truncated full dressing (TFD) prescription, resummation amounts to simply replacing $m_i^2 \rightarrow M_i^2|_\text{trunc.}$ on the level of the effective potential: $V_{\mathrm{eff}}^{\text{TFD}} = V_{\mathrm{eff}} \big|_{M_i^2|_\text{trunc.}}$, with $i=\pm$ labeling the finite-temperature mass eigenstates. In the high-temperature expansion, the resummed potential reduces to the one obtained with the Parwani prescription, namely 
\begin{equation}\label{eq:Veff_mixing_Parwani}
    V_{\mathrm{eff}}^{\text{TFD,Parwani}} = V_0 + \frac{T^2}{24} \left(\overline{M}_+^2 + \overline{M}_-^2 \right) - \frac{T}{12\pi} \left( \overline{M}_+^3 + \overline{M}_-^3 \right) - \frac{L}{64 \pi^2} \left(\overline{M}_+^4 + \overline{M}_-^4 \right) \,,
\end{equation}
or equivalently
\begin{align}
    V_{\mathrm{eff}}^{\text{TFD,Parwani}} ={}& \frac{1}{2}\left(- \mu_1^2 + c_1 T^2\right) \phi_1^2 + \frac{\lambda_1}{4} \phi_1^4 + \frac{1}{2}\left(- \mu_2^2 + c_2 T^2 \right) \phi_2^2 + \frac{\lambda_2}{4} \phi_2^4 + \frac{\lambda_{12}}{4} \phi_1^2 \phi_1^2 \nonumber \\
    & - \frac{T}{12\pi} \left( \overline{M}_+^3 + \overline{M}_-^3 \right) - \frac{L}{64 \pi^2} \left( \overline{M}_+^4 + \overline{M}_-^4 \right) \,.
\end{align}
Using the Arnold-Espinosa prescription, the thermal mass is only inserted in the $T \overline{M}^3$ terms.


\subsection{Partial dressing}
\label{sec:mixing_PD}

As discussed in \cref{sec:single_field,sec:multi_field_mixing}, truncated dressing suffers from various issues. First, it does not resum subleading super-daisy corrections. Second, it unavoidably relies on a high-temperature expansion of the thermal masses, which often is not justified. Third, mismatches between the treatment of various two-loop contributions lead to unphysical kinks in the effective potential. As we have discussed, partial dressing avoids these issues. One shortcoming, however, is that prior to this work it was unknown how to apply the partial dressing prescription in the case where multiple scalar fields acquire non-zero vacuum expectation values and mix. Here, we will demonstrate how partial dressing can be applied to the case of mixing scalar fields. 

We start with the gap equations. If we go beyond the leading term in the high-temperature expansion, the gap equations are promoted to a matrix equation
\begin{align}\label{eq:gap_equation_mixing}
\mathcal{M}^2_{T} = \left[
\begin{pmatrix}
\frac{\partial^2}{\partial\phi_1^2} & \frac{\partial^2}{\partial\phi_1\partial\phi_2} \\
\frac{\partial^2}{\partial\phi_1\partial\phi_2} & \frac{\partial^2}{\partial\phi_2^2} 
\end{pmatrix}
V_\text{eff}
\right]_{(m_\pm,s_{2\theta})\rightarrow(M_\pm,s_{2\Theta})},
\end{align}
where $M_\pm$ and $s_{2\Theta}=\sin 2\Theta$ are determined by diagonalizing $\mathcal{M}^2_{T}$. The second derivatives $\frac{\partial^2}{\partial\phi_i\partial\phi_j}V_\text{eff}$ directly correspond to the $\phi_i\phi_j$ two-point functions at zero momentum. The resulting mixing angle relates the original fields $\phi_{1,2}$ to the loop-corrected fields $\Phi_\pm$,
\begin{equation}
    \begin{pmatrix} \phi_1 \\ \phi_2 \end{pmatrix} = \mathcal{R}_\Theta \begin{pmatrix} \Phi_+ \\ \Phi_- \end{pmatrix} \,.
\end{equation}
These loop corrections include both zero-temperature as well as finite-temperature effects.

To solve \cref{eq:gap_equation_mixing} iteratively, it is important to express the right-hand side completely in terms of the masses and the mixing angle. This can be done either by calculating the second derivatives of the effective potential diagrammatically (i.e., in terms of self-energy Feynman diagrams) or by expressing the first and second derivatives of the (field-dependent) masses in terms of the masses and the mixing angle. For example,
\begin{equation}
    \frac{\partial m_+^2}{\partial \phi_1} = \left[ 6 \lambda_1 \sin^2 \theta(\phi_1,\phi_2) + \lambda_{12} \cos^2 \theta(\phi_1,\phi_2) \right] \phi_1 + \lambda_{12} \sin 2 \theta(\phi_1,\phi_2) \, \phi_2 \,.
\end{equation}
In the Feynman-diagrammatic approach, this angular dependence follows directly from the Feynman rules. For example, the coupling of $\phi_1$ to two $\Phi_-$, which appears in the $\phi_1$ tadpole corrections, is given by
\begin{align}
    c(\phi_1, \Phi_-,\Phi_-) ={}& c^2_{\Theta} c(\phi_1,\phi_1,\phi_1) - 2 c_{\Theta} s_{\Theta}  c(\phi_1,\phi_1,\phi_2) + s^2_{\Theta} c(\phi_1,\phi_2,\phi_2) = \nonumber\\
    ={}& \phi_1 \lambda_1 c^2_{\Theta} - 2 \phi_2 \lambda_{12} c_{\Theta} s_{\Theta} +  \phi_1\lambda_{12} s^2_{\Theta}
\end{align}
In the diagrammatic approach, it is furthermore straightforward to also add the dependence on the external momentum. In this case, the matrix $\mathcal{R}_\theta$ becomes non-unitary and can not be parameterized by a single mixing angle. We leave this for future work.

After the determination of the thermal masses and the thermal mixing angle, we insert them into the first derivatives of the effective potential $\partial_i V_{\text{eff}}$ (tadpoles). While for the case of vanishing mixing, we only need to consider one tadpole (the one for the non-vanishing field), which is then integrated to obtain the effective potential, an ambiguity arises in the case of non-vanishing mixing since in general $V_{\text{eff}}^{\text{TPD},1} \neq V_{\text{eff}}^{\text{TPD},2}$ (with $V_{\text{eff}}^{\text{TPD},i} = \int d\phi_i \, \partial_i V_{\text{eff}} \big|_{M_j^2}$).\footnote{$V_{\text{eff}}^{\text{TPD},1}$ correctly captures the $\phi_1$-dependent part of the effective potential but not the $\phi_2$-dependent part. In contrast, $V_{\text{eff}}^{\text{TPD},2}$ correctly captures the $\phi_2$-dependent part of the effective potential but not the $\phi_1$-dependent part.}

A reasonable solution would be to replace the derivative with a gradient $V_{\text{eff}}'\rightarrow \nabla V_{\text{eff}}$ and the integral over $\phi$ with a line integral to the position in field space $(\phi_1^*, \phi_2^*)$ where we intend to evaluate the potential $\int d\phi \rightarrow \int_\mathcal{C}d\vec{s}$. We therefore propose the following multi-field generalization
\begin{equation}
    V_{\text{eff}}^{\text{TPD}} = \int_\mathcal{C} d\vec{s} \cdot \nabla V_{\text{eff}}^{(1)} \big|_{(m_\pm,\theta)\rightarrow(M_\pm,\Theta)} \,,
\end{equation}
where $\nabla V_{\text{eff}} = \hat{\phi}_1 \frac{\partial V_{\text{eff}}}{\partial \phi_1} + \hat{\phi}_2 \frac{\partial V_{\text{eff}}}{\partial \phi_2}$ and the curve $\mathcal{C}$ connects the origin to $(\phi_1^*, \phi_2^*)$. As a consequence of Green's theorem and the fact that the curl of a gradient is zero, the exact form of $\mathcal{C}$ does not matter. For simplicity, we choose $\mathcal{C}$ to be a straight line, which we parameterize as $\vec{s}(t) = ( \phi_1^* t ,\, \phi_2^* t)$ with $t \in [0,1]$. The expression for the effective potential becomes
\begin{equation}\label{explicitVTPD}
    V_{\text{eff}}^{\text{TPD}} = \int_0^1 dt \, \left( \phi_1^* \, \frac{\partial V_{\text{eff}}^{(1)}}{\partial \phi_1} \bigg|_{(\phi_1^* \, t,\, \phi_2^* \, t)} +  \phi_2^* \, \frac{\partial V_{\text{eff}}^{(1)}}{\partial \phi_2} \bigg|_{(\phi_1^* \, t, \, \phi_2^* \, t)} \right) \Bigg|_{(M_\pm,s_{2\Theta})} \,,
\end{equation}
where the unresummed effective potential appearing on the right-hand side is $V_{\text{eff}}^{(1)} = V_0 + V_{1\text{-loop}}$, with $V_0$ in \cref{V0} and $V_{1\text{-loop}}$ in \cref{V1}. 

In the high-temperature expansion, the first derivative with respect to $\phi_1$ is
\begin{equation}
\begin{split}
    \frac{\partial V_{\text{eff}}}{\partial \phi_1}\bigg|_{\text{high-}T} & = -\mu_1^2 \phi_1 + \lambda_1 \phi_1^3 + \frac{\lambda_{12}}{2} \phi_1 \phi_2^2 \\
    & + \left( \frac{T^2}{24} - \frac{T m_+}{8\pi} - \frac{L m_+^2}{32 \pi^2} \right) \frac{d m_+^2}{d\phi_1} + \left( \frac{T^2}{24} - \frac{T m_-}{8\pi} - \frac{L m_-^2}{32 \pi^2} \right) \frac{d m_-^2}{d\phi_1} \,,
\end{split}
\end{equation}
where
\begin{equation}
    \frac{dm_\pm^2}{d\phi_1} = \left[ 6 \lambda_1 \sin^2 \theta + \lambda_{12} \cos^2 \theta \right] \phi_1 \pm \lambda_{12} \sin 2 \theta \phi_2 \,.
\end{equation}
A similar expression holds for $\partial V_{\text{eff}}/\partial \phi_2$ with $1 \leftrightarrow 2$. Next we should replace any instances of $m_\pm^2$ and $\theta$ with the thermal quantities $M_\pm^2$ and $\Theta$. Finally, we evaluate at $(\phi_1^* t, \phi_2^* t)$, multiply by $\phi_{1,2}^*$ respectively, take the sum, and integrate over $0 \leq t \leq 1$, in accordance with \cref{explicitVTPD}.

Due to neglecting the momentum dependence of the self-energy insertions (see \cref{subsec:tadpole}), partial dressing fails to correctly reproduce two-loop sunset diagrams since it can not account for the case of overlapping loop momenta. While for the non-mixing case, one can easily correct for this by multiplying the $\mathcal{I}_2$-contribution to the gap equations by a factor $2/3$, the correction is more subtle in the case of mixing scalar fields. We discuss this in detail in \cref{app:sunset}. As already known for the non-mixing case in the literature, we also find this correction to be numerically of minor importance for the case of mixing scalars.

\subsection{Numerical comparison}
\label{sec:2field_mixing_compare}

For our numerical comparison, we choose the following parameter point (closely following the parameter point chosen in \cref{subsec:numcompare}):
\begin{align}
    \lambda_1 &= \lambda_2 = 1/3, \;\;\lambda_{12} = 2,\;\;\mu_{1}^2 = -4 \tev^2,\;\; \mu_2^2 = -1 \tev^2\;.
\end{align}
These parameters are chosen so that both fields can develop a non-zero vacuum expectation value.


\subsubsection*{Thermal masses}

\begin{figure}
    \centering
    \includegraphics[width=.65\textwidth,trim={.5cm .5cm .5cm .5cm},clip]{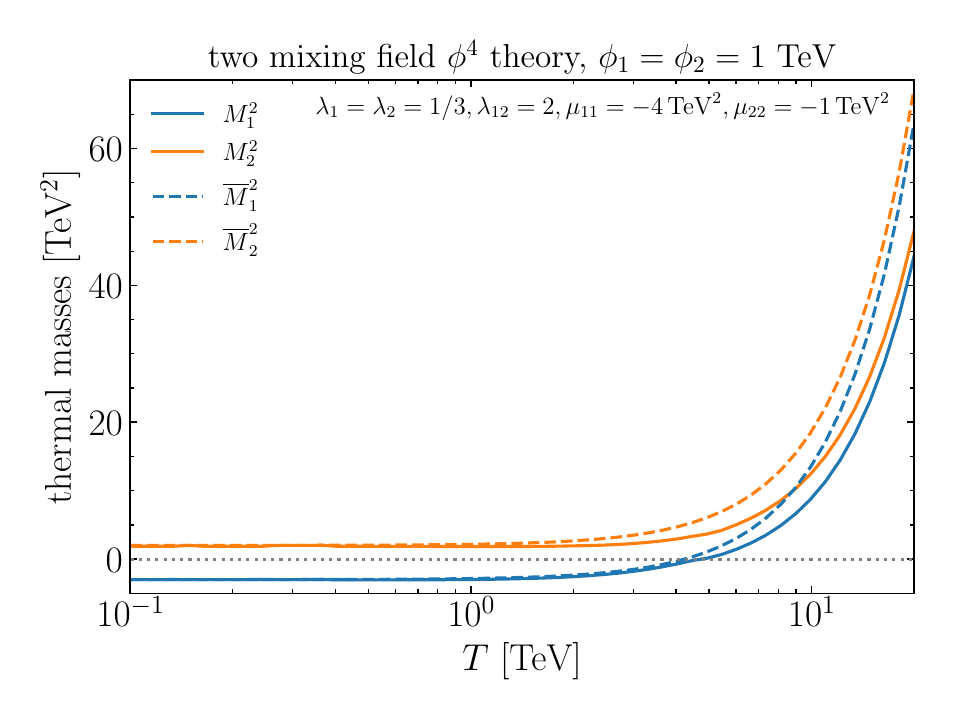}
    \caption{Thermal masses of $\phi_1$ and $\phi_2$ in the two-field $\phi^2$ theory as a function of the temperature calculated by solving the gap equations (solid) and the high-temperature expansion (dashed).}
    \label{fig:mass_iteration_2singlet}
\end{figure}

We start by investigating the thermal masses as a function of the temperature (see \cref{fig:mass_iteration_2singlet}) for $\phi_1 = \phi_2 = 1\tev$. While for low temperatures, the solutions of the gap equation agree well with the thermal masses in the high-temperature masses, a sizeable difference arises for temperatures close to the zero-temperature masses ($T\sim 5\tev$). For larger temperatures, the differences shrink again even though a visible difference remains. This behaviour is very similar to the results obtained in the EWSNR toy model (see \cref{sec:servant_model}). \cref{fig:mass_iteration_2singlet} again shows that $\beta_{1,2}$ are well below confirming the reliability of the partial dressing approximation.


\subsubsection*{Potential}

\begin{figure}
    \centering
    \includegraphics[width=.49\textwidth,trim={.5cm 0 .5cm 0},clip]{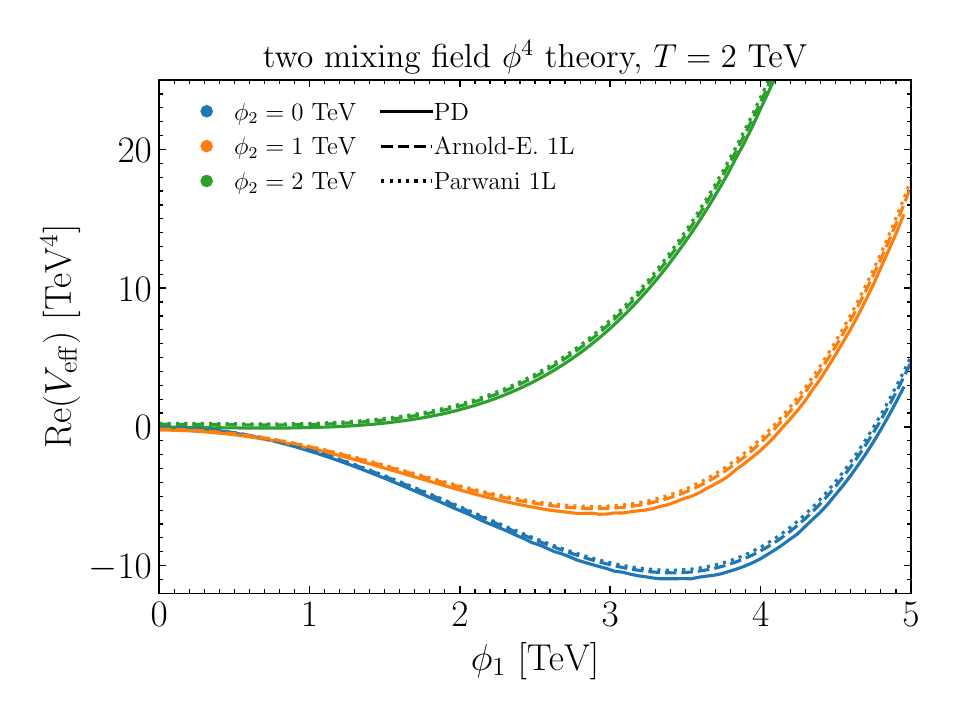}
    \includegraphics[width=.49\textwidth,trim={.5cm 0 .5cm 0},clip]{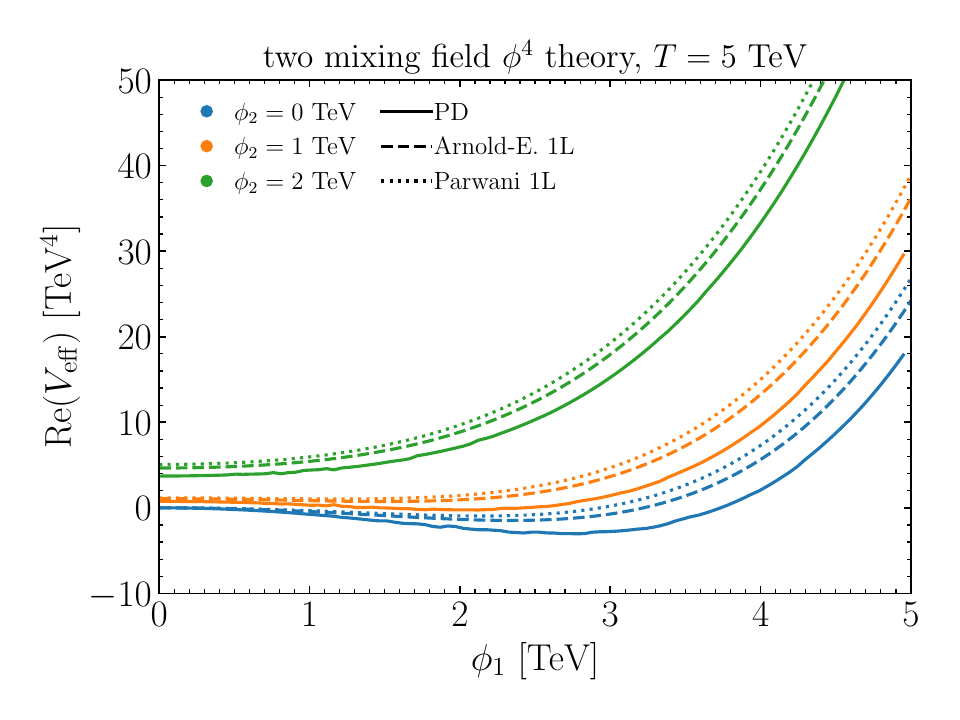}
    \includegraphics[width=.49\textwidth,trim={.5cm 0 .5cm 0},clip]{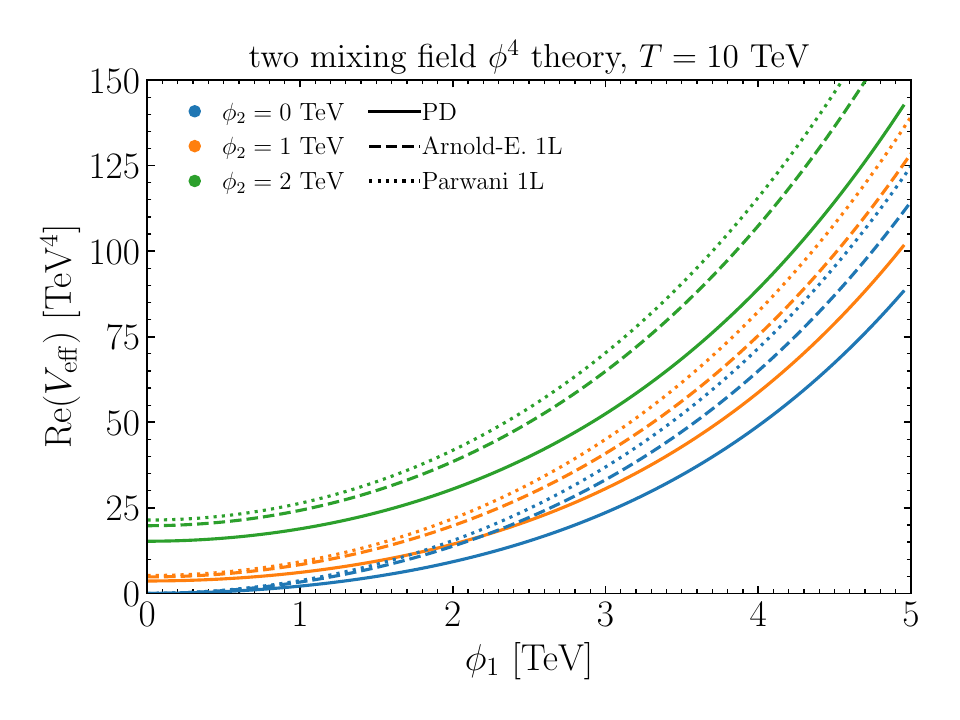}
    \caption{Real part of the effective potential in the two-field $\phi^4$ theory with mixing as a function of $\phi_1$ for different values of $\phi_2$. The temperature is chosen to be $2\tev$ (upper left), $5\gev$ (upper right), and $10\gev$ (bottom). Three different resummation methods are compared: partial dressing (solid), Arnold-Espinosa (dashed), and Parwani (dotted). The imaginary parts of the effective potential are negligible compared to the real part.}
    \label{fig:Veff_2singlet_mixing}
\end{figure}

Next, we investigate the real part of the effective potential itself in \cref{fig:Veff_2singlet_mixing}. For low temperatures (see upper left panel), partial dressing and Arnold-Espinosa/Parwani resummation yield very similar results regardless of the values for $\phi_1$ and $\phi_2$. This is expected since for low temperatures, the difference in the thermal masses and also the thermal corrections to the effective potential are small. For higher temperatures comparable to the zero-temperature masses of $\phi_1$ and $\phi_2$, visible differences between Arnold-Espinosa/Parwani resummation and partial dressing arise. In this regime, the high-temperature expansion for the calculation of the thermal masses is not a good approximation (see discussion above). Moreover, partial dressing includes the resummation of subleading thermal corrections, which give a sizeable contribution at the considered parameter point. This is evident from the fact that the two different Parwani and Arnold-Espinosa resummation show visible differences indicating the importance of subleading thermal corrections. For even larger temperatures (see bottom panel), the differences between the various resummation methods are further increased. While the thermal masses are in slightly better agreement, the subleading thermal corrections have a bigger impact resulting in an overall larger difference between the three resummation methods.

We finally note again that partial dressing fails to correctly reproduce two-loop sunset contributions. As investigated in detail in \cref{app:sunset}, the numerical impact of this effect is significantly smaller than the difference between partial dressing and Arnold-Espinosa/Parwani resummation.


\section{Conclusions}\label{sec:conclusions}

Accurate predictions for phase transitions are very important for the investigation of a wide range of physics phenomena. This necessitates a precise calculation of the effective potential at finite temperatures implying the need to resum large thermal corrections.

In the present work, we reviewed various resummation methods focusing on partial dressing as well as Arnold-Espinosa and Parwani resummation. Using a scalar toy model, we discussed at the one- and two-loop level that partial dressing is advantageous since it does not rely on the high-temperature expansion and also resums subleading thermal corrections.

As the next, we investigated EWSNR, for which large thermal corrections are expected, implying the need to include subleading corrections. While these are automatically included if using partial dressing, higher-loop corrections have to be explicitly calculated if using Arnold-Espinosa or Parwani resummation. We, moreover, demonstrated that the inclusion of two-loop corrections in the Arnold-Espinosa or Parwani approaches leads to unphysical kinks in the prediction for the effective potential. These kinks originate from the occurrence of large imaginary contributions to the effective potential caused by negative mass squares. While these contributions largely cancel in the partial dressing approach, the cancellation is incomplete in the Arnold-Espinosa or Parwani resummation approaches due to the inherent high-temperature expansion in parts of the calculation. 

The discussion of EWSNR concentrates on the case in which only one of the fields takes a non-zero value thereby implying the absence of mixing between the scalar fields. In the next step, we focused on the case of mixing scalar fields and showed how to promote the gap equation to a matrix equation and how to perform a path integration in the multi-dimensional field space. This allowed us to consistently implement partial dressing even for mixing fields, largely extending the applicability of this technique to a broader class of problems. Of special interest are many BSM extensions of the SM Higgs sector which are relevant for electroweak baryogenesis and the production of gravitational wave signals. 


\section*{Acknowledgments}
\sloppy{ 
We thank Thomas Biekötter, Pedro Bittar, Sven Heinemeyer, Seth Koren, Georg Weiglein, and an anonymous referee for useful discussions and remarks. Fermilab is operated by  Fermi  Research  Alliance, LLC under contract number DE-AC02-07CH11359 with the United States Department of Energy. M.C.\ and C.W.\ would like to thank the Aspen Center for Physics, which is supported by National Science Foundation grant No.~PHY-1607611, where part of this work has been done. C.W.\ has been partially supported by the U.S.~Department of Energy under contracts No.\ DEAC02- 06CH11357 at Argonne National Laboratory. The work of C.W.\ at the University of Chicago has also been supported by the DOE grant DE-SC0013642. H.B.\ acknowledges support by the Alexander von Humboldt foundation.}


\newpage
\appendix\label{sec:appendix}


\section{Thermal loop functions}
\label{sec:loop_funcs}

In this Appendix, we collect the various thermal loop functions used in this paper. We start with the bosonic thermal loop function appearing in the one-loop effective potential, which is given by
\begin{equation}
    J_B(y^2) = \int_0^\infty dx\, x^2 \ln \left( 1 - e^{- \sqrt{x^2 + y^2}} \right) \,.
\end{equation}
In the limits of small and large argument, this admits expansions
\begin{equation}\label{JBhighT}
    J_B(y^2 \ll 1) \simeq - \frac{\pi^4}{45} + \frac{\pi^2}{12} y^2 - \frac{\pi}{6} y^3 - \frac{1}{32} y^4 \log (y^2/a_B) + \mathcal{O}(y^6) \,,
\end{equation}
\begin{equation}\label{JBlowT}
    J_B(y^2 \gg 1) \simeq - \sum_{n=1} \frac{1}{n^2} y^2 K_2(yn) \,,
\end{equation}
where $a_B = 16 \pi^2 e^{3/2-2 \gamma_E}$ and $K_2$ is the modified Bessel function of the second kind.

In addition to the $J$ integral, we also need the one-loop vacuum integrals with up to three vertices. The one-vertex integral is defined by 
\begin{equation}\label{eq:Idef}
    \mathcal{I}[m] \equiv \mathcal{I}_1[m] \equiv \sumint_K \frac{1}{K^2 + m^2} \,.
\end{equation}
Its high-temperature expansion is given by
\begin{equation}\label{eq:IhighT}
    \mathcal{I}[m] \simeq \frac{1}{12} T^2 - \frac{1}{4\pi} m T - \frac{L_R}{16 \pi^2} m^2 + \frac{\zeta(3)}{128 \pi^4} \frac{m^4}{T^2} \,.
\end{equation}
Note that we have included the $\mathcal{O}(T^2)$ constant as well as kept terms up to $\mathcal{O}(1/T^2)$, since these lead to field-dependent contributions in $V_{2\text{-loop}}^A \sim \mathcal{I}[m]^2$ of $\mathcal{O}(T^0)$. There are also terms $\mathcal{O}(1/\epsilon)$ and $\mathcal{O}(\epsilon)$ in this expansion, which we omit for simplicity but which will nevertheless give a finite contribution to $V_{2\text{-loop}}^A \sim \mathcal{I}[m]^2$. See Ref.~\cite{ArnoldEspinosa:1993} for the complete expression. We also introduce the high-temperature expansion of the Arnold-Espinosa resummed $\mathcal{I}$-function, 
\begin{equation}\label{eq:AEI}
    \mathcal{I}_{\text{AE}}[m] \simeq \frac{1}{12} T^2 - \frac{1}{4\pi}\overline{M} T - \frac{L_R}{16\pi^2} m^2 + \frac{\zeta(3)}{128\pi^4} \frac{m^4}{T^2} \,.
\end{equation}

The vacuum integral with two propagators is defined via
\begin{equation}\label{eq:I2def}
    \mathcal{I}_2[m_1,m_2] \equiv \sumint_K \frac{1}{K^2 + m_1^2} \frac{1}{K^2 + m_2^2} \,.
\end{equation}
If both masses are different, we can write $\mathcal{I}_2$ in terms of $\mathcal{I}$ integrals
\begin{equation}\label{eq:I2red}
    \mathcal{I}_2[m_1,m_2] = \frac{\mathcal{I}[m_2] - \mathcal{I}[m_1]}{m_1^2 - m_2^2} \,.
\end{equation}
The high-temperature expansion follows immediately,
\begin{equation}\label{eq:I2highT}
    \mathcal{I}_2[m_1,m_2] \simeq \frac{1}{4\pi} \frac{T}{m_1 + m_2} + \ldots \,.
\end{equation}
We define the three-propagator vacuum integral via
\begin{equation}\label{eq:I3def}
    \mathcal{I}_3[m_1,m_2,m_3] \equiv \sumint_K \frac{1}{K^2 + m_1^2} \frac{1}{K^2 + m_2^2}\frac{1}{K^2 + m_3^2}\,.
\end{equation}
If all masses (or a subset) are different, we can write it in terms of $\mathcal{I}$ and $\mathcal{I}_2$ functions
\begin{align}\label{eq:I3red}
\mathcal{I}_3[m_1,m_2,m_3] ={}& \frac{\mathcal{I}[m_1]}{(m_1^2 - m_2^2)(m_1^2 - m_3^2)} - \frac{\mathcal{I}[m_2]}{(m_1^2 - m_2^2)(m_2^2 - m_3^2)} \nonumber\\
& + \frac{\mathcal{I}[m_3]}{(m_1^2 - m_3^2)(m_2^2 - m_3^2)}\,, \\
\mathcal{I}_3[m_1,m_1,m_2] ={}& - \frac{\mathcal{I}[m_1]}{(m_1^2 - m_2^2)^2} + \frac{\mathcal{I}[m_2]}{(m_1^2 - m_2^2)^2} - \frac{\mathcal{I}_2[m_1,m_1]}{m_1^2 - m_2^2}\,.
\end{align}
In addition to the one-loop integrals, also the two-loop bosonic sunset diagram appears. For three arbitrary masses --- $m_1$, $m_2$, $m_3$ ---, it is defined by
\begin{equation}
    \mathcal{H}[m_1, m_2, m_3] = \sumint_P \sumint_Q \frac{1}{(P^2 + m_1^2)(Q^2 + m_2^2)((P+Q)^2 + m_3^2)} \,.
\end{equation}
The high-temperature expansion of the bosonic sunset is rather involved and has been evaluated in \ccite{Ekstedt:2020},
\begin{equation}\label{HhighT}
\begin{split}
    \mathcal{H}[m_1, m_2, m_3] & \simeq \frac{T^2}{16 \pi^2} \left[ \ln \left( \frac{\mu_R}{m_1+m_2+m_3} \right) + \frac{1}{2} \right]\\
    & - \frac{T}{64 \pi^3} \left[\sum_{i=1,2,3} m_i \left( \ln \left( \frac{\mu_R^2}{4 m_i^2} \right) + L_R + 2 \right) \right]\\
    & - \frac{1}{256 \pi^4} \left[ \sum_{i=1,2,3} m_i^2 \left( L_R^2 + L_R - 2 \gamma_E^2 - 4 \gamma_1 + \frac{\pi^2}{4} + \frac{3}{2} \right) \right] + \ldots\,,
\end{split}
\end{equation}
with $\gamma_1 \simeq -0.0728$ the first Stieltjes constant.

We also need the bosonic sunset integral with one additional propagator, which we denote by $\widetilde{\mathcal{H}}$,
\begin{equation}
    \widetilde{\mathcal{H}}[m_1, m_2, m_3, m_4] = \sumint_P \sumint_Q \frac{1}{(P^2 + m_1^2)(P^2 + m_2^2)(Q^2 + m_3^2)((P+Q)^2 + m_4^2)}\,.
\end{equation}
If $m_1 \neq m_2$, it can be related to the normal sunset integral via
\begin{equation}
    \widetilde{\mathcal{H}}[m_1, m_2, m_3, m_4] = \frac{1}{m_1^2 - m_2^2}\left(\mathcal{H}[m_2, m_3, m_4] - \mathcal{H}[m_1, m_3, m_4]\right)
\end{equation}
For $m_1 = m_2$, it can be derived by a derivative of the normal sunset integral
\begin{equation}
\begin{split}
    \widetilde{\mathcal{H}}[m_1, m_1, m_3, m_4] &= - \frac{\partial}{\partial m_1^2}\mathcal{H}[m_1, m_3, m_4]
\end{split}
\end{equation}
The high-temperature expansion of $\widetilde{\mathcal{H}}$ is then given by
\begin{align}\label{HtildehighT}
    \widetilde{\mathcal{H}}[m_1, m_2, m_3, m_4] &= \frac{T^2}{16\pi^2}\frac{1}{m_1^2-m_2^2}\ln\left(\frac{m_1 + m_3 + m_4}{m_2 + m_3 + m_4}\right) + \ldots \,, \\
    \widetilde{\mathcal{H}}[m_1, m_1, m_3, m_4] &= \frac{T^2}{32\pi^2} \frac{1}{m_1}\frac{1}{m_1 + m_3 + m_4} + \ldots \,.
\end{align}


\section{Three-loop cross-check of partial dressing with mixing fields}

In this Appendix, we explicitly cross-check up to the three-loop level that partial dressing correctly takes into account all self-energy-like loop insertions in the presence of mixing (i.e., daisy-chain corrections).

To better separate the loop corrections from tree-level mixing, we work in a special basis. First, we expand the fields around the point $(\hat\phi_1,\hat\phi_2)$ at which we want to evaluate the effective potential,
\begin{align}
    \phi_i = \hat\phi_i + \tilde\phi_i
\end{align}
where $\tilde\phi_i$ are the new dynamical degrees of freedom, which are eventually set to zero. We then rotate to tree-level mass eigenstates,
\begin{align}\label{forref}
    \begin{pmatrix}\tilde\phi_1 \\ \tilde\phi_2\end{pmatrix}
    = \mathcal{R}_\alpha
    \begin{pmatrix}h_1 \\ h_2\end{pmatrix}\,.
\end{align}
The resulting potential has the form
\begin{align}
    V(h_1, h_2) ={}& \frac{1}{2}m_1 h_1^2 + \frac{1}{2} m_2 h_2^2 \nonumber\\
    & + \frac{1}{6} A_{111} h_1^3 + \frac{1}{2} A_{112} h_1^2 h_2 + \frac{1}{2} A_{122} h_1 h_2^2  + \frac{1}{6} A_{222} h_2^3 \nonumber\\
    & + \frac{1}{24}\lambda_{1111}h_1^4 + \frac{1}{6}\lambda_{1112}h_1^3 h_2 + \frac{1}{4}\lambda_{1122}h_1^2 h_2^2 \nonumber\\
    & + \frac{1}{6}\lambda_{1222}h_1 h_2^3 + \frac{1}{24}\lambda_{2222} h_2^4 \,,
\end{align}
where the parameters $A_{ijk}$ and $\lambda_{ijkl}$ are given in terms of the original parameters $\lambda_{1,12,2}$, $(\hat\phi_1,\hat\phi_2)$, and the mixing angle $\alpha$. Even though no tree-level mixing exists in this basis, mixing is reintroduced at the loop level via self-energy corrections. 


\subsection{Feynman-diagrammatic approach}

\begin{figure}
    \centering
    \includegraphics[width=.7\textwidth]{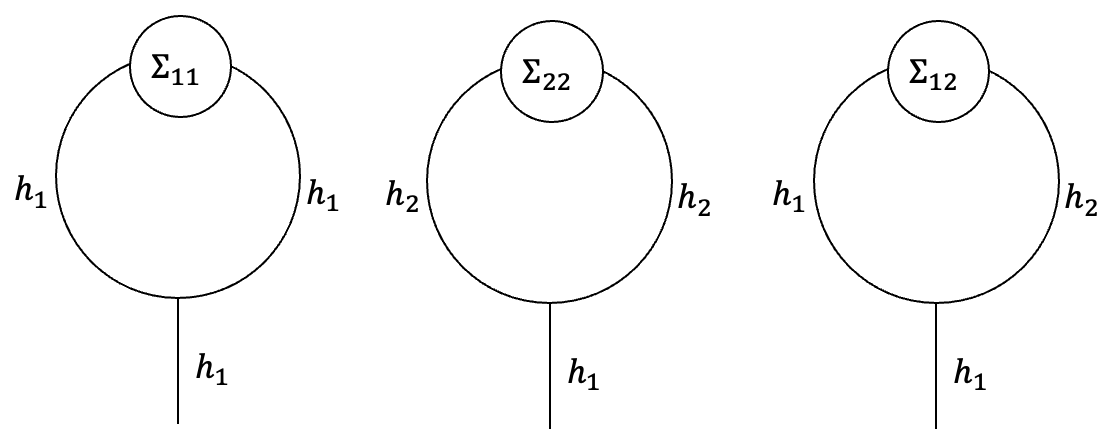}
    \includegraphics[width=1\textwidth]{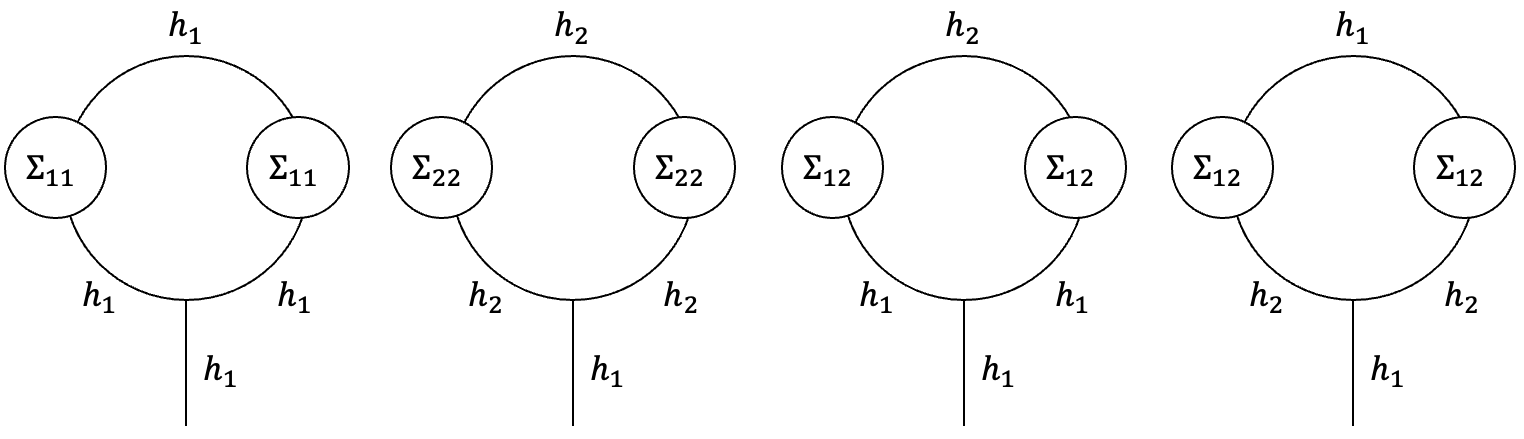}
    \caption{Daisy-chain-like self-energy corrections to the $h_1$ tadpole. One self energy is inserted in the upper row; two, in the lower row.}
    \label{fig:tadpole_daisy}
\end{figure}

Working in the tree-level mass eigenstate basis, we can calculate the daisy-chain-like loop corrections to the $h_1$ tadpoles explicitly using a Feynman-diagrammatic (FD) approach. They are given by
\begin{align}\label{eq:PD_crosscheck_TFD1}
    T_\text{FD}^{(1)} = \frac{1}{2} \left[A_{111} \mathcal{I}(m_1) + A_{122} \mathcal{I}(m_2) \right]
\end{align}
at the one-loop order,
\begin{align}\label{eq:PD_crosscheck_TFD2}
    T_\text{FD}^{(2)} = - \frac{1}{2} \left[A_{111} \Sigma_{11}^{(1)} \mathcal{I}_2(m_1,m_1) + A_{122} \Sigma_{22}^{(1)} \mathcal{I}_2(m_2,m_2) + 2 A_{112}\Sigma_{12}^{(1)} \mathcal{I}_2(m_1,m_2) \right]
\end{align}
at the two-loop order (see upper row of \cref{fig:tadpole_daisy}), and by
\begin{align}\label{eq:PD_crosscheck_TFD3}
    T_\text{FD}^{(3)} ={} \frac{1}{2} \bigg\{& A_{111} \bigg[\Sigma_{11}^{(2)} \mathcal{I}_2(m_1,m_1) + \left(\Sigma_{11}^{(1)}\right)^2 \mathcal{I}_3(m_1,m_1,m_1) \nonumber\\
    &\hspace{1.1cm} + \left(\Sigma_{12}^{(1)}\right)^2 \mathcal{I}_3(m_1,m_1,m_2)\bigg] \nonumber\\
    & +A_{122} \bigg[\Sigma_{22}^{(2)} \mathcal{I}_2(m_2,m_2) + \left(\Sigma_{22}^{(1)}\right)^2 \mathcal{I}_3(m_2,m_2,m_2) \nonumber\\
    &\hspace{1.5cm} + \left(\Sigma_{12}^{(1)}\right)^2 \mathcal{I}_3(m_1,m_2,m_2)\bigg] \nonumber\\
    & + 2 A_{112} \bigg[\Sigma_{12}^{(2)} \mathcal{I}_2(m_1,m_2) + \Sigma_{11}^{(1)}\Sigma_{12}^{(1)} \mathcal{I}_3(m_1,m_1,m_2) \nonumber\\
    &\hspace{1.7cm}+ \Sigma_{12}^{(1)}\Sigma_{22}^{(1)} \mathcal{I}_3(m_1,m_2,m_2)\bigg]\bigg\}
\end{align}
at the three-loop order (see upper and lower row of \cref{fig:tadpole_daisy}). Here, we neglected the momentum dependence of the (renormalized) self energies,\footnote{See discussion in \cref{app:sunset}.}
\begin{align}
    \Sigma_{ij} = \Sigma_{ij}(p^2 = 0)\,.
\end{align}
The superscripts are used to denote the loop order of the respective self-energy.


\subsection{Partial dressing}

For partial dressing, the first step is to determine the loop-corrected masses $M_{1,2}$ and the mixing angle $\beta$ relating the loop-corrected mass eigenstates (called $H_{1,2}$ in the following) to the tree-level mass eigenstates $h_{1,2}$. These can be obtained by solving the gap equations,
\begin{align}
    M_{1,2}^2 &= \frac{1}{2}\left(m_1^2 - \Sigma_{11} + m_2^2 - \Sigma_{22} \mp \sqrt{(m_1^2 - \Sigma_{11} - m_2^2 + \Sigma_{22})^2 + 4 \Sigma_{12}^2}\right), \label{eq:PD_crosscheck_M12}\\
    s_{2\alpha} &= \frac{2 \Sigma_{12}}{\sqrt{(m_1^2 - \Sigma_{11} - m_2^2 + \Sigma_{22})^2 + 4 \Sigma_{12}}} \label{eq:PD_crosscheck_S2A}
\end{align}
where here we again neglect the momentum dependence of the (renormalized) self energies,
\begin{align}
    \Sigma_{ij} = \Sigma_{ij}(p^2 = 0)\,.
\end{align}
Furthermore, we also do not take the intrinsic dependence of the self energies on the masses and the mixing angle into account. This dependence generates super-daisy contributions, which we do not consider in the discussion here.

The one-loop tadpole diagrams with inserted loop-corrected mass and mixing angle are then given by
\begin{align}\label{eq:PD_crosscheck_TPD}
    T_\text{PD}^{(1)} ={}& \frac{1}{2} \left[A_{h_1H_1H_1} \mathcal{I}(M_1) + A_{h_1H_2H_2} \mathcal{I}(M_2) \right] = \\
    ={}&\frac{1}{2}(c_\alpha^2 A_{111} + s_{2\alpha} A_{112} + s_\alpha^2 A_{122}) \mathcal{I}(M_1) \nonumber\\
    &+ \frac{1}{2}(c_\alpha^2 A_{122} - s_{2\alpha} A_{112} + s_\alpha^2 A_{111}) \mathcal{I}(M_2)\,.
\end{align}
In the last step, we rewrote the trilinear $(h_1,H_i,H_i)$ couplings in terms of the original $A_{ijk}$ couplings.

Expanding the expressions for the loop-corrected masses and mixing angle (see \cref{eq:PD_crosscheck_M12,eq:PD_crosscheck_S2A} up to the two-loop level, we obtain,
\begin{align}
    M_1^2 &= m_1^2 - \Sigma^{(1)}_{11} - \Sigma^{(2)}_{11} + \frac{\left(\Sigma^{(1)}_{12}\right)^2}{m_1^2 - m_2^2} + \ldots\,, \\
    M_2^2 &= m_2^2 - \Sigma^{(1)}_{22} - \Sigma^{(2)}_{22} + \frac{\left(\Sigma^{(1)}_{12}\right)^2}{m_2^2 - m_1^2} + \ldots\,, \\
    \alpha &= \frac{\Sigma^{(1)}_{12}}{m_2^2 - m_1^2} + \frac{\Sigma^{(2)}_{12}}{m_2^2 - m_1^2} + \frac{\Sigma^{(1)}_{12}\left(\Sigma^{(1)}_{22} - \Sigma^{(1)}_{11}\right)}{(m_1^2 - m_2^2)^2} + \ldots\,.
\end{align}
Since there is no mixing at the tree level, the mixing angle $\alpha$ is zero at the tree level.

Inserting these loop expansions into \cref{eq:PD_crosscheck_TPD}, we recover \cref{eq:PD_crosscheck_TFD1,eq:PD_crosscheck_TFD2,eq:PD_crosscheck_TFD3} after applying the recursion relations in \cref{eq:I2red,eq:I3red}. This explicitly confirms that self-energy insertions are correctly resumed up to the three-loop level. This cross-check can easily be extended to higher-loop order.


\section{Partial dressing and overlapping momenta}
\label{app:sunset}

As already noted in \ccite{Boyd:1993,Curtin:2016}, partial dressing miscounts two-loop sunset diagrams. For $N$ non-mixing scalar fields, this type of diagram is proportional to $\lambda^3\phi_1^2 N \frac{T^2}{m^2}$ and miscounted by a factor $3/2$. 

The reason for this miscounting is that partial dressing ignores the momentum dependence of the self-energy insertions. While there is no momentum dependence for diagrams involving quartic scalar couplings, this is not true for diagrams with triple scalar couplings forming a sunset diagram. For this type of diagram, the momenta of the loop integrals overlap resulting in the miscounting. In the non-mixing case, this miscounting can be fixed easily by inserting a factor of $2/3$ into the solution of the gap equation,
\begin{align}
    M^2 = m^2 + \frac{1}{4}\lambda T^2 - \frac{3}{4\pi} \lambda T M - \frac{3}{16\pi^2}\lambda M^2 L - \zeta \left[\frac{9}{4\pi}\lambda^2\phi^2 \frac{T}{M} + \frac{9}{8\pi^2}\lambda^2\phi^2\right] + \ldots\,,
\end{align}
where $\zeta = 2/3$.

\begin{figure}
    \centering
    \includegraphics[width=.2\textwidth]{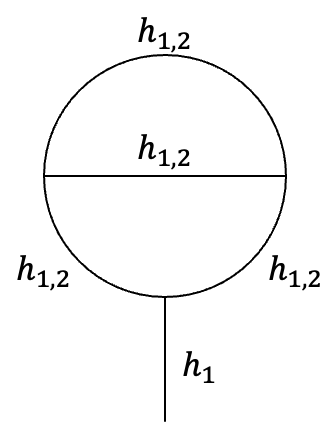}
    \caption{sunset corrections to the $h_1$ tadpole.}
    \label{fig:tadpole_sunset}
\end{figure}

In the presence of mixing scalar fields, this fix is slightly more complicated. The general sunset contribution to the $h_1$ tadpole (see \cref{fig:tadpole_sunset}) is given by (working again in the tree-level mass eigenstate basis)
\begin{align}
    T^{(2)}_\text{FD,sunset} ={}& \frac{1}{4}A_{111}\bigg[A_{111}^2\widetilde{\mathcal{H}}(m_1,m_1,m_1,m_1) + A_{122}^2\widetilde{\mathcal{H}}(m_1,m_1,m_2,m_2) \nonumber\\
    &\hspace{1.4cm} + 2 A_{112}^2\widetilde{\mathcal{H}}(m_1,m_1,m_1,m_2)\bigg] \nonumber\\
    & + \frac{1}{4}A_{122}\bigg[A_{222}^2\widetilde{\mathcal{H}}(m_2,m_2,m_2,m_2) + A_{112}^2\widetilde{\mathcal{H}}(m_2,m_2,m_1,m_1) \nonumber\\
    &\hspace{1.8cm} + 2 A_{122}^2\widetilde{\mathcal{H}}(m_2,m_2,m_1,m_2)\bigg]  \nonumber\\
    & + \frac{1}{2}A_{112}\bigg[A_{111} A_{112}\widetilde{\mathcal{H}}(m_1,m_2,m_1,m_1) + A_{122}A_{222}\widetilde{\mathcal{H}}(m_1,m_2,m_2,m_2) \nonumber\\
    &\hspace{1.8cm} + 2 A_{112}A_{122}\widetilde{\mathcal{H}}(m_1,m_2,m_1,m_2)\bigg]  \overset{\text{high-}T}{\simeq} \nonumber\\
    \simeq{}& \frac{1}{48\pi^2}\bigg[\frac{1}{6}A_{111}^3 \frac{T^2}{m_1^2}+ \frac{1}{2}A_{111} A_{122}^2 \frac{T^2}{m_1(m_1 + 2m_2)} + A_{111} A_{112}^2 \frac{T^2}{m_1(2m_1+m_2)} \nonumber\\
    & \hspace{1.1cm}+ \frac{1}{6}A_{122}A_{222}^2 \frac{T^2}{m_2^2} + \frac{1}{2}A_{122} A_{112}^2 \frac{T^2}{m_2(2m_1+m_2)} \nonumber\\
    &\hspace{1.1cm} + A_{122}^3 \frac{T^2}{m_2(m_1+2m_2)} + 2 A_{111} A_{112}^2 \frac{T^2}{m_1^2-m_2^2}\ln\frac{3m_1}{2m_1+m_2} \nonumber\\
    &\hspace{1.1cm} + 4 A_{112}^2 A_{122} \frac{T^2}{m_1^2-m_2^2}\ln\frac{2m_1 + m_2}{m_1+2m_2} \nonumber\\
    &\hspace{1.1cm} + 2 A_{112}A_{122} A_{222} \frac{T^2}{m_1^2-m_2^2}\ln\frac{m_1 + 2 m_2}{3m_2}\bigg] + \ldots\,.
\end{align}
Using partial dressing, this contribution is generated by one self-energy insertion into the one-loop tadpole (see upper row of \cref{fig:tadpole_daisy}). More specifically, it is generated by the $\mathcal{I}_2$ contributions to the self-energies.\footnote{The $\mathcal{I}$ contributions generate the daisy-chain-like corrections (see discussion above).} Since the momentum dependence of the self-energy insertions is neglected, the sunset-type contribution is given by
\begin{align}
    T^{(2)}_\text{PD,sunset} ={}& \frac{1}{2}\bigg[A_{111}\mathcal{I}_2(m_1,m_1)\Sigma_{11} + A_{122}\mathcal{I}_2(m_2,m_2)\Sigma_{22} + 2 A_{112}\mathcal{I}_2(m_1,m_2)\Sigma_{12}\bigg]\overset{\mathcal{I}_2^2\;\text{contr.}} {=}\nonumber \\
    ={}& \frac{1}{4}A_{111}\bigg[A_{111}^2(\mathcal{I}_2(m_1,m_1))^2 + A_{122}^2\mathcal{I}_2(m_1,m_1)\mathcal{I}_2(m_2,m_2) \nonumber\\
    &\hspace{1.4cm} + 2 A_{112}^2\mathcal{I}_2(m_1,m_1)\mathcal{I}_2(m_1,m_2)\bigg] \nonumber\\
    & + \frac{1}{4}A_{122}\bigg[A_{222}^2(\mathcal{I}_2(m_2,m_2))^2 + A_{112}^2\mathcal{I}_2(m_2,m_2)\mathcal{I}_2(m_1,m_1) \nonumber\\
    &\hspace{1.8cm} + 2 A_{122}^2\mathcal{I}_2(m_2,m_2)\mathcal{I}_2(m_1,m_2)\bigg] \nonumber\\
    & + \frac{1}{2}A_{112}\bigg[A_{111}A_{112}\mathcal{I}_2(m_1,m_2)\mathcal{I}_2(m_1,m_1) \nonumber\\
    &\hspace{1.8cm} + A_{122}A_{222}\mathcal{I}_2(m_1,m_2)\mathcal{I}_2(m_2,m_2) \nonumber\\
    &\hspace{1.8cm} + 2 A_{112}A_{122}\mathcal{I}_2(m_1,m_2)\mathcal{I}_2(m_1,m_2)\bigg]\overset{\text{high-}T}{\simeq} \nonumber\\
    \simeq{}& \frac{1}{48\pi^2}\bigg[\frac{1}{4}A_{111}^3 \frac{T^2}{m_1^2} + \frac{1}{4}A_{111} A_{122}^2 \frac{T^2}{m_1 m_2} + A_{111} A_{112}^2 \frac{T^2}{m_1 (m_1+m_2)} \nonumber\\
    & \hspace{1.1cm} + \frac{1}{4}A_{122}A_{222}^2 \frac{T^2}{m_2^2}  + \frac{1}{4}A_{112}^2 A_{122} \frac{T^2}{m_1 m_2} + A_{122}^3 \frac{T^2}{m_2(m_1+m_2)}\nonumber\\
    &\hspace{1.1cm} + A_{111}A_{112}^2 \frac{T^2}{m_1(m_1+m_2)} + A_{112}A_{122}A_{222} \frac{T^2}{m_2(m_1+m_2)}\nonumber\\
    &\hspace{1.1cm} + 4 A_{112}^2 A_{122} \frac{T^2}{(m_1+m_2)^2}\bigg] + \ldots\,.
\end{align}
It is clear that only for those contributions for which all internal masses are equal, the correction factor $2/3$ is valid. For the general case with multiple masses, each contribution has to be adjusted individually. One possibility to achieve this is to replace the $\mathcal{I}_2$ loop functions in the self-energies by ratios of $\widetilde{\mathcal{H}}$ and $\mathcal{I}_2$ functions. For example, we can replace
\begin{align}
    \Sigma_{11}\Big|_{\mathcal{I}_2\text{-contr.}} = \frac{1}{2}\left[A_{111}^2\mathcal{I}_2(m_1,m_1) + A_{122}^2\mathcal{I}_2(m_2,m_2)+ 2 A_{112}^2\mathcal{I}_2(m_1,m_2)\right]
\end{align}
by 
\begin{align}
    \widetilde\Sigma_{11}\Big|_{\mathcal{I}_2\text{-contr.}} ={}& \frac{1}{2\mathcal{I}_2(m_1,m_1)}\bigg[A_{111}^2\widetilde{\mathcal{H}}(m_1,m_1,m_1,m_1) + A_{122}^2\widetilde{\mathcal{H}}(m_1,m_1,m_2,m_2) \nonumber\\
    &\hspace{2.6cm} + 2 A_{112}^2\widetilde{\mathcal{H}}(m_1,m_1,m_1,m_2)\bigg]\,.
\end{align}
The analogous replacements have to be done for $\Sigma_{12}$ and $\Sigma_{22}$.

\begin{figure}
    \centering
    \includegraphics[width=.49\textwidth,trim={.5cm 0 .5cm 0},clip]{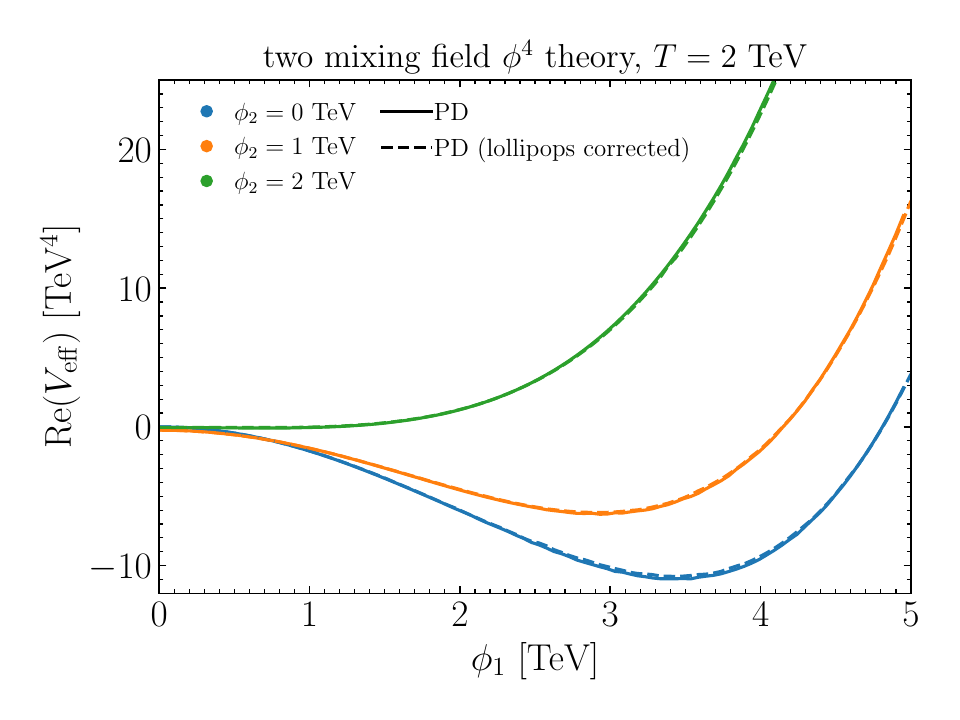}
    \includegraphics[width=.49\textwidth,trim={.5cm 0 .5cm 0},clip]{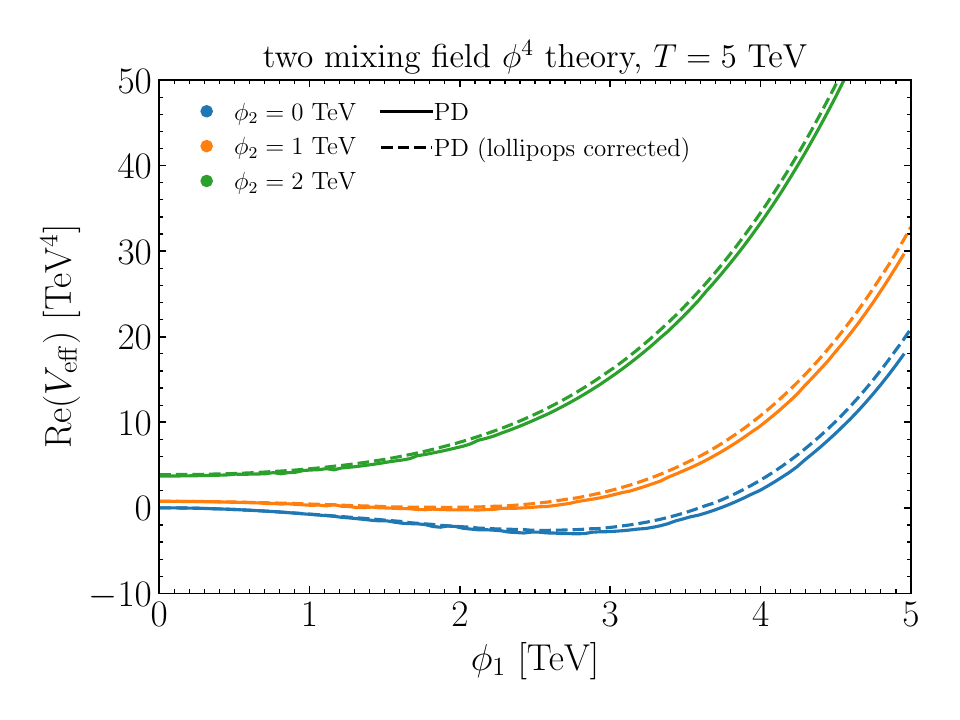}
    \includegraphics[width=.49\textwidth,trim={.5cm 0 .5cm 0},clip]{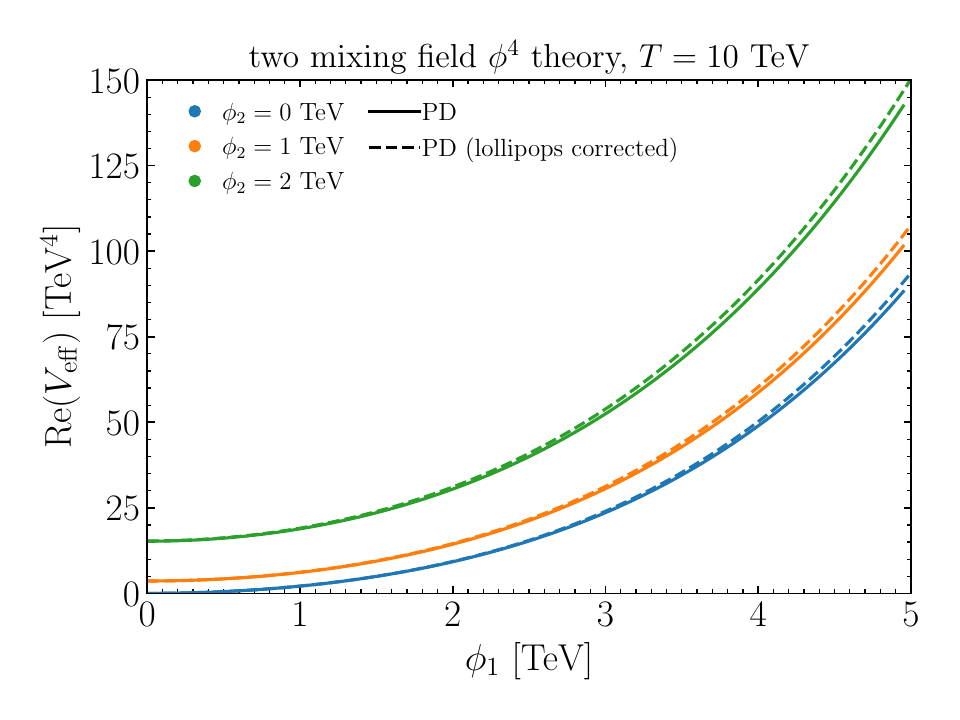}
    \caption{Same as \cref{fig:Veff_2singlet_mixing} but the partial dressing results without (solid) and with corrected sunset diagrams (dashed) are shown.}
    \label{fig:Veff_2singlet_sunset}
\end{figure}

We investigate the numerical impact of this correction in \cref{fig:Veff_2singlet_sunset}. Choosing the same benchmark point as in \cref{sec:2field_mixing_compare}, we compare the partial dressing result with (dashed) and without (solid) correcting the sunset contribution. We find the difference between the two results to be comparably smaller (i.e., significantly smaller than the difference between partial dressing and Arnold-Espinosa/Parwani resummation). For low temperatures and low field values, the difference is hardly visible, while for higher temperatures and field values a small difference occurs. Correcting the sunset contributions shifts the effective potential upwards slightly reducing the difference to Arnold-Espinosa/Parwani resummation (see \cref{fig:Veff_2singlet_mixing}).

\clearpage
\printbibliography

\end{document}

%% file: abbr.tex
\newcommand{\gev}{\;\text{GeV}\xspace}
\newcommand{\tev}{\;\text{TeV}\xspace}

\newcommand{\MS}{\ensuremath{\overline{\text{MS}}}\xspace}

\renewcommand{\order}[1]{\ensuremath{{\cal O}(#1)}}

\NewBibliographyString{refname}
\NewBibliographyString{refsname}
\DefineBibliographyStrings{english}{%
  refname = {Ref\adddot},
  refsname = {Refs\adddot}
}
\DeclareCiteCommand{\ccite}
  {%
  \ifnum\thecitetotal=1
    \bibstring{refname}%
  \else%
    \bibstring{refsname}%
  \fi%
  \addnbspace\bibopenbracket%
  \usebibmacro{cite:init}%
   \usebibmacro{prenote}}
  {\usebibmacro{citeindex}%
   \usebibmacro{cite:comp}}
  {}
  {\usebibmacro{cite:dump}%
   \usebibmacro{postnote}%
   \bibclosebracket}
\newrobustcmd*{\Ccite}{\bibsentence\ccite}